\documentclass[12pt]{article}

\usepackage[left=1in,top=1in,right=1in,bottom=1in,head=.1in,nofoot]{geometry}
\usepackage{setspace,url,bm,amsmath} % For double-spacing, URL font, math symbols
\usepackage{titlesec} % Section header formatting
\titlelabel{\thetitle.\quad} % Section header formatting
\usepackage[margin=20pt]{subcaption}
\usepackage{amsmath,amsfonts,amsthm,amssymb}
\usepackage{graphicx}
\usepackage[colorlinks=true,linkcolor=black,citecolor=blue,urlcolor=blue]{hyperref}
\usepackage[table]{xcolor}
\usepackage{multirow}
\usepackage{array}
\usepackage{booktabs}
\usepackage{threeparttable}
\usepackage{comment}
\usepackage{float}
\usepackage{natbib}
\usepackage{indentfirst}

\newtheorem{assumption}{Assumption}
\newtheorem*{theorem*}{Theorem}
\newtheorem{theorem}{Theorem}
\newtheorem{proposition}{Proposition}
\newtheorem{lemma}{Lemma}
\newtheorem{remark}{Remark}

\newtheorem{definition}{Definition}
\newtheorem{corollary}{Corollary}
\newtheorem*{corollary*}{Corollary}

\def\T{\text{T}}
\def\Var{\text{Var}}

\def\adj{\text{adj}}
\def\ancova{\text{ancova}}

\def\fe{\text{adj}}
\def\taufe{\hat \tau_\text{adj}}
\def\diff{\text{diff}}
\def\str{\text{interact}}
\def\taustr{\hat \tau_{\text{interact}} }
\def\simp{\text{adj}}
\def\tausimp{\hat \tau^*_{\text{}} }
\def\ancova{\text{adj}}
\def\tauancova{\hat \tau^*_{\text{adj}} }

\def\es{\text{es}}
\def\essed{\text{es2}}
\def\inter{\text{interact}}
\def\tauinter{\hat \tau^*_{\text{interact}}}

\def\bz{\bm{Z}}
\def\sumi{\sum_{i=1}^{n}}
\def\sumk{\sum_{k=1}^{K} }
\def\sumik{\sum_{i \in [k]}}
\def\nkt{n_{[k]1}}
\def\nkc{n_{[k]0}}
\def\pik{\pi_{[k]}}

\def\pnk{p_{n[k]}}

\def\tauk{ \tau_{[k]}}
\def\taukhat{\hat \tau_{[k]}}
\def\pk{p_{[k]}}
\def\qk{q_{[k]}}
\def\nk{n_{[k]}}

\def\nt{n_1}
\def\nc{n_0}

\def\YkThat{\bar{Y}_{[k]1}}
\def\YkChat{\bar{Y}_{[k]0}}
\def\Ykbar{\bar{Y}_{[k]}}
\def\Xkbar{\bar{\bx}_{[k]}}

\def\YT{\bar{Y}_1}
\def\YC{\bar{Y}_0}
\def\XT{\bar{\bm{X}_1}}
\def\XC{\bar{\bm{X}_0}}
\def\bx{\bm{X}}
\def\XkT{\bar{\bm{X}}_{[k]}}
\def\XkC{\bar{\bm{X}}_{[k]}}
\def\XkThat{\bar{\bm{X}}_{[k]1}}
\def\XkChat{\bar{\bm{X}}_{[k]0}}

\DeclareMathOperator*{\argmin}{arg\,min}

\begin{document}
\begin{singlespace}
%\title{\bf Impact of Misclassification on Covariate-Adaptive Randomization: Design and Inference Aspects}
\title{\bf Regression analysis for covariate-adaptive randomization: A robust and efficient inference perspective}

\author{
\small
{
Wei Ma$^{1}$, Fuyi Tu$^{1}$, Hanzhong Liu$^{2}$\thanks{\small{Correspondence: \texttt{lhz2016@tsinghua.edu.cn}}}
}
\\ \\
{\small $^{1}$ Institute of Statistics and Big Data, Renmin University of China, Beijing, China}\\
{\small $^{2}$ Center for Statistical Science, Department of Industrial Engineering, Tsinghua University, Beijing, China}
}

\date{}
\maketitle
\end{singlespace}

\thispagestyle{empty}
\vskip -8mm 

\begin{singlespace}
\begin{abstract}

Linear regression is arguably the most fundamental statistical model; however, the validity of its use in randomized clinical trials, despite being common practice, has never been crystal clear, particularly when stratified or covariate-adaptive randomization is used. In this paper, we investigate several of the most intuitive and commonly used regression models for estimating and inferring the treatment effect in randomized clinical trials. By allowing the regression model to be arbitrarily misspecified, we demonstrate that all these regression-based estimators robustly estimate the treatment effect, albeit with possibly different efficiency. We also propose consistent non-parametric variance estimators and compare their performances to those of the model-based variance estimators that are readily available in standard statistical software. Based on the results and taking into account both theoretical efficiency and practical feasibility, we make recommendations for the effective use of regression  under various scenarios. For equal allocation, it suffices to use the regression adjustment for the stratum covariates and additional baseline covariates, if available, with the usual ordinary-least-squares variance estimator. For unequal allocation, regression with treatment-by-covariate interactions should be used, together with our proposed variance estimators. These recommendations apply to simple and stratified randomization, and minimization, among others. We hope this work helps to clarify and promote the usage of regression in randomized clinical trials.

\vspace{12pt}
\end{abstract}

\end{singlespace}

\newpage

\clearpage
\setcounter{page}{1}

\allowdisplaybreaks
%\subsection{Introduction}
\baselineskip=24pt

\begin{singlespace}

\section{Introduction}\label{sec1}
The use of linear regression models is standard practice for estimating the treatment effect in randomized clinical trials, regardless of whether simple randomization, a covariate-adaptive allocation, or other approaches are used. However, concerns have been raised regarding the validity of the resulting inferences, because the ``usual'' assumptions for linear regression, such as linearity, normality, and homoskedasticity, may not be fulfilled, and certainly cannot be verified during the development of a statistical analysis plan prior to initiation of a trial.
Consequently, the use of regression in randomized clinical trials must be justified by demonstrating that valid inference can be obtained even when the regression model is arbitrarily misspecified. Although this is not the case in general (e.g., in observational studies), regression does exhibit the desired robustness thanks to randomization.

The robustness of regression is evident under simple randomization. The seminal work of \cite{Yang2001} examined three commonly used regression models, namely, simple regression, which is equivalent to the difference-in-means estimator, and two regression models that adjust for covariates with and without covariate-by-treatment interactions. The authors proved that all three ordinary-least-squares (OLS) estimators are consistent and they argued that the regression with covariate-by-treatment interactions produces the most efficient estimator. From a practical standpoint, \citet{Wang2019} pointed out that the OLS variance estimator is consistent under equal allocation for regression without interactions, which supports the use of  standard statistical software (e.g., €œlm in R or proc reg in SAS) for constructing confidence intervals and tests.

% Some subsequent work \citep{Leon2003, Tsiatis2008} has viewed the two regression estimators as augmenting the difference in-means-estimator by covariate adjustment following the spirit in survey sampling literature \citep[e.g.,][]{Cassel1976, Sarndal2003}.

Similar results have also been obtained under the ``finite population'' framework. \citet{Freedman2008}, being apparently unaware of the work by \cite{Yang2001}, criticized regression because the adjustment (without interactions) can actually reduce the asymptotic precision compared with that obtained using the difference-in-means estimator. The issue was later addressed in the influential paper by \citet{lin2013}: the regression with interactions cannot hurt, and often improves, the asymptotic precision. \cite{lin2013} also showed that the Huber--White  variance estimator is either consistent or asymptotically conservative.

Compared with simple randomization, the robustness of regression in stratified randomization, or more generally, covariate-adaptive randomization, appears to be more elusive, despite its extensive use in clinical trials. Covariate-adaptive randomization aims to balance treatment allocation among covariates. For example, stratified randomization defines a set of strata based on one or more covariates and performs a separate randomization, most commonly blocked randomization, within each stratum. Minimization was proposed to achieve balance over covariates€™ margins \citep{Taves1974, Pocock1975}. This approach has been generalized for use in simultaneously controlling various types of imbalances (within-stratum, within-covariate-margin, and overall) \citep{Hu2012, Hu2020}.
For a more comprehensive review, see, for example, \cite{Hu2014} and \cite{Rosenberger2015}. The use of covariate-adaptive randomization predominates in clinical trials. According to a recent survey of 224 randomized trials published in leading medical journals in 2014, over 80\% of them had used covariate-adaptive randomization \citep{Lin2015}.

The past decade has witnessed significant advances in inference under covariate-adaptive randomization. Various valid tests have been proposed using different randomization methods and model assumptions \citep{Shao2010, Shao2013, Ma2015, Ma2019, Wang2020}. However, one drawback of most of these studies is that the data generation model must be correctly specified. More recently, the topic of robust inference under covariate-adaptive randomization has drawn much attention \citep[e.g.,][]{Bugni2018, Bugni2019, Ye2018robust, Wang2019arXiv}. Most notably, \citet{Bugni2018} studied the two-sample t-test and the regression that adjusts for stratification indicators under stratified randomization, without assuming the true data generation model. Under the finite population framework, \citet{Liu2019} also proposed a regression-adjusted treatment effect estimator under stratified randomization that achieves strong balance€.

In additional to the stratum covariates, additional baseline covariates may be available to improve efficiency. However, although the stratification indicators are generally adjusted in the analysis of clinical trial data as required by regulatory guidelines \citep{ICH1998, EMA2015}, the inferential properties remain largely unknown when additional covariates are adjusted in the regression analysis. As pointed out in \citet{Wang2019}, ``It is an open question, to the best of our knowledge, as to what happens when more variables than the stratification indicators are included in the ANCOVA model under such randomization schemes [covariate-adaptive randomization], in terms of consistency of the ANCOVA estimator and how to compute its asymptotic variance under arbitrary model misspecification.''

Because covariate-adaptive randomization is so common in clinical trials, it is crucial to clarify and provide guidance on the use of regression models for covariate-adaptive randomized trials, with the acknowledgment that the models may be misspecified and that additional baseline covariates may be available. In particular, the following two questions demand definite answers: (i) Which linear regression model can be used to ensure robust and efficient estimation of the treatment effect? (ii) Is a consistent variance estimator  available and easily accessible to facilitate valid inference?

This paper seeks to address these questions. We evaluate six commonly used regression models for estimating and inferring treatment effects under a broad range of covariate-adaptive randomizations. These models cover the scenarios in which only stratum covariates are used for analysis as well as when additional baseline covariates are also adjusted. Based on the asymptotic results, the most efficient estimator is identified. Some special cases, such as equal allocation and strong balance, are discussed in detail. Moreover, we propose non-parametric consistent variance estimators for all of the regression estimators under consideration. We also examine the OLS variance estimator, which  is commonly used in practice, and indicate when it can be used for valid inference. Finally, we make practical recommendations for the use of regression under covariate-adaptive randomization.

The rest of the paper is organized as follows. Section 2 introduces the framework and notation for covariate-adaptive randomization. Section 3 investigates the theoretical properties of three widely used regression estimators for the treatment effect. Section 4 introduces another three regression estimators that adjust for additional baseline covariates and investigates their asymptotic properties. Section 5 studies the optimality of these regression estimators. Section 6 and Section 7  provide the results of a simulation study and a real data application, respectively. Section 8 presents a short discussion. All of the proofs are given in the Appendix A. 

% \newpage

\section{Framework and notation}

% Let $B_i = B(\bx_i)$, $i=1,\dots,n$ be the stratum label, and we assume that each unit is assigned to each stratum  $\pk = P(B_i = k) > 0$ for $k=1,\dots,K$. 

Consider a covariate-adaptive randomized experiment with $n$ units. For each unit $i=1,\dots,n$, let $A_i$ denote the treatment assignment with $A_i=1$ for the treatment and $A_i=0$ for the control, and $\bx_i$ denote a $p$-dimensional vector of baseline covariates. The experimental units are stratified into $K$ strata based on baseline covariates $\bx_i$ using a function $B$: $\text{supp}(\bx_i) \rightarrow \{1,\dots,K\}$. For example, the units are stratified according to gender, grade or location. For simplicity, we assume that  units are assigned to each stratum  with positive probability, i.e.,  $\pk = P(B_i = k) > 0$, $i=1,\dots,n$, $k=1,\dots,K$, where $B_i = B(\bx_i)$ is the stratum label. In stratum $k$, let $[k]$ be the index set of units, let $\nk = \sum_{i \in [k] } 1$, $\nkt = \sum_{i \in [k] } A_i$, $\nkc = \sum_{i \in [k] } (1 - A_i ) $ denote the number of units, the number of treated units, and the number of control units, respectively, and let $p_{n[k]}  = \nk / n$ and $\pik = \nkt / \nk $  denote the proportions of stratum sizes and the proportions of treated units, respectively. The theoretical property of covariate-adaptive randomization critically depends  on the difference between  the actual  treatment assignment and  the target treatment proportion $\pi \in (0,1)$:
%Further denote by $Q_n$ the distribution of $\{ \bm{W}_i= (Y_i(1), Y_i(0), \bx_i ), \  i = 1,\dots, n \}$, and  $P_n$  the distribution of the observed data $\{ (Y_i, \bx_i, A_i): i = 1,\dots, n \} $. 
$$
D_{n[k]} = \sumi (A_i - \pi ) I_{i \in [k] },  \quad  k = 1,\dots,K,
$$
where  $I_{i \in [k]} $ is an indicator function which equals one if $ i \in [k]$ and zero otherwise.

% \quad \pik = \frac{\nkt}{\nk},
% Further denote $[k]$  the index of units in stratum $k$ and $I_{i \in [k]} $ the indicator function which equals one if $ i \in [k]$ and zero otherwise. 

We use  the Neyman--Rubin potential outcomes model \citep{Neyman:1923,Rubin:1974} to define the treatment effect. For unit $i$,  this model assumes that there are two potential outcomes, $Y_i(1)$ and $Y_i(0)$, when exposed to the treatment and to the control, respectively. Under the stable unit treatment value assumption \citep{Rubin:1980}, the observed outcome $Y_i$ is a function of the treatment assignment  and the potential outcomes:
$
Y_i = A_iY_i(1)+(1-A_i)Y_i(0).
$
Our goal is to estimate and infer the treatment effect $\tau = E \{ Y_i(1) - Y_i(0) \}$ based on the observed data $\{ (Y_i, \bx_i, A_i): i = 1,\dots, n \} $. 

In this paper, we make the following assumptions regarding the data generation process and covariate-adaptive randomization.  We denote the sets of random variables with bounded $q$th ($q\geq 2$) moments and positive stratum-specific variances as $\mathcal{L}_q = \{  (V_1,\dots, V_m) :  E( |V_j | ^q)  < \infty, \ j=1,\dots,m \} $ and $ \mathcal{R}_2 = \{  (V_1,\dots, V_m) :  \Var \{ V_j - E( V_j | B_j )  \}   > 0 , \ j=1,\dots,m \} $, 
respectively. Let $A^{(n)} = (A_1,\dots, A_n)$, $B^{(n)} = (B_1,\dots, B_n)$, and $\bm{W}^{(n)} = ( \bm{W}_1,\dots,\bm{W}_n)$ where $ \bm{W}_i = (Y_i(1), Y_i(0), \bx_i)$, $i=1,\dots,n$. We assume that the  covariance  $ E[ \{ \bx_i - E(\bx_i | B_i ) \} \{ \bx_i - E(\bx_i | B_i ) \} ^\T ]  $ is (strictly) positive-definite.

%and make the following assumptions on $\bm{W_i} = (Y_i(1),Y_i(0),\bx_i)$ and  $A_i$.
%We assume that $\bm{W}_i = (Y_i(1),Y_i(0),\bx_i)$ are  i.i.d. samples from $Q$.

\begin{assumption} 
\label{assum::Q}
 %$E[V^2] < \infty$ and $\Var[ V - E[ V | S_i ]  ] > 0$, for $V = Y_i(1), \ Y_i(0), \ \bx$.
$\bm{W}_i \in \mathcal{L}_2  \cap \mathcal{R}_2  $, $i=1,\dots,n$, are independent and identically distributed (i.i.d.) samples from the population distribution of $\bm{W} = (Y(1), Y(0), \bx )$.
\end{assumption}

% The treatment assignment mechanism satisfies

\begin{assumption}
\label{assum::A1}
$\bm{W}^{(n)} \perp A^{(n)} | B^{(n)}$.
\end{assumption}

\begin{assumption}
\label{assum::A2}
$\big \{  \{ n^{-1/2}   D_{n[k]} \}_{k = 1,\dots,K} \big | B^{(n)} \big \}  \xrightarrow{d} \mathcal{N}(0, \Sigma_{D} )  $ a.s., where $\Sigma_{D} = \text{diag} \{ \pk \qk : k = 1,\dots,K  \}$ with $0 \leq \qk \leq \pi ( 1 - \pi ) $ for $k=1,\dots,K$.
\end{assumption}

\begin{assumption}
\label{assum::A3}
$D_{n[k]} = O_p( \sqrt{n} )$, for $k=1,\dots,K$.
\end{assumption}

%Assumption~\ref{assum::A1} requires that the treatment assignment is independent of the potential outcomes and covariates given the stratification. Assumption~\ref{assum::A3} was proposed by \cite{Bugni2018} for inference under covariate-adaptive randomization. Examples satisfying this assumption include simple random sampling, biased-coin design \citep{Efron1971}, adaptive biased-coin design \citep{Wei1978}, and stratified block randomization \citep{zelen1974randomization}. If $\qk = 0$, $k=1,\dots,K$,  we call the randomization achieves strong balance. Assumption~\ref{assum::A3} is weaker than Assumption~\ref{assum::A2}, for example, the Pocock and Simon's minimization \citep{Pocock1975} satisfies the former but  not  the latter. In fact,  almost all covariate-adaptive randomizations satisfy Assumption~\ref{assum::A3}, which is enough to obtain the consistency and asymptotic normality of our recommended regression-based treatment effect estimators. 

Assumption~\ref{assum::A1} requires that the treatment assignment be independent of the potential outcomes and covariates given the stratification.  Assumption~\ref{assum::A2} was proposed by \cite{Bugni2018} to study inference under covariate-adaptive randomization. This assumption characterizes the asymptotic behavior of jointly independent imbalances within strata and thus is particularly relevant to stratified randomization. As simple randomization leads to $\qk=\pi ( 1 - \pi)$, it is reasonable to expect  that a smaller value of $q_k$ can be achieved by stratified randomization. For example, when used within each stratum, Wei's urn design \citep{Wei1978} leads to a $q_k$ value between zero and $\pi ( 1 - \pi)$, whereas blocked randomization \citep{zelen1974randomization} and Efron's biased coin design \citep{Efron1971} can  reduce  $q_k$ to zero. In the latter case, we say that the randomization achieves strong balance \citep{Bugni2018}. Moreover, Assumption~\ref{assum::A3} is weaker than Assumption~\ref{assum::A2} as no (asymptotic) independence is required between different strata, and it allows us to consider the use of a covariate-adaptive randomization method that has a complicated dependence structure across strata, such as Pocock and Simon's minimization \citep{Pocock1975} and the class of designs proposed by \cite{Hu2012}. This assumption is also satisfied when the asymptotic distribution is not normal, but is instead, for example, is a truncated normal distribution \citep{li2016asymptotic}. We emphasize that  Assumption~\ref{assum::A3} is quite general and is satisfied by most, if not all, covariate-adaptive randomization methods. 

% In the following sections, $r_i(a)$ can be the potential outcomes $Y_i(a)$ or transformed potential outcomes. 

{\bf Notation}. We need more notations to proceed. For a random variable $V$, let $\tilde V = V - E(V | B)$ be the $V$ centered at its stratum-specific mean, $\mu_{V} = E(V)$ be its mean, and $\sigma^2_{V} = \Var(V)$ be its variance. For potential outcomes $r_i(a)$ such as $Y_i(a)$ or their transformations ($i=1,\dots,n$, $a=0,1$) and stratum $k$ ($k=1,\cdots,K$),  denote the population mean and the mean within stratum $k$ as $  \mu_{r}(a) = E \{ r_i(a) \} $ and $\mu_{[k]r}(a) = E \{ r_i(a) | B_i = k \}$, respectively. The corresponding sample means for the treatment group are denoted as
$ \bar{r}_1 = ( 1 / \nt ) \sumi A_i r_i(1) $ and $ \bar{r}_{[k]1} = ( 1 / \nkt)  \sumik A_i r_i(1).$
Similarly, we define the population and stratum-specific sample means for the control group as
$
\bar{r}_0 = ( 1 / \nc )  \sumi (1 - A_i ) r_i(0)$ and $ \bar{r}_{[k]0} = ( 1 / \nkc ) \sumik (1 - A_i ) r_i(0).
$
The population and sample means for the covariates are denoted as $\mu_{\bx}, \mu_{[k] \bx}, \XT, \XkThat, \XC$, and $ \XkChat$. The population variance of $r_i(a)$ is denoted as $\sigma^2_{r(a)}$. The asymptotic variances of the treatment effect estimators depend on the following quantities:
\[
\varsigma^2_{\tilde r}(\pi) = \frac{1}{\pi} \sigma^2_{\tilde r(1)} + \frac{1}{ 1 - \pi } \sigma^2_{\tilde r(0)},  \quad
\varsigma^2_{H r} = \sumk \pk \Big[  \{ \mu_{[k]r} (1) - \mu_{r}(1) \}  -  \{ \mu_{[k]r}(0) - \mu_{r}(0) \}  \Big]^2,
\]
\[
\varsigma^2_{A r}(\pi) = \sumk \pk \qk \Big\{  \frac{ \mu_{[k]r} (1) - \mu_{r}(1) }{\pi} + \frac{ \mu_{[k]r}(0) - \mu_{r}(0) }{ 1 - \pi} \Big\}^2,
\]
\[
\varsigma^2_{\pi r} = \frac{( 1 - 2\pi ) ^2 }{ \pi^2 ( 1 - \pi )^2 }  \sumk \pk \qk \Big[  \{ \mu_{[k]r} (1) - \mu_{r}(1) \}  -  \{ \mu_{[k]r}(0) - \mu_{r}(0) \}  \Big]^2.
\]
We denote their sample analog as
$$
\hat \varsigma^2_{\tilde r}(\pi) = \frac{1}{\pi}  \sumk \pnk  \Big(  \frac{1}{\nkt} \sumik  A_i  ( r_i -  \bar{r}_{[k]1} ) ^2  \Big) + \frac{1}{1 - \pi} \sumk \pnk  \Big\{  \frac{1}{\nkc} \sumik (1 - A_i ) ( r_i -  \bar{r}_{[k]0} )^2   \Big\},
$$ 
$$
\hat  \varsigma^2_{H r} = \sumk \pnk \big\{ ( \bar{r}_{[k]1}  - \bar{r}_1  )  - (  \bar{r}_{[k]0}  - \bar{r}_0 )  \big\}^2, \quad 
\hat \varsigma^2_{A r}(\pi) = \sumk \pnk \qk \Big(  \frac{ \bar{r}_{[k]1}  - \bar{r}_1 }{ \pi } + \frac{  \bar{r}_{[k]0}  - \bar{r}_0 }{ 1 - \pi } \Big)^2,
$$
$$
\hat \varsigma^2_{\pi r} = \frac{( 1 - 2\pi ) ^2 }{ \pi^2 ( 1 - \pi )^2 }  \sumk \pnk \qk \Big[  \{ \bar{r}_{[k]1}  - \bar{r}_1  \}  -  \{  \bar{r}_{[k]0}  - \bar{r}_0 \}  \Big]^2.
$$
Let  $ \Sigma_{RQ} = E[ \{R - E(R) \} \{ Q - E(Q)  \}^\T ]  $ be the covariance between two random vectors $R$ and $Q$.

%For a given $p$-dimensional coefficients vector $\gamma$, define the potential outcomes $r_i(\gamma, a) = Y_i(a) - \bx_i^\T \gamma$, $a=0,1$.

% Following the notations in \cite{Bugni2018}, 

\section{Adjustment for stratum covariates}

%In this section, we study the theoretical properties, such as unbiasedness, consistency, asymptotic normality and efficiency, of six widely used regression-based point and variance estimators for the treatment effect $\tau$. The first three regressions do not use the additional covariates $\bx_i$ while the last three do. 

In this section, we discuss the theoretical properties, i.e., unbiasedness, consistency, asymptotic normality, and efficiency, of three widely used regression-based point and variance estimators with respect to the treatment effect $\tau$. We do not consider the adjustment of additional covariates $\bx_i$, which is deferred to the next section.
In this section, we  bring together many pieces scattered throughout the literature and offer an integrated summary within the linear regression model framework. To complement the content found in the literature, we consider  the validity of the OLS variance estimator, which is often overlooked but is essential from the practical point of view.

\subsection{Difference-in-means}

Using the difference in the sample means is perhaps the most straightforward and intuitive approach for estimating the treatment effect. Unadjusted methods such as this, which also include the two-sample t-test, are commonly used for analyzing data from randomized clinical trials due to their simplicity, transparency, and robustness to model-misspecification, among other practical reasons \citep{Shao2010,  lin2013}. Although failure to  adjust  baseline covariates may reduce the  efficiency and is contrary to regulatory guidelines for analyzing clinical trial data with covariate-adaptive randomization \citep{ICH1998, EMA2015}, we start with the difference-in-means estimator because of its popularity in practice and its theoretical importance in its own right. It also serves as a benchmark for evaluating  subsequent estimators.

The  difference-in-means estimator $\hat \tau = \YT - \YC $ is equal to the OLS estimator of the coefficient of $A_i$ in the following regression:
\begin{equation}
\label{reg::diff}
Y_i \sim \alpha +  A_i \tau. \nonumber
\end{equation}
Let $\hat \sigma^2$ be the usual OLS variance estimator of $\hat \tau$. We have the following proposition.

% when regressing $Y_i$ on the treatment indicator $A_i$ with intercept.

% and the usual OLS variance estimator $\hat \sigma^2$ converges in probability to $\varsigma^2_{ Y}( 1 - \pi)$. 

\begin{proposition}
\label{prop::difference-in-mean}
Under Assumptions \ref{assum::Q} -- \ref{assum::A2}, 
\begin{equation}
E ( \hat \tau ) = \tau, \quad \sqrt{n} ( \hat \tau - \tau ) \xrightarrow{d} \mathcal{N}(0, \varsigma^2_{\tilde Y}(\pi) + \varsigma^2_{HY} + \varsigma^2_{AY}(\pi) ), \quad n \hat \sigma^2  \xrightarrow{P}   \varsigma^2_{ Y}( 1 - \pi),   \nonumber
\end{equation}
and the asymptotic variance can be consistently estimated by $\hat  \varsigma^2_{\tilde Y}(\pi) +  \hat \varsigma^2_{HY} + \hat \varsigma^2_{AY}(\pi)  $. 
%$$
%\varsigma^2_{\tilde Y}(\pi) + \varsigma^2_{HY} + \varsigma^2_{AY}(\pi) + \sumk \pk \{ \pi ( 1 - \pi ) - \qk \} \Big\{ \frac{ \mu_{[k]Y} (1) - \mu_{Y}(1) }{\pi} + \frac{ \mu_{[k]Y}(0) - \mu_{Y}(0) }{ 1 - \pi }  \Big\}^2.
%$$
\end{proposition}

%Thus, $\hat \sigma^2$ is consistent under the following two conditions: (1) $\pi = 1/2$ or $\sigma^2_{Y(1)} = \sigma^2_{Y(0)}$, and (2) simple randomization ($\qk = \pi ( 1 - \pi ) $) or the stratification is irrelevant to the potential outcomes in the sense that $\mu_{[k]Y} (a) = \mu_{Y}(a)$,  $k = 1,\dots,K$, $a = 0, 1$. Generally, $\hat \sigma^2$ can be anti-conservative, resulting in an invalid test with type I error larger than the preassigned level. 

\begin{remark}
The asymptotic normality of $\hat \tau$ and the consistent variance estimator were obtained by \cite{Bugni2018}. Proposition~\ref{prop::difference-in-mean} complements their results by  providing the probability limit of the usual OLS variance estimator. 
\end{remark}

By Lemma 1 in the Appendix A, the difference between the limit of the usual OLS estimator and the true asymptotic variance,  $ \varsigma^2_{ Y}( 1 - \pi) - \{   \varsigma^2_{\tilde Y}(\pi) + \varsigma^2_{HY} + \varsigma^2_{AY}(\pi)  \} $, is equal to
$$
\varsigma^2_{ Y}( 1 - \pi) - \varsigma^2_{ Y}( \pi) +  \sumk \pk \{ \pi ( 1 - \pi ) - \qk \} \Big\{ \frac{ \mu_{[k]Y} (1) - \mu_{Y}(1) }{\pi} + \frac{ \mu_{[k]Y}(0) - \mu_{Y}(0) }{ 1 - \pi }  \Big\}^2.
$$
Thus, generally, the consistency of $\hat \sigma^2$ is not guaranteed, and it is possible that $\hat \sigma^2$ is even anti-conservative, which could result in an inflated type I error rate. However, an important observation is that, under simple randomization (where the summation term vanishes),  $\hat \sigma^2$ is consistent under equal allocation ($\pi = 1/2$) or homogeneity ($\sigma^2_{Y(1)} = \sigma^2_{Y(0)}$). This result may support the use of OLS variance estimator in certain scenarios. Note that, the variance estimator with the assumption of homoskedasticity is also commonly used, especially for the two-sample t-tests. We do not explore these estimators as they are not easily fitted into the regression framework;  interested readers are referred to \citet{Shao2010} and  \citet{Bugni2018}.

%an invalid test with type I error larger than the preassigned level

\begin{remark}
The estimator $ \hat \varsigma^2_{\tilde Y}(\pi)$ differs  slightly (in terms of the weights $\pnk$) from that proposed by \cite{Bugni2018}, but has the same limit. Based on the asymptotic normality and the consistent variance estimator, we can construct a valid confidence interval or test for the treatment effect $\tau$. 
\end{remark}

%\begin{remark}
%The population variance $ \varsigma^2_{\tilde Y}(\pi)$, $\varsigma^2_{HY}$, $ \varsigma^2_{AY}(\pi)$ can be consistently estimated by the corresponding sample quantities, that is,
%$$
%\hat \varsigma^2_{\tilde Y}(\pi) = \frac{1}{\pi}  \sumk \pnk  \left(  \frac{1}{\nkt} \sumik  A_i  ( Y_i -  \YkThat ) ^2  \right) + \frac{1}{1 - \pi} \sumk \pnk  \left(  \frac{1}{\nkc} \sumik (1 - A_i ) (Y_i -  \YkChat )^2   \right),
%$$ 
%$$
%\hat  \varsigma^2_{HY} = \sumk \pnk \big\{ ( \YkThat - \YT )  - ( \YkChat - \YC )  \big\}^2, \quad 
%\hat \varsigma^2_{AY}(\pi) = \sumk \pnk \qk \left(  \frac{ \YkThat - \YT }{ \pi } + \frac{  \YkChat - \YC }{ 1 - \pi } \right)^2.
%$$
%\end{remark}

%\begin{proposition}[\citep{Bugni2018}]
% \label{prop::variance}

\subsection{Regression adjustment without interaction}

It is generally recognized that adjusting for baseline covariates can help to remedy the imperfect balance with respect to the covariates in a completely randomized experiment and thus improve the precision in estimating the treatment effect \citep[e.g.,][]{Cochran1957, Cox1982ancova}. Efficiency gain has also been realized both numerically \citep[e.g.,][]{Birkett1985, Forsythe1987} and  theoretically  \citep[e.g.,][]{Shao2010, Ma2015, Bugni2018}, when the covariates are well balanced by covariates-adaptive randomization,. As noted in the previous section, regulatory guidelines also recommend that the covariates used in the randomization be adjusted in the subsequent analysis. Given the extensive use of regression models in clinical trials and other randomized comparative studies, it is critical to clarify the theoretical properties of the regression-based estimators under covariate-adaptive randomization. The most basic way to do so is to include the stratification indicators in the linear regression model, i.e., 
\begin{equation}
\label{reg::fe}
Y_i \sim \alpha +  A_i \tau + \sum_{k=1}^{K-1} \alpha_k  I_{i \in [k] }. 
\end{equation}
Let $\taufe $ and $\hat \sigma^2_{\fe}$ be the OLS point and variance estimators of $\tau$ and define the following weights
$$
\omega_{[k]} = \frac{\pik ( 1 - \pik ) p_{n[k]} }{  \sum_{k'=1}^{K}  \pi_{[k']} ( 1 - \pi_{[k']} ) p_{n[k']} }, \quad k = 1,\dots,K.
$$
Denote the stratum-specific treatment effect in stratum $k$ and its difference-in-means estimator as
$$
\tauk = E \{ Y_i(1) - Y_i(0) | B_i = k \} = \mu_{[k]Y}(1)  - \mu_{[k]Y}(0), \quad  \taukhat =  \YkThat - \YkChat.
$$

% $\taukhat =  \YkThat - \YkChat$ be its difference-in-means estimator. 
%Let $\tauk = E[Y_i(1) - Y_i(0) | B_i = k ]  = \mu_{[k]Y}(1)  - \mu_{[k]Y}(0)$ be the average treatment effect within stratum $k$ and let $\taukhat =  \YkThat - \YkChat$ be its difference-in-means estimator. 

%  \cite{Bugni2018}  obtained the following result. 

\begin{proposition}
\label{prop::fe}
The estimator $\taufe$ has the formula $\taufe = \sumk \omega_{[k]} \taukhat $. Under Assumptions \ref{assum::Q} -- \ref{assum::A2},
\begin{equation}
E ( \taufe ) =   \sumk E [\omega_{[k]} ] \tauk, \quad \sqrt{n} ( \taufe - \tau ) \xrightarrow{d}  \mathcal{N} (0, \varsigma^2_{\tilde Y}(\pi) + \varsigma^2_{HY} + \varsigma^2_{\pi Y} ), \quad n \hat \sigma^2_{\fe}  \xrightarrow{P} \varsigma^2_{\tilde Y} ( 1 - \pi ) +  \varsigma_{HY}^2, \nonumber
\end{equation}
and the asymptotic variance can be consistently estimated by $\hat \varsigma^2_{\tilde Y}(\pi) + \hat \varsigma^2_{HY} + \hat \varsigma^2_{\pi Y} $. Furthermore, when $\pi = 1/2$, the conclusions hold if  Assumption~\ref{assum::A2} is replaced by Assumption~\ref{assum::A3}.
\end{proposition}

%$\hat \sigma^2_{\fe}$ converges in probability to  $ \varsigma^2_{\tilde Y} ( 1 - \pi ) +  \varsigma_{HY}^2. $ Furthermore, $\hat \varsigma^2_{\tilde Y}(\pi) + \hat \varsigma^2_{HY} + \hat \varsigma^2_{\pi Y} $ is a consistent variance estimator. 

%{\color{red} Have some trouble proving the unbiasedness of $\taufe$ $\dots$}

% if we replace Assumption~\ref{assum::A2} by the weaker Assumption~\ref{assum::A3}.

\begin{remark}
\cite{Bugni2018} obtained the asymptotic normality of $\taufe$, the consistent variance estimator, and the conservativeness of the Huber--White variance estimator under Assumptions \ref{assum::Q} -- \ref{assum::A2}. We complement their results by providing the probability limit of the usual OLS variance estimator widely used in clinical trials. The fact that Assumption ~\ref{assum::A2} can be weakened to Assumption~\ref{assum::A3}, when $\pi = 1/2$,  is also a new result.
\end{remark}

{\color{black} The performances of $\taufe $ and $\hat \sigma^2_{\fe}$ depends on whether the treatment allocation is balanced. For equal allocation ($\pi = 1/2$),  because  $\varsigma^2_{\pi Y} =0$ and $\varsigma^2_{AY}(\pi)  \geq 0$,  the asymptotic variance of $\taufe$ is less than or equal to that of $\hat \tau$. That is, the regression  improves, or at least does not hurt, the precision of estimating and inferring the treatment effect $\tau$. Moreover, $\varsigma^2_{\pi Y} =0$ and $ \varsigma^2_{\tilde Y}(\pi) =  \varsigma^2_{\tilde Y}(1 - \pi)$  imply that  $\hat \sigma^2_{\fe}$ is a consistent estimator for the asymptotic variance of $\taufe$.  More importantly, Proposition~\ref{prop::fe} holds under Assumption~\ref{assum::A3}. Thus, for  most if not all covariate-adaptive randomized trials, it is valid and efficient to use the usual OLS point and variance estimators obtained using regression \eqref{reg::fe}. The validity of the usual OLS point and variance estimators was confirmed by \citet{Yang2001} and \citet{Wang2019}, but under simple randomization.

% Moreover, since $\varsigma^2_{AY}(\pi)  \geq 0$,  the asymptotic variance of $\taufe$ is less than or equal to that of $\hat \tau$. That is, the regression  improves, at least does not hurt, precision for estimating and inferring $\tau$. 

For unequal allocation ($\pi \neq 1/2$), $\hat \sigma^2_{\fe}$ can be anti-conservative even for a strong balance treatment assignment, which  results in an inflated type I error rate. Even if we  use the consistent variance estimator $\hat \varsigma^2_{\tilde Y}(\pi) + \hat \varsigma^2_{HY} + \hat \varsigma^2_{\pi Y} $  to construct a valid confidence interval or test, we should keep in mind that the performance of $\taufe$ can be worse than that of the difference-in-means estimator.
}

% Therefore, regression~\eqref{reg::fe} is a better choice than regression \eqref{reg::diff} in terms of both point and variance estimations.

%\begin{remark}
%As pointed out by \cite{Bugni2018}, the asymptotic variances of 
%\end{remark}

%  \cite{Freedman2008}  and \cite{freedman2008randomization} criticized the use of regression analysis in randomized experiments by showing that the resulting sample average treatment effect estimator can have larger asymptotic variance than the difference-in-means estimator when $\pi \neq 1/2$.  Freedman's criticism can be addressed by adding  interactions in the regression \citep{lin2013}.}

\subsection{Regression adjustment with interaction}

In this section, we study the regression model with stratification-by-treatment interactions under covariate-adaptive randomization. This is  motivated in part by  results reported in  the literature, which state that adding the interaction terms can further improve  precision when estimating the treatment effect under various randomization methods  \citep{Yang2001, lin2013, Bugni2019, Liu2019}. 
	
The third regression under consideration is
\begin{equation}
\label{reg::str}
Y_i \sim \alpha +  A_i \tau + \sum_{k=1}^{K-1} \alpha_k  I_{i \in [k] } + \sum_{k=1}^{K-1} \nu_k A_i (  I_{i \in [k] } - \pnk ).
\end{equation}
The OLS estimator of the coefficient of $A_i$ is denoted as $\taustr$. {\color{black} Note that,  we center the stratum indicator $I_{i \in [k] }$ at its sample mean $\pnk$ in the interactions to ensure that $\taustr$ can be interpreted as the treatment effect, which is equivalent to the fully saturated regression estimator in \citet{Bugni2019}.} In fact, $\taustr$ has a more intuitive expression that can be considered to be a stratified difference-in-means estimator, as shown in the proof of the following Proposition~\ref{thm::str}:
\begin{equation}
\label{est::str}
\taustr = \sumk p_{n[k]} ( \YkThat - \YkChat  ) = \sumk \pnk \taukhat, 
\end{equation}
where $\taukhat $ is the difference-in-means estimator for $\tauk$ in stratum $k$. As $\tau = \sumk \pk \tauk$, $\taustr$ is a natural plug-in estimator. Let $\hat \sigma^2_{\str}$ be the OLS variance estimator of $\taustr$. 

%We have the following proposition.

\begin{proposition}
\label{thm::str}
Under Assumptions \ref{assum::Q}, \ref{assum::A1} and \ref{assum::A3},
$$
E ( \taustr ) = \tau, \quad \sqrt{n}( \taustr - \tau)  \xrightarrow{d} \mathcal{N} (0, \varsigma^2_{\tilde Y} (\pi) + \varsigma^2_{H Y} ), \quad n \hat \sigma^2_{\str}  \xrightarrow{P} \varsigma^2_{\tilde Y} ( 1 - \pi ), 
$$
and the asymptotic variance can be consistently estimated by $\hat \varsigma^2_{\tilde Y}(\pi) + \hat \varsigma^2_{HY} $.

%and the OLS variance estimator $\hat \sigma^2_{\str}$ converges in probability to $ \varsigma^2_{\tilde Y} ( 1 - \pi )$. Furthermore, $\hat \varsigma^2_{\tilde Y}(\pi) + \hat \varsigma^2_{HY} $ is a consistent variance estimator. 

\end{proposition}

\begin{remark}
\label{remark4}
The asymptotic normality of $\taustr$ and the consistent variance estimator were obtained by \cite{Bugni2019} under a weaker condition, $ \nkt / \nk \xrightarrow{P} \pi$,  rather than Assumption~\ref{assum::A3}. {\color{black} We use Assumption~\ref{assum::A3} for two reasons: first, it is satisfied by almost all covariate-adaptive randomizations, and second, we require Assumption~\ref{assum::A3}  to obtain the validity of the regression without interactions methods discussed in Section 3.2 and Section 4.2  for the case of equal allocation.}
\end{remark}

Based on the asymptotic normality and using the consistent variance estimator, we can construct a valid confidence interval or test for $\tau$. Moreover, Proposition~\ref{thm::str} implies that the OLS variance estimator $\hat \sigma^2_{\str}$ can be anti-conservative even when $\pi = 1/2$, so we should not use it. As in many clinical trials, $\pi = 1/2$, so we discuss this special case in more details. 

\begin{corollary}
\label{cor::strfe}
When $\pi = 1/2$, under Assumptions \ref{assum::Q}, \ref{assum::A1} and \ref{assum::A3}, $\taustr$ and $\taufe$ are asymptotically equivalent, with both asymptotic variances being smaller than or equal to that of $\hat \tau$. Furthermore, both $n \hat \sigma^2_{\fe}$ and  $\hat \varsigma^2_{\tilde Y}(\pi) + \hat \varsigma^2_{HY} $ are consistent variance estimators. 
\end{corollary}

% Generally, the asymptotic normality of $\taufe$ requires Assumption~\ref{assum::A2}. However, when $\pi = 1/2$, Assumption~\ref{assum::A2} can be weakened to Assumption~\ref{assum::A3}.    

Corollary~\ref{cor::strfe} implies that for almost all covariate-adaptive randomization with equal allocation, it is valid and efficient to use regression \eqref{reg::str} to estimate and infer $\tau$.

%\begin{remark}
%Under super-population framework, $\taufe$ generally has a small finite-sample bias, which vanishes quickly,  for estimating the population average treatment effect $\tau$, but both $\hat \tau$ and $\taustr$ are unbiased even when the target treatment proportion $\pi$ varies across strata. While under finite-population framework,  in general, only $\taustr$  is unbiased for the sample average treatment effect $\bar{Y}(1) - \bar{Y}(0) $ when $\pi$ varies across strata.
%\end{remark}

%Taking both simplicity and efficiency into account, we make the following recommendations:
%\begin{itemize}
%\item When the treatment assignment is strong balance, we recommend $\hat \tau$ as point estimator and $\hat \varsigma^2_{\tilde Y}(\pi) + \hat \varsigma^2_{HY} $ as variance estimator,
%\item When the treatment assignment is not strong balance but $\pi = 1/2$, we recommend using the strata fixed effect regression with the OLS point estimator $\taufe$ and variance estimator $\hat \sigma^2_{\fe}$. If unbiasedness is a major concern, we recommend $\taustr$ and $\hat \varsigma^2_{\tilde Y}(\pi) + \hat \varsigma^2_{HY}$ as point and variance estimators,
%\item When the treatment assignment is not strong balance and $\pi \neq 1/2$, we recommend $\taustr$ and $\hat \varsigma^2_{\tilde Y}(\pi) + \hat \varsigma^2_{HY} $ as point and variance estimators.
%\end{itemize}

% for estimating the population average treatment effect $\tau$,

\subsection{Summary and recommendation}

\begin{table}[ht]
\centering
\caption{\label{tab::var} Summary of asymptotic results for various regression-based treatment effect estimators.}
	\begin{threeparttable}
		\setlength{\tabcolsep}{4pt}{

\begin{tabular}{cclllc}
\hline
 Regression & Target & & &  Proposed consistent  &  Is OLS variance  \\ 
adjustment & allocation&   Estimator&  Asymptotic variance &  variance estimator & estimator valid?    \\ \hline
Stratification & 1/2 & $\hat \tau$   & $  \varsigma^2_{\tilde Y}(\pi) + \varsigma^2_{HY} + \varsigma^2_{AY}(\pi)$  & $ \hat  \varsigma^2_{\tilde Y}(\pi) +  \hat \varsigma^2_{HY} + \hat \varsigma^2_{AY}(\pi) $ & No   \\ 

only &  & $\taufe$  & $ \varsigma^2_{\tilde Y}(\pi) + \varsigma^2_{HY}$ &   $ \hat \varsigma^2_{\tilde Y}(\pi) + \hat \varsigma^2_{HY} $ & Yes \\

&   & $\taustr$ & $ \varsigma^2_{\tilde Y}(\pi) + \varsigma^2_{HY}$ &   $ \hat \varsigma^2_{\tilde Y}(\pi) + \hat \varsigma^2_{HY} $ &  No\\

& $ \neq 1/2 $ & $\hat \tau$     & $  \varsigma^2_{\tilde Y}(\pi) + \varsigma^2_{HY} + \varsigma^2_{AY}(\pi)$  & $ \hat  \varsigma^2_{\tilde Y}(\pi) +  \hat \varsigma^2_{HY} + \hat \varsigma^2_{AY}(\pi) $ & No \\

 &  & $\taufe$  & $ \varsigma^2_{\tilde Y}(\pi) + \varsigma^2_{HY} + \varsigma^2_{\pi Y} $   &   $ \hat \varsigma^2_{\tilde Y}(\pi) + \hat \varsigma^2_{HY} + \hat \varsigma^2_{\pi Y} $ & No\\ 

&  & $\taustr$  & $ \varsigma^2_{\tilde Y}(\pi) + \varsigma^2_{HY}$ &   $ \hat \varsigma^2_{\tilde Y}(\pi) + \hat \varsigma^2_{HY} $ & No\\ \hline

Stratification  & 1/2& $\tausimp$   & $  \varsigma^2_{\tilde r}(\pi) + \varsigma^2_{Hr} + \varsigma^2_{Ar}(\pi)$  & $ \hat  \varsigma^2_{\tilde r}(\pi) +  \hat \varsigma^2_{Hr} + \hat \varsigma^2_{Ar}(\pi) $ & No  \\ 

and &  & $\tauancova$& $ \varsigma^2_{\tilde r_{\ancova} }(\pi) + \varsigma^2_{H r_{\ancova} }$ &   $ \hat \varsigma^2_{\tilde r_{\es}}(\pi) + \hat \varsigma^2_{H r_{\es} } $ & Yes \\ 

additional  & & $\tauinter$ &  $ \varsigma^2_{\tilde r_{\inter} }(\pi) + \varsigma^2_{H r_{\inter} }$ &   $ \hat \varsigma^2_{\tilde r_{\essed} }(\pi) + \hat \varsigma^2_{H r_{\essed} } $ & No \\

covariates & $ \neq 1/2 $ &$\tausimp$     & $  \varsigma^2_{\tilde r}(\pi) + \varsigma^2_{Hr} + \varsigma^2_{Ar}(\pi)$  & $ \hat  \varsigma^2_{\tilde r}(\pi) +  \hat \varsigma^2_{Hr} + \hat \varsigma^2_{Ar}(\pi) $ & No\\

 &  & $\tauancova$& $  \varsigma^2_{\tilde r_{\ancova} }(\pi) + \varsigma^2_{H r_{\ancova} } + \varsigma^2_{\pi r_{\ancova} } $   &   $ \hat \varsigma^2_{\tilde r_{\es}}(\pi) + \hat \varsigma^2_{H r_{\es} } + \hat \varsigma^2_{\pi r_{\es} } $ & No \\ 

 & & $\tauinter$ & $ \varsigma^2_{\tilde r_{\inter} }(\pi) + \varsigma^2_{H r_{\inter} }$ &   $ \hat \varsigma^2_{\tilde r_{\essed} }(\pi) + \hat \varsigma^2_{H r_{\essed} } $ & No \\ \hline

\end{tabular}}
\begin{tablenotes}
\item Note: All listed estimators are consistent and asymptotically normal.
\end{tablenotes}
\end{threeparttable}
\end{table}

\begin{table}[ht]
\centering
\caption{\label{tab::recom} Recommendation for regression-based treatment effect estimators.}
{\small
\begin{tabular}{lccc}
\hline
Regression adjustment & Target allocation & Point estimator  & Variance estimator \\  \hline

Stratification only & $ 1/2 $ &  $\taufe$   & $ \hat \sigma^2_{\fe}$  \\

&  $ \neq 1/2$ & $\taustr$  &  $ \{ \hat \varsigma^2_{\tilde Y}(\pi) + \hat \varsigma^2_{HY} \} / n$ \\ \hline

Stratification and & $1/2 $ & $\tauancova$  &  $ \hat \sigma^{2*}_{\ancova}$   \\ 

additional covariates&  $\neq 1/2 $ & $\tauinter$  &  $ \{ \hat  \varsigma^2_{\tilde r_{\essed} }( \pi ) + \hat \varsigma^2_{H r_{\essed} } \} / n $ \\ \hline
 
\end{tabular}}
\end{table}

The asymptotic variances and  variance estimators of $\hat \tau$, $\taufe$, and $\taustr$ are summarized in Table~\ref{tab::var}. As $\varsigma^2_{AY}(\pi)$ and $\varsigma^2_{\pi Y}$ are non-decreasing functions of $\qk$,  $\hat \tau $ and  $\taufe$ under covariate-adaptive randomization are always no worse than those obtained under simple randomization. Thus, covariate-adaptive randomization can improve efficiency compared to simple randomization, even when adjusting for  stratification indicators in the analysis.  Generally, the asymptotic variances of $\hat \tau$ and $\taufe$ are not ordered unambiguously when $\pi \neq 1/2$.  Moreover, as $ \varsigma^2_{\pi Y} \geq 0$ and $\varsigma^2_{AY}(\pi)  \geq 0$, the stratified difference-in-means estimator $\taustr$ has the smallest asymptotic variance. When $\pi = 1/2$, it holds that $\varsigma^2_{\pi Y} = 0$, thus $\taufe$ and $\taustr$ are asymptotically equivalent, both being no worse than $\hat \tau$. Furthermore, when the treatment assignment achieves strong balance ($\qk = 0, \ k=1,\dots,K$), $ \varsigma^2_{AY}(\pi) = \varsigma^2_{\pi Y} = 0 $, which implies that  $\hat \tau $, $\taufe$ and $\taustr$ are asymptotically equivalent. Taking into account both  simplicity and efficiency, our recommendations are summarized in Table~\ref{tab::recom}. Note that, the asymptotic variances of $\taustr$ and, when $\pi = 1/2$, $\taufe$ do not depend on  a particular covariate-adaptive randomization procedure as long as the randomization satisfies Assumption \ref{assum::A3}. Thus, the recommendations apply for most commonly used covariate-adaptive randomization methods, including stratified randomization and minimization.

% , which does not depend on particular covariate-adaptive randomization procedure as long as the randomization satisfies Assumptions~\ref{assum::A1} and \ref{assum::A3}.

% Stratification and randomization ensures that the covariates in the treatment and control groups are the same on average, however, for a particular treatment assignment, the probability of covariates exhibiting imbalance between treatment and control groups can be large \citep{fisher1926, morgan2012rerandomization}.  Regression adjustment is a common strategy to adjust for the remaining imbalance in the covariates \citep{Liu2019,LiDing2020}.

% We study the consistency and asymptotic normality of these estimators and how to estimate the asymptotic variance. 

\section{Adjustment for additional baseline covariates}

Apart from the stratification indicators, the covariates $\bx_i$ may contain additional information with which to better estimate the treatment effect. Regression adjustment is a common strategy for adjusting for additional covariates to improve the precision \citep{Liu2019,LiDing2020}.  In the following, we discuss three widely used regression methods that adjust for additional covariates, which correspond to the three regressions discussed in Section 3 with the addition of  $\bx_i$.  Note that if a discrete covariate has already been adjusted in the regression as a stratification indicator, it is not, per se, an additional covariate. In general, we should remove the covariates that can be linearly represented by the stratification indicators from $\bx_i$. For simplicity, however, we continue to use the same notation. Before formally  stating the asymptotic results, we define several population-level regression (or projection) coefficients: for $a=0,1$,
\begin{equation}
\gamma(a) = \argmin_{\gamma \in R^p} E \big [ Y_i(a) - E \{ Y_i(a) \}  - \{ \bx_i - E(\bx_i) \}^\T \gamma \big ]^2 = \Sigma^{-1}_{\bx \bx} \Sigma_{\bx Y(a)}, \nonumber
\end{equation}
\begin{equation}
\label{eqn::tildebeta}
\beta(a) = \argmin_{ \beta \in R^p} E \big [  Y_i(a) - E \{ Y_i(a) | B_i \} -  \{ \bx_i - E ( \bx_i | B_i )  \} ^\T \beta  \big ]^2 = \Sigma_{\tilde \bx \tilde \bx }^{-1} \Sigma_{\tilde \bx Y(a)},   \nonumber
\end{equation}
%$$
%\gamma  = \pi \gamma(1) + ( 1 - \pi )  \gamma(0) , \quad \beta_{\inter} =  ( 1 - \pi ) \gamma(1) +  \pi  \gamma(0),
%$$
%$$
%\beta_{\ancova}  = \pi  \beta(1) +  ( 1 - \pi ) \beta(0), \quad  \beta_{\inter} = (1 - \pi) \beta(1) + \pi \beta(0),
%$$
where $\gamma(a)$ is the population regression coefficient for regressing $Y_i(a)$ on $\bx_i$ with an intercept, and $\beta(a) $ is the population regression coefficient  for regressing $\tilde Y_i(a)$ $(= Y_i(a) -  E \{ Y_i(a) | B_i \} )$ on $\tilde \bx_i$ $(= \bx_i - E ( \bx_i | B_i ))$.

%$\gamma $ and $\beta_{\inter}$ are linear combinations of $\gamma(1)$ and $\gamma(0)$, and $\beta_{\ancova} $ and $ \beta_{\inter} $ are linear combinations of $\beta(1)$ and $\beta(0)$.

\subsection{Difference-in-means}

%The first regression using covariates $\bx_i$ is to simply regress $Y_i$ on $A_i$ and $\bx_i$ with intercept:
%\begin{equation}
%Y_i \sim \alpha +  A_i \tau + \bx_i ^\T \beta.
%\end{equation}

First, we consider the regression method that regresses $Y_i$ on $A_i$ and $\bx_i$ with an intercept but without stratification indicators:
\begin{equation}
\label{eqn::reg-simp}
Y_i \sim \alpha +  A_i \tau + \bx_i ^\T \gamma.
\end{equation}
Because making no adjustment to the stratum covariates is contrary to regulatory guidelines and general practice, and usually decreases efficiency (which becomes more evident shortly), we only include just a short discussion for completeness. Denote $\tausimp$  as the OLS point  estimator of $\tau$ in regression \eqref{eqn::reg-simp}, and define the transformed outcomes $ r_{i} (a) = Y_i(a)  -  \bx_i ^\T  \gamma$, where $\gamma =  \pi \gamma(1) + ( 1 - \pi )  \gamma(0) $. We have the following theorem.

%or equivalently decompose $Y_i(z)$ as follows:
%\begin{equation}
%Y_i(z) = E[Y_i(z)] + ( \bx_i - \mu_{ \bx } ) ^\T  \beta (z) + r_i(z).
%\end{equation}

\begin{theorem} 
\label{thm::simp}
Suppose that $( r_{i}(1), r_{i}(0) ) \in  \mathcal{R}_2$ and Assumptions~\ref{assum::Q} -- \ref{assum::A2} hold, then
$$
\sqrt{n} ( \tausimp - \tau ) \xrightarrow{d} \mathcal{N}  (0,  \varsigma^2_{\tilde r}(\pi) + \varsigma^2_{Hr} + \varsigma^2_{A r}(\pi)  ).
$$
Furthermore, the difference between the asymptotic variances of $\tausimp$ and $\hat \tau$  is
\begin{eqnarray}
& & \Delta_{  * - \diff }    =   -  \frac{1}{\pi ( 1 - \pi ) }  \times \Big [ \gamma^\T  \Sigma_{ \bx \bx } \gamma - 2 ( 2\pi - 1 ) \gamma^\T  \Sigma_{ \bx \bx }  \{ \gamma(1) - \gamma(0)  \} + \nonumber \\
&& \gamma^\T  \sumk \pk \Big\{ 1 - \frac{\qk}{\pi ( 1 - \pi )} \Big\}   ( \mu_{[k] \bx }  - \mu_{\bx} )  ( \mu_{[k] \bx }  - \mu_{\bx} )^\T   \gamma - \nonumber \\
&& 2  \gamma^\T  \sumk \pk \Big\{ 1 - \frac{\qk}{\pi ( 1 - \pi )} \Big\}   ( \mu_{[k] \bx }  - \mu_{\bx} ) \{ ( 1 - \pi ) ( \mu_{[k] Y }(1) - \mu_{Y}(1) ) + \pi ( \mu_{[k] Y }(0) - \mu_{Y}(0) )   \}    \Big]. \nonumber 
\end{eqnarray}
%The difference of the asymptotic variances of $\tausimp$ and $\taufe$ is
%$$
%\Delta_{ * - \fe}  =   - \frac{1}{\pi ( 1 - \pi ) }  \gamma^\T  \Sigma_{ \tilde \bx  \tilde \bx } \gamma  +  \frac{2}{\pi ( 1 - \pi ) }  \gamma^\T  \Sigma_{\tilde \bx \tilde \bx }  (  \gamma - \beta_{\ancova} ) + \varsigma^2_{A r} - \varsigma^2_{\pi Y}.
%$$

%\begin{eqnarray*}
%\Delta_{\simp - \text{fe}} &  =  & -  \frac{1}{\pi ( 1 - \pi ) }  \gamma^\T  \Sigma_{ \bx \bx } \gamma + \frac{2 ( 2\pi - 1 )}{\pi ( 1 - \pi )} \gamma^\T  \Sigma_{ \bx \bx }  \{ \gamma(1) - \gamma(0)  \} \nonumber \\
%&& -  \sumk \pk \{ \pi ( 1 - \pi ) -  \qk \} \Big(   \frac{1}{\pi} E[ m_{r1}(\bx_i) | B_i = k ] + \frac{1}{1 - \pi } E [ m_{r0}(\bx_i) | B_i = k ]   \Big)^2.
%\end{eqnarray*}

% Since the asymptotic variance $ \varsigma^2_{\tilde r}(\pi) + \varsigma^2_{Hr} + \varsigma^2_{A r}(\pi) $  is increasing with $\qk$, 

\end{theorem}

{\color{black}

\begin{remark}
Similar to $\hat \tau$, the first two terms of the asymptotic variance of $\tausimp$ are irrelevant to $\qk$, whereas the third term increases with $\qk$. Thus,  the covariate-adaptive randomization with $0\leq \qk < \pi ( 1 - \pi )$ is always no worse than simple randomization ($\qk = \pi ( 1 - \pi )$) if we use regression \eqref{eqn::reg-simp} under both randomization procedures. 
\end{remark}

\begin{remark}
For simple randomization with equal allocation $(\pi = 1/2)$, 
$ \Delta_{  * - \diff }   =   -  \gamma^\T  \Sigma_{ \bx \bx } \gamma / \{ \pi ( 1 - \pi ) \}  \leq 0.$
Thus, $\tausimp$ is generally more efficient than $\hat \tau$, which reproduces the results in \cite{Yang2001}. For general covariate-adaptive randomization with $0 \leq \qk < \pi ( 1 - \pi ) $, $\tausimp$ may hurt the precision even when $\pi = 1/2$, as compared with $\hat \tau$. Later, we show that adding stratification indicators (when $\pi = 1/2$) or adding both  stratification indicators and treatment-by-covariate interactions (when $\pi \neq 1/2$) generally improves efficiency, so we do not recommend $\tausimp$.

\end{remark}

}

% Moreover, compared to $\taufe$,  $\tausimp$ may hurt precision even when .   Regression~\eqref{}

% \Big(  E[ ( \bx - E[\bx | B] )  ( \bx - E[\bx | B] )^\T]  \Big) 
%  \Big(  E[ ( \bx - E[\bx | B] ) ( \pi  \{ Y(1) - E[Y(1) | B] \} + ( 1 - \pi )  \{ Y(0) - E[Y(0) | B] \} ] \Big )

\subsection{Regression adjustment without interaction}
The second most widely used regression analysis method that adjusts for additional covariates $\bx_i$ in randomized experiments including clinical trials is to regresses $Y_i$ on $A_i$, $ I_{i \in [k] }$ and $\bx_i$:
\begin{equation}
\label{reg::ancova}
Y_i \sim \alpha +  A_i \tau + \sum_{k=1}^{K-1} \alpha_k  I_{i \in [k] }  + \bx_i ^\T \beta.
\end{equation}
We refer to this regression method as analysis of covariance (ANCOVA). Let $\tauancova$ and $\hat \sigma^{2*}_{\ancova} $ be the OLS point and variance estimators of $\tau$, respectively, and let $\hat \beta_{\ancova}$ be the OLS estimator of $\beta$. Define the transformed outcomes 
$
 r_{i, \ancova}(a) = Y_i(a)   -  \bx_i ^\T \beta_{\ancova},
$
where $ \beta_{\ancova} =  \pi  \beta(1) +  ( 1 - \pi ) \beta(0)$, and denote the residuals $r_{i,\es}(a) = Y_i(a) - \bx_i ^\T \hat \beta_{\ancova} $.

%$$
% r_{i, \ancova}(z) = Y_i(z)   -  \bx_i ^\T \beta_{\ancova}.
%$$

\begin{theorem}
\label{thm::ancova}
Suppose that $( r_{i,\ancova}(1), r_{i,\ancova}(0) ) \in \mathcal{R}_2$ and Assumptions~\ref{assum::Q} -- \ref{assum::A2} hold, then
$$
\sqrt{n} ( \tauancova - \tau ) \xrightarrow{d} \mathcal{N} (0,  \varsigma^2_{\tilde r_{\ancova}}(\pi) + \varsigma^2_{Hr_{\ancova}} + \varsigma^2_{\pi r_{\ancova}} ), \quad n \hat \sigma^{2*}_{\ancova} \xrightarrow{P} \varsigma^2_{\tilde r_{\ancova} }(1 - \pi ) + \varsigma^2_{H r_{\ancova}},
$$
the asymptotic variance can be consistently estimated by $\hat  \varsigma^2_{\tilde r_{\es} }( \pi ) + \hat \varsigma^2_{H r_{\es} } + \hat \varsigma^2_{\pi r_{\es}} $, and the difference between the asymptotic variances of $\tauancova$ and $\taufe$ is
\begin{eqnarray}
\Delta_{\ancova * - \fe} =    - \frac{1}{ \pi ( 1 - \pi ) } \beta_{\ancova}^\T \Sigma_{\tilde \bx \tilde \bx} \beta_{\ancova} +  \frac{ 2 (2 \pi - 1 ) }{ \pi ( 1 - \pi ) } \beta_{\ancova}^\T \Sigma_{\tilde \bx \tilde \bx}  \{  \beta(1)  - \beta(0)   \}. \nonumber 
\end{eqnarray}
 Furthermore, when $\pi = 1/2$, the conclusions hold if  Assumption~\ref{assum::A2} is replaced by Assumption~\ref{assum::A3}.
%and the difference of the asymptotic variances of $\tauancova$ and $\hat \tau$ is
%$$
%\Delta_{\ancova * - \text{diff}} =   \Delta_{\ancova * - \fe} + \varsigma^2_{\pi Y} - \varsigma^2_{AY}(\pi).
%$$

\end{theorem}

{\color{black}
% under equal allocation and unequal allocation, respectively
% the asymptotic variance of $\tauancova$ is no greater than those of $\hat \tau$, $\taufe$ and $\taustr$, that is, improves, at least does not hurt, precision when compared to

Based on the above results, we can compare the efficiencies of  different estimators. First, we consider the case of equal allocation ($ \pi = 1/2$). In this case,
$
\Delta_{\ancova * - \fe} =    -  \beta_{\ancova}^\T \Sigma_{\tilde \bx \tilde \bx} \beta_{\ancova} / \{ \pi ( 1 - \pi ) \} \leq 0,
$
thus, the asymptotic variance of $\tauancova$ is no greater than that of $\taufe$. In Corollary~\ref{cor::strfe} we showed that $\taufe$ and $\taustr$ are asymptotically equivalent and they are more efficient than (or at least as efficient as)  $\hat \tau$. Therefore, $\tauancova$ is more efficient than (or at least as efficient as)  $\hat \tau$, $\taufe$, and $\taustr$. Later, we show later that in general,  $\tauancova$ is also more efficient than $\tausimp$ (see the following Corollary~\ref{cor::ancovainter} and Theorem~\ref{thm::optimal}). Moreover, as $
 \varsigma^2_{\pi r_{\ancova}}  = 0$ and $  \varsigma^2_{\tilde r_{\ancova} }(1 - \pi ) =  \varsigma^2_{\tilde r_{\ancova} }( \pi ),
$
the usual OLS variance estimator $\hat \sigma^{2*}_{\ancova}$ is consistent. Therefore, it is valid to infer $\tau$ based on the result obtained from standard statistical packages that perform regression~\eqref{reg::ancova}.

However, for unequal allocation ($ \pi \neq 1/2$), $\tauancova$  may hurt the precision when compared with that of $\taufe$, and $\hat \sigma^{2*}_{\ancova}$ can be anti-conservative. The same phenomenon has been observed when estimating the sample average treatment effect in completely randomized experiments under the finite population framework \citep{Freedman2008}.
}

%Finally,  for  simple randomization, Theorems~\ref{thm::simp} and \ref{thm::ancova} imply that the difference of the asymptotic variances of $\tauancova$ and $\tausimp$ is
%$$
%\Delta_{\ancova * -  * } = \Delta_{\ancova * - \fe} - \Delta_{ * - \adj } = - \frac{1}{ \pi ( 1 - \pi ) } ( \beta_{\ancova} - \gamma )^\T \Sigma_{\tilde \bx \tilde \bx}  ( \beta_{\ancova} - \gamma )^\T -  \varsigma^2_{A r }(\pi) \leq 0,
%$$
%thus, $\tauancova$ is more efficient than  (as least as efficient as) $\tausimp$.

% for strong balance treatment assignment with $\pi = 1/2$, it holds that $ \varsigma^2_{\pi Y} =  \varsigma^2_{AY}(\pi) = 0$, thus,  the asymptotic variance of $\tauancova$ is no greater than that of estimator under simple randomization with or without using covariates in the analysis. 

% Otherwise, they are not ordered unambiguously.

\subsection{Regression adjustment with interaction}

% \subsubsection{Second-order Interaction Regression Estimator}

% \subsubsection{Asymptotic Normality}

% , otherwise, the resulting estimator can be biased even asymptotically. Denote $\tauinter$ and $\hat \sigma^{2*}_{\inter}$ as the OLS point and variance estimators of $\tau$ and 

%\cite{Freedman2008}  and \cite{freedman2008randomization} criticized the use of regression analysis in completely randomized experiments by showing that the resulting sample average treatment effect estimator can have larger asymptotic variance than the difference-in-means estimator when $\pi \neq 1/2$.  Freedman's criticism can be addressed by adding the treatment-by-covariate interactions in the regression \citep{lin2013}, which motivates us to consider the following regression:

In this section, we consider a regression that follows the same principle as that described in Section 3.3 while adding additional covariates $\bx_i$:
\begin{equation}
\label{eqn::reg_inter}
Y_i \sim \alpha +  A_i \tau + \sum_{k=1}^{K-1} \alpha_k  I_{i \in [k] } + \sum_{k=1}^{K-1} \nu_k A_i (  I_{i \in [k] } - p_{n[k]} ) + \bx_i ^\T \beta  + A_i ( \bx_i -   \bar{\bx} ) ^\T \xi.
\end{equation}
Denote $\tauinter$ and $\hat \sigma^{2*}_{\inter}$ as the OLS point and variance estimators of $\tau$, respectively. Note that we center the stratification indicators and covariates at their sample means in the interactions to ensure that $\tau$ can be interpreted as the treatment effect. As shown in our proof, regression~\eqref{eqn::reg_inter} is equivalent to running two regressions. First, we separately regress the outcomes $Y_i$ on the stratification indicators $I_{i \in [k] }$ and covariates $\bx_i$ in the treatment and control groups, and obtain the OLS estimators $\hat \beta(1)$ and $\hat \beta(0)$ of the coefficients of $\bx_i$. Then:
$$
\tauinter =  \sumk \pnk \Big[  \Big\{   \YkThat - ( \XkThat -  \XkT )^\T \hat \beta(1)   \Big\}  - \Big\{  \YkChat - ( \XkChat -  \XkC )^\T \hat \beta(0)   \Big\}  \Big].
$$

%In view of \eqref{est::str}, we can see $\tauinter$ as augmenting the stratified difference-in-means estimator by adjusting for the additional covariates $\bx_i$ within each stratum. 

\begin{remark}
By comparing the above expression to \eqref{est::str}, $\tauinter$ can be viewed as augmenting $\taustr$ by further adjusting for the additional covariates $\bx_i$ when estimating the stratum-specific treatment effect. This covariate adjustment has also been used for estimators under simple or stratified randomization \citep{Tsiatis2008, Liu2019}.
\end{remark}

Define the transformed outcomes $r_{i,\inter}(a) =  Y_i(a) - \bx_i^\T \beta_{\inter}$ and residuals $ r_{i,\essed}(a) = Y_i(a) - \bx_i^\T \hat \beta_{\inter}$, where $ \beta_{\inter} = (1 - \pi) \beta(1) + \pi \beta(0) $ and $\hat \beta_{\inter} = ( 1 - \pik ) \hat \beta(1) + \pik \hat \beta(0)$.

%\begin{eqnarray}
%\label{reg::control1}
%Y_i(0) \sim  \sumk \alpha_{k0} I_{i \in [k]} +  \bx_i ^\T  \gamma(0), \quad A_i = 0. \nonumber
%\end{eqnarray}

%\begin{eqnarray}
%&& \frac{ \sigma^2_{\tilde r(1)} }{ 1 - \pi } + \frac{ \sigma^2_{\tilde r(0)} }{ \pi }  \nonumber \\
%& = & \varsigma^2_{\tilde r_{\inter}}( 1 - \pi )  - \frac{1}{1 - \pi}  \{ \beta(1)  -  \beta_{\inter} \} ^\T \Sigma_{\tilde \bx \tilde \bx} \{ \beta(1)  -  \beta_{\inter} \} - \frac{1}{\pi}  \{ \beta(0)  -  \beta_{\inter} \} ^\T \Sigma_{\tilde \bx \tilde \bx} \{ \beta(0)  -  \beta_{\inter} \}. \nonumber
%\end{eqnarray}

% \quad r_i(a) = Y_i(a) - \bx_i^\T \beta(a)

%Let $\hat \beta_{\inter}$ be the OLS estimator of $$

% \quad \hat \sigma^{2*}_{\inter} \xrightarrow{P}  \varsigma^2_{\tilde r}( 1 - \pi ) ,

\begin{theorem}
\label{thm::inter}
Suppose that $( r_{i,\inter}(1), r_{i,\inter}(0) ) \in \mathcal{R}_2$ and Assumptions~\ref{assum::Q}, \ref{assum::A1} and \ref{assum::A3} hold, then
$$
\sqrt{n} ( \tauinter - \tau ) \xrightarrow{d} \mathcal{N}  (0,  \varsigma^2_{\tilde r_{\inter}}(\pi) + \varsigma^2_{Hr_{\inter}}  ), 
$$
$$
n \hat \sigma^{2*}_{\inter} \xrightarrow{P}  \varsigma^2_{\tilde r_{\inter}}( 1 - \pi )  - \Big \{  \frac{1}{\pi ( 1 - \pi )} - 3  \Big \}   \{ \beta(1)  -  \beta(0) \} ^\T \Sigma_{\tilde \bx \tilde \bx} \{ \beta(1)  -  \beta(0) \},
$$
and the asymptotic variance can be consistently estimated by $\hat  \varsigma^2_{\tilde r_{\essed} }( \pi ) + \hat \varsigma^2_{H r_{\essed} } $. Furthermore, the difference of the asymptotic variances of $\tauinter$ and $\taustr$ is
\begin{equation}
\Delta_{\inter * - \str }  =    -  \frac{1}{\pi ( 1 - \pi ) }  \beta_{\inter}^\T \Sigma_{\tilde \bx \tilde \bx} \beta_{\inter} \leq 0, \nonumber
\end{equation}
and the difference between the asymptotic variances of $\tauinter$ and $\tauancova$ is
\begin{equation}
\Delta_{\inter *  - \ancova *} =  -  \frac{1}{\pi ( 1 - \pi ) }    ( \beta_{\inter} -  \beta_{\ancova} ) ^\T \Sigma_{\tilde \bx \tilde \bx}  ( \beta_{\inter} -  \beta_{\ancova} ) ^\T   - \varsigma^2_{\pi r_{\ancova}} \leq 0. \nonumber
\end{equation}
\end{theorem}

%\begin{remark}
%By the proof of Theorem~\ref{thm::inter}, the second-order interaction regression  is equivalent to regressing $Y_i$ on $ I_{i \in [k] }$ and  $\bx_i$ in the treatment and control groups respectively and obtaining the OLS estimator of the coefficients of $\bx_i$, $\hat \gamma(1)$ and $\hat \gamma(0)$, then 
%$$
%\tauinter =  \sumk \pnk \Big[  \Big\{   \YkThat - ( \XkThat -  \XkT )^\T \hat \gamma(1)   \Big\}  - \Big\{  \YkChat - ( \XkChat -  \XkC )^\T \hat \gamma(0)   \Big\}  \Big].
%$$
%\end{remark}
\begin{remark}
Theorem~\ref{thm::inter} still holds if we replace Assumption~\ref{assum::A3} by an even weaker condition $\pik  \xrightarrow{P} \pi$; however, we still use Assumption~\ref{assum::A3} in the theorem for the same reason as stated in  Remark~\ref{remark4}.
%We use Assumption~\ref{assum::A3} since it holds for almost all covariate-adaptive randomizations.
\end{remark}

% We use Assumption~\ref{assum::A3} for two reasons: first, it is satisfied by almost all covariate-adaptive randomizations, and second, we require Assumption~\ref{assum::A3}  to obtain the validity of the regression without interactions methods  in Sections 3.2 and 4.2 for the case of equal allocation ($\pi = 1/2$).

Theorem~\ref{thm::inter} implies that the regression with treatment-by-covariate interactions results in an estimator $\tauinter$ whose asymptotic variance is no greater than that of $\taustr$, the most efficient estimator among $\hat \tau$, $\taufe$, and $\taustr$. That is, using additional covariates in an interaction regression will improve, or at least  not hurt, the efficiency. As $\Delta_{\inter * - \ancova *}  \leq 0$,  the  regression with treatment-by-covariate interactions is generally more efficient than ANCOVA. As $1/\{ \pi ( 1 - \pi ) \} - 3 \geq 1$,   the OLS variance estimator $\hat \sigma^{2*}_{\inter}$ may underestimate the asymptotic variance even when $\pi = 1/2$ and thus should not be used. Theorem~\ref{thm::inter}  provides a consistent variance estimator that can be used to construct a valid confidence interval or test for $\tau$. We note that the asymptotic variance of $\tauinter$ and the proposed variance estimator are invariant with respect to randomization methods.

% covariate-adaptive randomization-free.  

% is invariant under different randomization methods

\begin{corollary}
\label{cor::ancovainter}
When $\pi = 1/2$, under Assumptions \ref{assum::Q}, \ref{assum::A1}, and \ref{assum::A3}, $\tauancova$ and $\tauinter$ are asymptotically equivalent and their asymptotic variances are smaller than or equal to that of $\taustr$. Furthermore, both $n \hat \sigma^{2*}_{\ancova}$ and  $\hat \varsigma^2_{\tilde r_{\essed} }(\pi) + \hat \varsigma^2_{H r_{\essed} } $ are consistent variance estimators. 
\end{corollary}

Corollary~\ref{cor::ancovainter} implies that, when $\pi = 1/2$,  $\tauancova$ is as efficient as $\tauinter$, neither of which hurts the efficiency, as compared with $\hat \tau$, $\taufe$, and $\taustr$, and it is valid and efficient to use the OLS point and variance estimators in ANCOVA. {\color{black} For equal allocation,  we recommend the ANCOVA for two reasons: first, it has significantly fewer  parameters than the regression with treatment-by-covariate interactions, and second,  it is easily implemented by standard statistical packages.}

% All the above three regression adjusted estimators $\tausimp$, $\tauancova$ and $\tauinter$ have small finite-sample bias, but their biases vanish quickly, at the rate of $n^{-1/2}$. 

\subsection{Summary and recommendation} 

{\color{black} At the end of this section, we present a summary of the asymptotic variances and variance estimators  in Table~\ref{tab::var}, which  provides answers to the open question raised by \citet{Wang2019} (stated in the introduction). Based on the above arguments, when taking into account both simplicity and efficiency, Table~\ref{tab::recom} shows our recommendations for regression estimators that adjust for additional covariates $\bx_i$. As in Section 3.4, these recommendations apply to most of the widely used covariate-adaptive randomization methods, including stratified randomization and minimization.}

 \section{$\mathcal{S}$-optimal}

% that  converge in probability to finite limits $\eta(1)$ and $\eta(0)$,

In this section, we will show that $\tauinter$ is optimal among the following class of  estimators: 
$$ \mathcal{S} \triangleq \{ \hat \tau^* ( \hat \eta(1), \hat \eta(0)  ), \hat \tau^*_{\ancova} ( \hat \eta(1), \hat \eta(0)  ) , \hat \tau^*_{\inter} ( \hat \eta(1), \hat \eta(0)  )   \}, $$ 
where $\hat \eta(1)$ and $\hat \eta(0)$ are some adjusted vectors  and
$$
\hat \tau^* ( \hat \eta(1), \hat \eta(0)  ) = \{ \YT -  ( \XT - \bar{\bx} ) ^\T \hat \eta(1)  \}   -  \{ \YC -  ( \XC - \bar{\bx} ) ^\T \hat \eta(0) \},
$$
$$
\tauancova (  \hat \eta(1), \hat \eta(0) ) = \sumk \omega_{[k]} \big[  \{  \YkThat - ( \XkThat - \Xkbar ) ^\T  \hat \eta(1) \}    -   \{ \YkChat -  ( \XkChat - \Xkbar ) ^\T \hat \eta(0) \}  \big ],
$$
$$
\tauinter (  \hat \eta(1), \hat \eta(0) ) = \sumk \pnk \big[  \{  \YkThat - ( \XkThat - \Xkbar ) ^\T  \hat \eta(1) \}    -   \{ \YkChat -  ( \XkChat - \Xkbar ) ^\T \hat \eta(0) \}  \big ].
$$

\begin{definition}
Given the randomization mechanism, $\tauinter$ is $\mathcal{S}$-optimal if it has the smallest asymptotic variance among the estimators in $\mathcal{S}$.
\end{definition}
The definition of the $\mathcal{S}$-optimal estimator appeared in \cite{LiDing2020} for completely randomized experiments, which we extend to covariate-adaptive randomization. {\color{black} In additional to $\tausimp$, $\tauancova$, and $\tauinter$, $\mathcal{S}$ contains other regression estimators, such as those with added interactions $A_i ( \bx_i - \bar{\bx} ) $ in regression~\eqref{eqn::reg-simp} or deleted interactions $A_i ( \bx_i -  \bar{\bx} )$ from regression~\eqref{eqn::reg_inter}. Our next theorem shows that $\tauinter$ is $\mathcal{S}$-optimal, so those regressions need not be  investigated. } Define $r_{i, \text{gen}}(a) = Y_i(a) - \bx_i^\T \{ ( 1 - \pi ) \eta(1) + \pi \eta(0)  \}$, $a=0,1$.
%In completely randomized clinical trials, \cite{LiDing2020} called this type of estimator $\mathcal{S}$-optimal.

% are asymptotically equivalent to $\hat \tau_{\diff} ( \eta, \eta )$, $\tauancova ( \eta_{\ancova}, \eta_{\ancova} )$, and $\tauinter ( \eta_{\inter}, \eta_{\inter})$, respectively. 

% the difference-in-means estimator for the potential outcomes $r_{i, \simp}(a) = Y_i(a) - \bx_i^\T  \eta $, $a=0,1$,
%$$
%\hat \tau_{\diff} ( \eta ) = ( \YT -  \XT ^\T  \eta  )   -  ( \YC -  \XC ^\T  \eta  ) .
%$$
%The ANCOVA estimator $\tauancova$ is asymptotically equivalent to the fixed effect estimator for the potential outcomes $r_{i, \ancova}(a) = Y_i(a) - \bx_i^\T  \eta_{\ancova} $:
%$$
%\tauancova (  \eta_{\ancova} ) = \sumk \omega_{[k]}  \{  ( \YkThat - \XkThat ^\T  \eta_{\ancova} )    -   ( \YkChat - \XkChat ^\T  \eta_{\ancova} )  ].
%$$
%The second-order interaction regression estimator $\tauinter$ is asymptotically equivalent to the stratified difference-in-means estimator for the potential outcomes $r_{i, \inter}(a) = Y_i(a) - \bx_i^\T  \eta_{\inter} $:
%$$
%\tauinter (  \eta_{\inter} ) = \sumk \pnk  \{  ( \YkThat - \XkThat ^\T  \eta_{\inter} )    -   ( \YkChat - \XkChat ^\T  \eta_{\inter} )  ].
%$$

% $\hat \tau_{\diff} ( \eta ) $, $\tauancova ( \eta ) $ and $\tauinter (\eta)$ for any fixed coefficients vector $\eta \in R^p$,

\begin{theorem}
\label{thm::optimal}
Suppose that Assumptions~\ref{assum::Q} -- \ref{assum::A2} hold, $\hat \eta(1) \xrightarrow{P} \eta(1)$, $\hat \eta(0) \xrightarrow{P} \eta(0)$, and $( r_{i,\text{gen}}(1), r_{i,\text{gen} }(0) ) \in \mathcal{R}_2$, then, $\tauinter $ is $\mathcal{S}$-optimal.
\end{theorem}

\begin{remark}
We note that $\tauinter$ is not optimal among all regression adjusted estimators for at least two reasons. First,  there exists weights $\tilde \omega_{[k]}$ such that  $ \sumk \tilde \omega_{[k]} [ \{   \YkThat - \XkThat ^\T  \hat \beta (1) \}    -   \{ \YkChat - \XkChat ^\T  \hat \beta (0) \}  ] $ has a smaller asymptotic variance than $\tauinter$. However, these weights  $\tilde \omega_{[k]}$  depend on the unknown potential outcomes in a complicated form. We leave  to future work the investigation of semi-parametric efficiency under covariate-adaptive randomization. Second,  to further improve efficiency, we can use the stratum-specific adjusted vectors $\hat \beta_{[k]}(1)$ and $\hat \beta_{[k]}(0)$ within stratum $k$  instead of the common adjusted vectors $\hat \beta(1)$ and $\hat \beta(0)$ for all strata.  Equivalently,  the use of higher-order interactions, such as $A_i I_{i \in [k] } \bx_i$,  can further improve efficiency in theory; however,  evidence suggested that this may lead to inferior performance when there exists small strata \citep{Liu2019}.

\end{remark}

% Theorem~\ref{thm::optimal} implies that  $\tauinter$ is the most efficient estimator among the estimators in $\mathcal{S}$. 

% $ \beta_{\inter} =  \beta_{\ancova} $ and $\varsigma^2_{\pi r_{\ancova}} = 0$, thus,

%As the same as finite-population setting, the interaction regression results in  an estimator $\tauinter$ whose asymptotic variance is equal to or typically less than those of the difference-in-means and fixed effect estimators. That is, asymptotically, $\tauinter$ improves, at least does not hurt, precision for estimating population average treatment effect.

% \section{Optimality of Linear Regression Adjustment}

%  potential outcomes $r_i(\gamma, a) = Y_i(a) - \bx_i^\T \gamma$, $a=0,1$.

% Compared with the other two regression analysis methods, $\tauinter$ has the smallest asymptotic variance. For a given $p$-dimensional coefficients vector $\gamma$, define

\section{Simulation Study}
\label{SimStudy}

In this section, we report the results of a simulation study in which the empirical performances of the six proposed estimators were examined with respect to treatment effect estimation and asymptotic variance. For comparison, the typical OLS and Huber--White variance estimators are also considered. For $a \in \{0,1\}$ and  $1 \leq i \leq n$, the potential outcomes are generated using the equation:
$
Y_i(a) = \mu_a + g_a(\bx_i) + \sigma_a(\bx_i)\varepsilon_{a,i},
$
where $\bx_i$, $g_a(\bx_i)$, and $\sigma_a(\bx_i)$ are specified below for three different models. In each of the models, $(\bx_i, \varepsilon_{0,i}, \varepsilon_{1,i}), 1 \leq i \leq n$ are i.i.d, and both $\varepsilon_{0,i}$ and $\varepsilon_{1,i}$ follow the standard normal distribution. The number of units $n$ is $1000$.

Model 1 (Linear model): $\bx_i$ is a five-dimensional vector, $$g_0(\bx_i) = g_1(\bx_i) = \sum\limits_{j=1}^5 \beta_j X_{ij},$$ where $X_{i1} \sim \textup{Beta}(2,2)$, $X_{i2}$ takes values in $\{1, 2, 3, 4\}$ with equal probability, $X_{i3} \sim \textup{Unif}[-2,2]$, $X_{i4}$ takes values in $\{1, 2, 3\}$ with respective probabilities of $0.3,  0.6,  0.1$ and $X_{i5} \sim \mathcal{N}(0,1)$, and they are all independent of each other. $\sigma_0(\bx_i) = 1, \ \sigma_1(\bx_i) = 3, \ \beta = (2, 8, 10, 3, 6)^\T$. $X_{i2}$ and $X_{i4}$ are used for randomization, and  $X_{i1}$ and $X_{i3}$ are used as the additional covariates.

Model 2 (Non-linear model with additional information): $\bx_i$ is a four-dimensional vector,
$$\begin{aligned}g_0(\bx_i) &= \alpha_1 X_{i1} + \log(\alpha_3 X_{i1} \log(X_{i3}+1)+1) + \alpha_4 e^{X_{i4}}, \\
g_1(\bx_i) &= \alpha_2 X_{i2}^2 + \log(\alpha_3 X_{i1} \log(X_{i3}+1)+1),
\end{aligned}$$
 where $X_{i1} \sim \textup{Gamma}(2,1)$, $X_{i2}$ takes values in \{1, 2, 3\} with respective probabilities of $0.3, 0.6, 0.1$, $X_{i3} \sim \textup{Poisson}(3), \ X_{i4} \sim \textup{Beta}(2,2)$, and they are all independent of each other. $\sigma_0(\bx_i) = 2, \ \sigma_1(\bx_i) = 1, \ \alpha = (5, 10, 3, 20)^\T$. $X_{i1}$ and $X_{i2}$ are used for randomization, and the stratum of $X_{i1}$ is determined by its relative value of $2.5$. $X_{i1}$ and $X_{i3}$ are used as the additional covariates.

Model 3 (Non-linear model with dependent covariates and errors): $\bx_i$ is a four-dimensional vector,
$$\begin{aligned}
g_0(\bx_i) &= \sum\limits_{j=1}^4 \beta_j X_{ij}, \\
g_1(\bx_i) &= \beta_1 \log(X_{i1})X_{i4},
\end{aligned}$$
 where $X_{i1} \sim \textup{Beta}(3,4)$, $X_{i2} \sim \textup{Unif}[-2,2]$, $X_{i3} = X_{i1} X_{i2}$, $X_{i4}$ takes values in $\{3, 5\}$ with respective probabilities of $0.6, 0.4$. $X_{i1}, \ X_{i2}, \ X_{i4}$ are independent of each other. $\sigma_0(\bx_i) = X_{i3S}, \ \sigma_1(\bx_i) = 2X_{i2S},$ and $ \beta = (20, 7, 5, 6)^\T$, where $X_{i2S}$ is the stratified variable of $X_{i2}$, if $X_{i2} >1, X_{i2S} = 2$, and $X_{i2S} = 1$ otherwise, and $X_{i3S}$ is the stratified variable of $X_{i3}$, if $X_{i3} >0, X_{i3S} = 2$, and $X_{i3S} = 1$ otherwise. $X_{i2S}$ and $X_{i4}$ are used for randomization, $X_{i1}$ and $X_{i3}$ are used as the additional covariates.

Here we present the simulation results of the six estimators under simple randomization, the stratified block randomization \citep{zelen1974randomization}, and Pocock and Simon's minimization \citep{Pocock1975} in Table \ref{equalsim} (equal allocation) and Table \ref{unequalsim} (unequal allocation).  The block size used in the stratified block randomization was $6$. The biased-coin probability and the weight used in the Pocock and Simon's minimization were $0.75$ and $(0.5,0.5)^\T$, respectively.  The bias, standard deviation (SD) of the treatment effect estimators, standard error (SE) estimators, and  empirical coverage probabilities (CP) were computed using $10^4$ replications. For bias calculation, the true treatment effect is defined as $\mu_1 - \mu_0 + E\{g_1(\bx_i)\} - E\{g_0(\bx_i)\}$ and was 
evaluated numerically by integration.
Note that, under Pocock and Simon's minimization, the standard error estimators of $\taustr$ and $\tauinter$, and when $\pi = 1/2$, $\taufe$ and $\tauancova$, are calculated according to the known theory, but they are not available for other estimators.  Additional simulation results under biased-coin design \citep{Efron1971}, adaptive biased-coin design \citep{Wei1978}, and the randomization method proposed by \citet{Hu2012} can be found in the Appendix B.

Table~\ref{equalsim} presents the simulation results under equal allocation. First, we can see that the six treatment effect estimators all have small finite-sample biases. Under simple randomization, the standard deviations of $\taufe$ and $\taustr$ are comparable and smaller than that of $\hat{\tau}$, with the reduction varying from approximately $3\%$ to $28\%$ for different models. Regarding stratified block randomization, which achieves  strong balance,  the standard deviations of these three estimators are nearly the same, which is consistent with the theoretical results. Under Pocock and Simon's minimization, the first three estimators also tend to have similar standard deviations. Moreover, by adding the stratification indicators into the regression, almost identical standard deviations are obtained by the different randomization methods, regardless of whether the interaction terms are included.

\begin{table}[H]
	\centering
	\caption{Simulated biases, standard deviations, standard errors, and coverage probabilities for different estimators and randomization methods under equal allocation ($\pi=1/2$).}\label{equalsim}
	\vskip 5mm
	\begin{threeparttable}
		\setlength{\tabcolsep}{2pt}
\resizebox{\textwidth}{55mm}{
		\begin{tabular}{llcccccccccccccccccccccccc}
		\cline{1-26}
		&  & \multicolumn{8}{c}{Simple Randomization} & \multicolumn{8}{c}{Stratified Block Randomization} & \multicolumn{8}{c}{Minimization} \\ \cline{3-26}
		&
		&
		\multicolumn{1}{c}{Bias} &
		\multicolumn{1}{c}{SD} &
		\multicolumn{3}{c}{SE} &
        \multicolumn{3}{c}{CP} &
		\multicolumn{1}{c}{Bias} &
		\multicolumn{1}{c}{SD} &
		\multicolumn{3}{c}{SE} &
        \multicolumn{3}{c}{CP} &
		\multicolumn{1}{c}{Bias} &
		\multicolumn{1}{c}{SD} &
		\multicolumn{3}{c}{SE} &
        \multicolumn{3}{c}{CP}  \\ \cline{5-7} \cline{8-10} \cline{13-15} \cline{16-18} \cline{21-23} \cline{24-26}
		Model &
		Estimator &
		\multicolumn{1}{c}{} &
		\multicolumn{1}{c}{} &
		\multicolumn{1}{c}{NEW} &
		\multicolumn{1}{c}{OLS} &
		\multicolumn{1}{c}{HW} &
		\multicolumn{1}{c}{NEW} &
		\multicolumn{1}{c}{OLS} &
		\multicolumn{1}{c}{HW} &
		\multicolumn{1}{c}{} &
		\multicolumn{1}{c}{} &
		\multicolumn{1}{c}{NEW} &
		\multicolumn{1}{c}{OLS} &
		\multicolumn{1}{c}{HW} &
		\multicolumn{1}{c}{NEW} &
		\multicolumn{1}{c}{OLS} &
		\multicolumn{1}{c}{HW} &
		\multicolumn{1}{c}{} &
		\multicolumn{1}{c}{} &
		\multicolumn{1}{c}{NEW} &
		\multicolumn{1}{c}{OLS} &
		\multicolumn{1}{c}{HW} &
		\multicolumn{1}{c}{NEW} &
		\multicolumn{1}{c}{OLS} &
		\multicolumn{1}{c}{HW}  \\ \hline
1&$\hat{\tau}$ & -0.04 & 1.01 & 1.01 & 1.02 & 1.01 & 0.95 & 0.95 & 0.95 & 0.00 & 0.83 & 0.83 & 1.02 & 1.01 & 0.95 & 0.99 & 0.99 & -0.01 & 0.85 & - & 1.02 & 1.01 & - & 0.97 & 0.97 \\
  &$\taufe$ & -0.02 & 0.83 & 0.83 & 0.84 & 0.83 & 0.95 & 0.95 & 0.95 & 0.01 & 0.83 & 0.83 & 0.84 & 0.83 & 0.95 & 0.95 & 0.95 & -0.00 & 0.85 & 0.83 & 0.84 & 0.83 & 0.94 & 0.94 & 0.94 \\
  &$\taustr$ & -0.02 & 0.83 & 0.83 & 0.84 & 0.83 & 0.95 & 0.95 & 0.95 & 0.01 & 0.83 & 0.83 & 0.84 & 0.83 & 0.95 & 0.95 & 0.95 & -0.00 & 0.85 & 0.83 & 0.84 & 0.83 & 0.94 & 0.95 & 0.94 \\
  &$\tausimp$ & -0.01 & 0.73 & 0.70 & 0.71 & 0.70 & 0.95 & 0.95 & 0.95 & -0.01 & 0.39 & 0.40 & 0.71 & 0.70 & 0.95 & 1.00 & 1.00 & -0.01 & 0.42 & - & 0.71 & 0.70 & - & 1.00 & 1.00 \\
  &$\tauancova$ & 0.01 & 0.42 & 0.40 & 0.41 & 0.40 & 0.95 & 0.95 & 0.95 & -0.00 & 0.39 & 0.40 & 0.41 & 0.40 & 0.95 & 0.95 & 0.95 & -0.01 & 0.42 & 0.40 & 0.41 & 0.40 & 0.94 & 0.94 & 0.94 \\
  &$\tauinter$ & 0.01 & 0.42 & 0.40 & 0.41 & 0.40 & 0.95 & 0.95 & 0.94 & -0.00 & 0.39 & 0.40 & 0.41 & 0.40 & 0.95 & 0.95 & 0.95 & -0.00 & 0.42 & 0.40 & 0.41 & 0.40 & 0.94 & 0.94 & 0.94 \\

  2&$\hat{\tau}$ & 0.01 & 1.13 & 1.12 & 1.12 & 1.12 & 0.94 & 0.94 & 0.94 & -0.03 & 0.85 & 0.84 & 1.12 & 1.12 & 0.94 & 0.99 & 0.99 & -0.05 & 0.86 & - & 1.12 & 1.12 & - & 0.98 & 0.98 \\
  &$\taufe$ & -0.02 & 0.84 & 0.84 & 0.84 & 0.85 & 0.95 & 0.95 & 0.95 & 0.01 & 0.85 & 0.84 & 0.84 & 0.84 & 0.94 & 0.94 & 0.94 & -0.07 & 0.86 & 0.84 & 0.84 & 0.84 & 0.95 & 0.95 & 0.95 \\
  &$\taustr$ & 0.03 & 0.82 & 0.84 & 0.41 & 0.41 & 0.95 & 0.67 & 0.66 & 0.01 & 0.85 & 0.84 & 0.41 & 0.41 & 0.94 & 0.66 & 0.65 & -0.04 & 0.86 & 0.84 & 0.41 & 0.41 & 0.95 & 0.64 & 0.64 \\
  &$\tausimp$ & 0.02 & 1.11 & 1.09 & 1.09 & 1.09 & 0.94 & 0.94 & 0.94 & -0.01 & 0.84 & 0.82 & 1.08 & 1.09 & 0.94 & 0.99 & 0.99 & -0.03 & 0.85 & - & 1.09 & 1.09 & - & 0.99 & 0.99 \\
  &$\tauancova$ & -0.01 & 0.83 & 0.82 & 0.82 & 0.83 & 0.95 & 0.95 & 0.95 & 0.03 & 0.83 & 0.82 & 0.83 & 0.82 & 0.94 & 0.94 & 0.94 & -0.05 & 0.85 & 0.82 & 0.83 & 0.83 & 0.95 & 0.95 & 0.95 \\
  &$\tauinter$ & 0.02 & 0.81 & 0.82 & 0.35 & 0.35 & 0.95 & 0.60 & 0.59 & 0.02 & 0.83 & 0.82 & 0.35 & 0.35 & 0.94 & 0.60 & 0.60 & -0.04 & 0.84 & 0.82 & 0.35 & 0.35 & 0.95 & 0.59 & 0.59 \\
  3&$\hat{\tau}$ & 0.00 & 1.99 & 2.00 & 2.00 & 2.00 & 0.95 & 0.95 & 0.95 & -0.03 & 1.92 & 1.94 & 2.00 & 2.00 & 0.95 & 0.96 & 0.96 & -0.07 & 1.81 & - & 2.00 & 2.00 & - & 0.96 & 0.96 \\
  &$\taufe$ & 0.03 & 1.91 & 1.94 & 1.95 & 1.95 & 0.96 & 0.96 & 0.96 & -0.03 & 1.92 & 1.94 & 1.94 & 1.94 & 0.95 & 0.95 & 0.95 & -0.06 & 1.81 & 1.94 & 1.94 & 1.94 & 0.96 & 0.96 & 0.96 \\
  &$\taustr$ & -0.01 & 1.91 & 1.94 & 1.77 & 1.76 & 0.96 & 0.93 & 0.93 & -0.03 & 1.92 & 1.94 & 1.76 & 1.76 & 0.95 & 0.92 & 0.92 & -0.08 & 1.81 & 1.94 & 1.76 & 1.76 & 0.96 & 0.93 & 0.93 \\
  &$\tausimp$ & -0.05 & 1.58 & 1.54 & 1.55 & 1.55 & 0.95 & 0.95 & 0.95 & -0.02 & 1.45 & 1.48 & 1.54 & 1.54 & 0.96 & 0.97 & 0.97 & -0.00 & 1.44 & - & 1.55 & 1.54 & - & 0.96 & 0.96 \\
  &$\tauancova$ & -0.02 & 1.52 & 1.49 & 1.49 & 1.50 & 0.95 & 0.95 & 0.95 & -0.02 & 1.45 & 1.48 & 1.49 & 1.49 & 0.96 & 0.96 & 0.96 & 0.01 & 1.44 & 1.49 & 1.49 & 1.49 & 0.95 & 0.95 & 0.95 \\
  &$\tauinter$ & -0.04 & 1.50 & 1.49 & 0.70 & 0.70 & 0.95 & 0.62 & 0.62 & 0.01 & 1.45 & 1.48 & 0.70 & 0.69 & 0.96 & 0.64 & 0.64 & 0.02 & 1.44 & 1.49 & 0.70 & 0.70 & 0.95 & 0.66 & 0.65 \\
   \hline
		\end{tabular}}
\begin{tablenotes}
\item Note: SD, standard deviation; SE, standard error; CP, coverage probability; HW, the \\ Huber--White variance estimator; NEW: the proposed non-parametric variance estimator.
%\item Abbreviation: NA, not available.
\end{tablenotes}
\end{threeparttable}
\end{table}

Making use of additional covariates in the regression can further improve the precision when estimating the treatment effect, as shown in Table~\ref{equalsim}. More precisely, compared with the standard deviations of $\taustr$, the standard deviations of $\tauinter$ are $1.22 \% - 51.81\%$ smaller under simple randomization, and $2.38\%-51.81\%$ smaller under stratified block randomization and Pocock and Simon's minimization.  Note that the relationship among the last three estimators is the same as that among the first three estimators.

Table~\ref{equalsim} also confirms that the proposed variance estimators coincide with the simulated variance of the treatment effect estimators and lead to the expected 95\% coverage probability. Moreover, the OLS and Huber--White variance estimators are valid for $\taufe$ and $\tauancova$, so in this situation, we can rely on the output of standard statistical packages, such as lm in R package stat, to obtain a valid inference. The OLS and Huber--White variance estimators are also valid for Model $1$, but the model assumptions are very strong and may not be satisfied in practice. For other scenarios, however, the OLS and Huber--White variance estimators are generally not valid and can be either larger or smaller than the true values. In particular, both the variance estimators tend to be anti-conservative for $\taustr$, which concurs with the results of \citet{Bugni2019}.

Table~\ref{unequalsim} shows the simulation results under unequal allocation ($\pi$ = 2/3). Our findings can be summarized as follows. First, both $\taufe$ and $\tauancova$ appear to have larger biases, whereas the biases of the other four estimators are negligible.
Second, unlike equal allocation, adjustment of the stratification indicators does not guarantee a gain in efficiency when the stratification-by-treatment interactions are not included; the standard deviation of $\taufe$ is larger than that of $\hat{\tau}$ under model $3$. However, adding the interactions always reduces the standard deviation. For example, under simple randomization, the standard deviations of $\taustr$ are $2.22\% -40.97\%$ and $1.14\% -18.27\%$ smaller than those of $\hat{\tau}$ and $\taufe$, respectively. Similar conclusions can be drawn for the last three estimators. Third, adjusting for additional covariates can further improve the efficiency. Among the six estimators, $\tauinter$ is the most efficient, and its variance is invariant under different randomization methods, which reflects the benefit of adding treatment-by-covariate interactions under unequal allocation. 
Finally, the OLS and Huber--White variance estimators of $\taustr$ or $\tauinter$ are no longer similar, and they are not valid under most of the considered scenarios. In extreme cases, the coverage probability when using the OLS variance estimator is as small as $0.10$. The proposed variance estimators, in contrast, are still valid and lead to the expected $95$\% coverage probability under all scenarios.

\begin{table}[H]
	\centering
	\caption{Simulated biases, standard deviations, standard errors, and coverage probabilities for different estimators and randomization methods under unequal allocation ($\pi=2/3$).}\label{unequalsim}
	\vskip 5mm
	\begin{threeparttable}
		\setlength{\tabcolsep}{2pt}
\resizebox{\textwidth}{55mm}{
		\begin{tabular}{llcccccccccccccccccccccccc}
		\cline{1-26}
		&  & \multicolumn{8}{c}{Simple Randomization} & \multicolumn{8}{c}{Stratified Block Randomization} & \multicolumn{8}{c}{Minimization} \\ \cline{3-26}
		&
		&
		\multicolumn{1}{c}{Bias} &
		\multicolumn{1}{c}{SD} &
		\multicolumn{3}{c}{SE} &
        \multicolumn{3}{c}{CP} &
		\multicolumn{1}{c}{Bias} &
		\multicolumn{1}{c}{SD} &
		\multicolumn{3}{c}{SE} &
        \multicolumn{3}{c}{CP} &
		\multicolumn{1}{c}{Bias} &
		\multicolumn{1}{c}{SD} &
		\multicolumn{3}{c}{SE} &
        \multicolumn{3}{c}{CP}  \\ \cline{5-7} \cline{8-10} \cline{13-15} \cline{16-18} \cline{21-23} \cline{24-26}
		Model &
		Estimator &
		\multicolumn{1}{c}{} &
		\multicolumn{1}{c}{} &
		\multicolumn{1}{c}{NEW} &
		\multicolumn{1}{c}{OLS} &
		\multicolumn{1}{c}{HW} &
		\multicolumn{1}{c}{NEW} &
		\multicolumn{1}{c}{OLS} &
		\multicolumn{1}{c}{HW} &
		\multicolumn{1}{c}{} &
		\multicolumn{1}{c}{} &
		\multicolumn{1}{c}{NEW} &
		\multicolumn{1}{c}{OLS} &
		\multicolumn{1}{c}{HW} &
		\multicolumn{1}{c}{NEW} &
		\multicolumn{1}{c}{OLS} &
		\multicolumn{1}{c}{HW} &
		\multicolumn{1}{c}{} &
		\multicolumn{1}{c}{} &
		\multicolumn{1}{c}{NEW} &
		\multicolumn{1}{c}{OLS} &
		\multicolumn{1}{c}{HW} &
		\multicolumn{1}{c}{NEW} &
		\multicolumn{1}{c}{OLS} &
		\multicolumn{1}{c}{HW}  \\ \hline
1&$\hat{\tau}$ & 0.01 & 1.08 & 1.07 & 1.08 & 1.07 & 0.95 & 0.96 & 0.95 & -0.05 & 0.89 & 0.87 & 1.07 & 1.07 & 0.95 & 0.98 & 0.98 & -0.05 & 0.86 & - & 1.08 & 1.07 & - & 0.99 & 0.99 \\
  &$\taufe$ & 0.01 & 0.91 & 0.88 & 0.90 & 0.88 & 0.93 & 0.94 & 0.94 & -0.03 & 0.89 & 0.87 & 0.88 & 0.87 & 0.95 & 0.95 & 0.95 & -0.05 & 0.84 & - & 0.89 & 0.88 & - & 0.96 & 0.96 \\
  &$\taustr$ & 0.01 & 0.91 & 0.87 & 0.90 & 0.88 & 0.94 & 0.94 & 0.94 & -0.03 & 0.89 & 0.87 & 0.88 & 0.86 & 0.95 & 0.95 & 0.95 & -0.05 & 0.85 & 0.87 & 0.89 & 0.87 & 0.96 & 0.96 & 0.96 \\
  &$\tausimp$ & -0.01 & 0.74 & 0.74 & 0.75 & 0.74 & 0.95 & 0.96 & 0.95 & -0.03 & 0.43 & 0.42 & 0.75 & 0.74 & 0.95 & 1.00 & 1.00 & 0.00 & 0.43 & - & 0.75 & 0.74 & - & 1.00 & 1.00 \\
  &$\tauancova$ & -0.00 & 0.42 & 0.42 & 0.44 & 0.42 & 0.95 & 0.96 & 0.95 & -0.02 & 0.42 & 0.42 & 0.43 & 0.42 & 0.95 & 0.96 & 0.95 & 0.01 & 0.41 & - & 0.44 & 0.42 & - & 0.97 & 0.96 \\
  &$\tauinter$ & -0.00 & 0.42 & 0.42 & 0.44 & 0.42 & 0.95 & 0.96 & 0.95 & -0.02 & 0.43 & 0.42 & 0.43 & 0.41 & 0.95 & 0.96 & 0.94 & 0.01 & 0.41 & 0.42 & 0.44 & 0.42 & 0.96 & 0.97 & 0.96 \\
 
  2&$\hat{\tau}$ & -0.01 & 1.04 & 1.06 & 1.30 & 1.06 & 0.95 & 0.98 & 0.95 & -0.11 & 0.89 & 0.89 & 1.29 & 1.06 & 0.94 & 0.99 & 0.98 & -0.07 & 0.92 & - & 1.30 & 1.05 & - & 1.00 & 0.98 \\
  &$\taufe$ & -0.09 & 1.00 & 1.02 & 0.81 & 1.03 & 0.96 & 0.88 & 0.96 & 0.03 & 0.89 & 0.89 & 0.82 & 1.02 & 0.95 & 0.94 & 0.97 & -0.01 & 0.92 & - & 0.82 & 1.02 & - & 0.92 & 0.97 \\
  &$\taustr$ & -0.02 & 0.85 & 0.88 & 0.36 & 0.50 & 0.96 & 0.59 & 0.75 & -0.02 & 0.89 & 0.89 & 0.36 & 0.49 & 0.95 & 0.59 & 0.72 & -0.01 & 0.91 & 0.88 & 0.36 & 0.49 & 0.95 & 0.56 & 0.71 \\
  &$\tausimp$ & 0.00 & 0.98 & 1.01 & 1.29 & 1.01 & 0.95 & 0.99 & 0.95 & -0.09 & 0.88 & 0.86 & 1.28 & 1.01 & 0.95 & 0.99 & 0.97 & -0.05 & 0.89 & - & 1.28 & 1.01 & - & 1.00 & 0.97 \\
  &$\tauancova$ & -0.08 & 0.98 & 1.01 & 0.80 & 1.01 & 0.95 & 0.90 & 0.95 & 0.03 & 0.87 & 0.86 & 0.81 & 1.00 & 0.95 & 0.93 & 0.98 & 0.00 & 0.90 & - & 0.81 & 1.00 & - & 0.92 & 0.97 \\
  &$\tauinter$ & -0.02 & 0.82 & 0.86 & 0.31 & 0.43 & 0.95 & 0.53 & 0.69 & -0.02 & 0.87 & 0.86 & 0.31 & 0.42 & 0.95 & 0.52 & 0.67 & -0.01 & 0.88 & 0.86 & 0.31 & 0.43 & 0.94 & 0.52 & 0.67 \\
  3&$\hat{\tau}$ & -0.01 & 1.86 & 1.80 & 2.40 & 1.80 & 0.95 & 0.99 & 0.95 & -0.04 & 1.75 & 1.76 & 2.39 & 1.80 & 0.95 & 0.99 & 0.95 & -0.00 & 1.76 & - & 2.39 & 1.80 & - & 0.99 & 0.96 \\
  &$\taufe$ & 0.06 & 1.94 & 1.85 & 2.30 & 1.86 & 0.94 & 0.98 & 0.94 & -0.04 & 1.75 & 1.76 & 2.28 & 1.85 & 0.95 & 0.98 & 0.96 & 0.02 & 1.76 & - & 2.28 & 1.85 & - & 0.99 & 0.97 \\
  &$\taustr$ & -0.01 & 1.82 & 1.76 & 2.15 & 1.56 & 0.94 & 0.98 & 0.92 & -0.01 & 1.75 & 1.76 & 2.13 & 1.56 & 0.95 & 0.98 & 0.92 & 0.02 & 1.75 & 1.75 & 2.13 & 1.56 & 0.96 & 0.98 & 0.92 \\
  &$\tausimp$ & -0.02 & 1.63 & 1.64 & 1.72 & 1.64 & 0.96 & 0.97 & 0.96 & -0.05 & 1.61 & 1.62 & 1.71 & 1.64 & 0.95 & 0.96 & 0.95 & -0.04 & 1.64 & - & 1.71 & 1.64 & - & 0.96 & 0.95 \\
  &$\tauancova$ & 0.07 & 1.75 & 1.72 & 1.57 & 1.73 & 0.95 & 0.92 & 0.95 & -0.08 & 1.61 & 1.62 & 1.56 & 1.71 & 0.95 & 0.95 & 0.96 & -0.03 & 1.64 & - & 1.56 & 1.72 & - & 0.93 & 0.96 \\
  &$\tauinter$ & 0.01 & 1.48 & 1.45 & 0.85 & 0.62 & 0.95 & 0.74 & 0.57 & -0.01 & 1.46 & 1.45 & 0.84 & 0.62 & 0.95 & 0.73 & 0.59 & 0.02 & 1.45 & 1.45 & 0.85 & 0.62 & 0.95 & 0.75 & 0.59 \\
   \hline
		\end{tabular}}
\begin{tablenotes}
\item Note: SD, standard deviation; SE, standard error; CP, coverage probability; HW, the \\ Huber--White  variance estimator; NEW: the proposed non-parametric variance estimator.
%\item Abbreviation: NA, not available.
\end{tablenotes}
\end{threeparttable}
\end{table}

\section{Clinical Trial Example}\label{SynData}

In this section, we consider a clinical trial example based on the synthetic data of the Nefazodone CBASP trial, which was conducted to compare the efficacy of three treatments for chronic depression \citep{Keller2000}. We focus on two of the treatments, Nefazodone and the combination of Nefazodone and the cognitive behavioral-analysis system of psychotherapy (CBASP). The process used to generate the synthetic data is detailed in the Appendix C. The analysis was performed according to the recommendations given in Table~\ref{tab::recom}, and the results are shown in Table~\ref{tab::syndata}.

As seen in Table~\ref{tab::syndata}, for simple randomization, adjusting for stratification indicators and additional baseline covariates clearly improves the estimation precision. Compared with $\hat\tau$, the variance reductions for $\taufe$ and $\tauancova$ are about $5$\% and $8$\%, respectively, under equal allocation, and the reductions for $\taustr$ and $\tauinter$ are more significant under unequal allocation, i.e., approximately $7$\% and $12$\%, respectively.
With respect to stratified block randomization, which achieves strong balance, the variance reduction of $\taufe$ under equal allocation is close to 0 and that of $\taustr$ under unequal allocation is exactly 0, as expected. Moreover, the variance reductions of $\tauancova$ and $\tauinter$ are moderate, approximately $2.50$\% and $6$\% under equal allocation and unequal allocation, respectively.
Finally, we note that the performance of each of the recommended estimators is comparable for different randomization methods.

\begin{table}[H]
\centering
\caption{Estimates, 95\% confidence intervals, and variance reductions under simple randomization and stratified block randomization for synthetic Nefazodone CBASP trial data.}\label{tab::syndata}
\resizebox{\textwidth}{22mm}{
\begin{tabular}{l|cccc|cccc}
		\cline{1-9}
Randomization		& \multicolumn{4}{c}{Equal Allocation ($\pi = 1/2$)} & \multicolumn{4}{c}{Unequal Allocation ($\pi = 2/3$)} \\ \cline{2-9}
Methods &  Estimator & Estimate & 95\% CI & Variance Reduction&  Estimator & Estimate & 95\% CI & Variance Reduction \\ \cline{1-9}
Simple & $\hat{\tau}$ & -4.69 & (-5.68, -3.70) & --- & $\hat{\tau}$ & -4.70 & (-5.75, -3.65) & --- \\
Randomization & $\taufe$ & -4.69 & (-5.66, -3.72) & 4.97\%& $\taustr$ & -4.70 & (-5.71, -3.68) & 6.69\% \\
 & $\tauancova$ & -4.70 & (-5.65, -3.74) & 7.99\% &  $\tauinter$ & -4.70 & (-5.69, -3.72) & 12.48\% \\
 \hline
Stratified & $\hat{\tau}$ & -4.69 & (-5.65, -3.73) & --- & $\hat{\tau}$ & -4.69& (-5.71, -3.67) & --- \\
Block & $\taufe$ & -4.69 & (-5.66, -3.73) & -0.81\% & $\taustr$ & -4.69 & (-5.71, -3.67) & 0\% \\
Randomization & $\tauancova$ & -4.69 & (-5.64, -3.74) & 2.50\%& $\tauinter$  & -4.69 & (-5.68, -3.70) & 6.15\%\\
 \hline
\end{tabular}}
\begin{tablenotes}
\item Note: CI, confidence interval.
\end{tablenotes}
\end{table}

\section{Discussion}

Linear regression is widely used to analyze the results of randomized clinical trials and other experiments with the hope of improving efficiency. In this article, we investigated the theoretical properties of six widely used regression-based treatment effect estimators under covariate-adaptive randomization, without imposing any modeling assumptions on the potential outcomes and covariates. We showed that these estimators are all consistent and asymptotically normal, albeit with possibly different efficiency. We provided non-parametric variance estimators to construct a valid confidence interval or test for the treatment effect and discussed the conditions under which the usual OLS variance estimator is valid. 

%We also compared the efficiency of these methods and found the following: (1) for equal allocation, $\taufe$ and  $\taustr$ are asymptotically equivalent, and the usual OLS variance estimator is valid for $\taufe$ but not for $\taustr$; similar conclusions hold for $\tauancova$ and $\tauinter$, and (2) for unequal allocation, generally, only $\taustr$ and $\tauinter$ can guarantee a gain in efficiency, and the latter is the most efficient. The usual OLS variance estimator fails in this case and our proposed variance estimator should be used in its stead. 

Taking into account both simplicity and efficiency, our final recommendations to practitioners are as follows.  For equal allocation, we recommend using regression (without interactions), which adjusts for the stratification indicators and additional covariates if available, and the use of the usual OLS variance estimator. For unequal allocation, we recommend using regression with interactions and the proposed non-parametric variance estimator.

This paper focused on using linear regression to estimate and infer the treatment effect under covariate-adaptive randomization with two treatments. It would be interesting to extend the results to more complicated settings, such as linear regression analysis for multiple treatments, for binary outcomes using logistic regression, and when there are missing values. Moreover, in this paper, we assumed that the number of strata and the number of covariates were fixed. It would be of interest to generalize the results to high-dimensional settings. Finally, although our results provide insights for using more general classes of models, theoretical results have yet to be established, especially regarding semi-parametric efficiency, which merits further investigation.

\section*{Acknowledgments}

Dr. Wei Ma's research was supported by grant 11801559 from the National Natural Science Foundation of China. Dr. Hanzhong Liu's research was supported by grant 11701316 from the National Natural Science Foundation of China.

\bibliographystyle{apalike}
\bibliography{causal}

%\section*{Supporting Information}
%Web Appendices A, B and C referenced in Sections \ref{sec1}, \ref{SimStudy} and \ref{SynData}, and the R code to reproduce the simulations are available with this paper at the Biometrics website of the Wiley Online Library.

\appendix

\section{Proof of main results}

\subsection{Useful lemmas}

We first introduce some useful lemmas obtained from \cite{Bugni2018,Bugni2019}. Several additional lemmas will be given later.

\begin{lemma}
\label{lem::equality}
For any  potential (transformed) outcomes $r_i(a)$, $i=1,\dots,n$, $a=0,1$, 
\begin{eqnarray}
& & \varsigma^2_{\tilde r}(\pi) + \varsigma^2_{H r} + \varsigma^2_{A r} (\pi) =  \varsigma^2_{ r}(\pi)  - \sumk \pk \{ \pi ( 1 - \pi ) - \qk \} \Big\{ \frac{ \mu_{[k]r} (1) - \mu_{r}(1) }{\pi}  + \frac{ \mu_{[k]r} (0) - \mu_{r}(0) }{ 1 - \pi }   \Big\}^2. \nonumber
\end{eqnarray}
\end{lemma}

\begin{lemma}
\label{lem::b3}
Suppose that Assumptions~\ref{assum::Q}, \ref{assum::A1} and \ref{assum::A3} hold, then, for i.i.d. random variables $V_i$, $i=1,\dots,n$, satisfying $E( |V_i| ) < \infty$, we have
$$
\frac{1}{n} \sumi A_i V_i  \xrightarrow{P} \pi  E(V_1).
$$
\end{lemma}

\begin{lemma}
\label{lem::proportion}
Under Assumptions \ref{assum::Q}, \ref{assum::A1} and \ref{assum::A3}, we have
$$
\frac{\nt}{n} \xrightarrow{P} \pi, \quad \frac{\nkt}{\nk} \xrightarrow{P} \pi, \quad \frac{\nkt}{n}  \xrightarrow{P} \pi \pk, \quad \pnk = \frac{\nk}{n}  \xrightarrow{P} \pk,
$$
$$
\frac{\nc}{n} \xrightarrow{P} 1 - \pi, \quad \frac{\nkc}{\nk} \xrightarrow{P} 1 - \pi,  \quad \frac{\nkc}{n}  \xrightarrow{P} \pi \pk.
$$
\end{lemma}

Lemma~\ref{lem::equality} can be found in the proof of Theorem 1 of \cite{Bugni2018}. Lemma~\ref{lem::b3} is a special case of Lemma C.4 of \cite{Bugni2019}. Lemma~\ref{lem::proportion} can be obtained directly from Lemma~\ref{lem::b3} and the strong low of large numbers. We omit the proofs of these lemmas.

% \section{Proofs of the main results}

\subsection{Proof of Proposition~\ref{prop::difference-in-mean}}
\begin{proof}
The asymptotic normality of $\hat \tau$ and the consistency of the non-parametric variance estimator  have been proved by \cite{Bugni2018}, so we omit it. We will only prove the unbiasedness and the probability limit of the variance estimator $\hat \sigma^2$. We first compute the conditional expectation $E \{ \YT  | B^{(n)}, A^{(n)} \}$.
\begin{eqnarray}
E \{  \YT | B^{(n)}, A^{(n)} \} & = &  \frac{1}{\nt} \sumi  A_i E \{ Y_i(1)  | B^{(n)}, A^{(n)}   \} \nonumber \\
& = &  \frac{1}{\nt} \sumi  A_i  E \{ Y_i(1)   | B^{(n)}  \}  \nonumber \\
& = &  E \{ Y_i(1)   | B^{(n)}   \}  \frac{1}{\nt}  \sumi  A_i  \nonumber \\
& = & E \{ Y_i(1)   | B_i \} ,  \nonumber
\end{eqnarray}
where the second equality is because of Assumption~\ref{assum::A1} that $\bm{W}^{(n)} \perp A^{(n)} | B^{(n)}$, and the last equality is because $W_i = ( Y_i(1), Y_i(0), \bx_i)$ are i.i.d. and
$
(1/\nt)  \sumi  A_i   = 1.
$
Similarly, 
$$
E \{ \YC | B^{(n)}, A^{(n)} \} =   E \{ Y_i(0)   | B_i \}.
$$ 
Thus,
$$
E ( \hat \tau ) = E ( \YT - \YC ) = E[   E\{ Y_i(1)   | B_i \} -   E \{ Y_i(0)   | B_i \} ] = E\{ Y_i(1) - Y_i(0) \} = \tau.
$$

Next, we obtain the probability limit of $\hat \sigma^2$. Let $\bz_{\diff}$ be the $n \times 2$ design matrix with the $i$th row $(A_i, 1)$. By the property of OLS,  $n \hat \sigma^2$ is the (1,1) element of the following matrix
\begin{eqnarray}
\left( \frac{1}{n-2} \sumi \hat r_i^2 \right)  \left( \frac{1}{n}  \bz_{\diff}^\T \bz_{\diff}   \right)^{-1}, \nonumber
\end{eqnarray}
where $ \hat r_i$ are the residuals. By Lemma~\ref{lem::proportion}, we have
\begin{eqnarray}
\label{eqn::z-diff}
 \left( \frac{1}{n}  \bz_{\diff}^\T \bz_{\diff}   \right)^{-1}  =  \left[
\begin{array}{cc}
\nt/n & \nt / n \\
\nt/n & 1
\end{array}
\right]^{-1} \ \xrightarrow{P} \ \frac{1}{ \pi ( 1 - \pi ) } \left[
\begin{array}{cc}
1 & - \pi \\
-\pi & \pi
\end{array}
\right].
\end{eqnarray}
The residuals $ \hat r_i$ will not change if we run the following regression:
\begin{equation*}
\label{reg::diff2}
Y_i \sim   A_i \tau_1 + ( 1  - A_i ) \tau_0 .
\end{equation*}
By the property of OLS, it holds that
$$
\frac{1}{n-2} \sumi \hat r_i^2 = \frac{1}{n-2} \sumi A_i ( Y_i - \YT )^2 + \frac{1}{n-2} \sumi ( 1 - A_i ) ( Y_i - \YC )^2 = \frac{\nt - 1}{n-2} \hat \sigma^2_{Y(1)} + \frac{\nc - 1}{ n-2 } \hat \sigma^2_{Y(0)},
$$
where 
$$
\hat \sigma^2_{Y(1)} = \frac{1}{\nt - 1} \sumi A_i ( Y_i - \YT )^2, \quad \hat \sigma^2_{Y(0)} = \frac{1}{\nc - 1} \sumi ( 1 - A_i ) ( Y_i - \YC )^2.
$$
By Lemmas~\ref{lem::b3} and \ref{lem::proportion} (see also the proof of Theorem 4.1 of \cite{Bugni2018}), 
$$
\hat \sigma^2_{Y(a)}  \xrightarrow{P} \sigma^2_{Y(a)}, \quad a = 0, 1.
$$
Therefore,
\begin{equation}
\label{eqn::eps-diff}
\frac{1}{n-2} \sumi \hat r_i^2  \ \xrightarrow{P} \  \pi   \sigma^2_{Y(1)} + ( 1 - \pi )  \sigma^2_{Y(0)}.
\end{equation}
Combing \eqref{eqn::z-diff} and \eqref{eqn::eps-diff}, 
$$
n \hat \sigma^2  \ \xrightarrow{P} \  \frac{ \sigma^2_{Y(1)} }{ 1 - \pi } + \frac{ \sigma^2_{Y(0)} }{  \pi } = \varsigma^2_{Y}( 1 - \pi).
$$
%By Lemma~\ref{lem::equality},
%$$
%\varsigma^2_{Y}(\pi) =  \varsigma^2_{\tilde Y}(\pi) + \varsigma^2_{H Y} + \varsigma^2_{A Y}(\pi)  +  \sumk \pk \{ \pi ( 1 - \pi ) - \qk \} \Big\{ \frac{ \mu_{[k]Y} (1) - \mu_{Y}(1) }{\pi} + \frac{ \mu_{[k]Y} (0) - \mu_{Y}(0) }{ 1 - \pi }  \Big\}^2.
%$$
\end{proof}

\subsection{Proof of Proposition~\ref{prop::fe}}

\begin{proof}
The formula and asymptotic normality of $\taufe$, and the consistency of the non-parametric variance estimator  have been obtained in \cite{Bugni2018}; see Theorem 4.3 and Theorem 4.4. Note that, by carefully examining the proof of these two theorems, when $\pi = 1/2$, the third term in the asymptotic variance ($\varsigma^2_{\pi Y}$) requiring the asymptotically independent normal distribution of $D_{n[k]}$ vanishes,  and  the conclusions hold if we replace Assumption~\ref{assum::A2} by the weaker Assumption~\ref{assum::A3}.  In the following, we will calculate the bias of $\taufe$ and investigate the property of the usual OLS variance estimator under Assumptions~\ref{assum::Q}, \ref{assum::A1} and \ref{assum::A3}. 

We first compute the conditional expectation $E \{ \YkThat  | B^{(n)}, A^{(n)} \}$.
\begin{eqnarray}
\label{eqn::Eykt}
E \{  \YkThat | B^{(n)}, A^{(n)} \} & = &  \frac{1}{\nkt} \sumi  A_i E \{ Y_i(1)  I_{i \in [k] } | B^{(n)}, A^{(n)}   \} \nonumber \\
& = &  \frac{1}{\nkt} \sumi  A_i  I_{i \in [k] } E \{ Y_i(1)   | B^{(n)}   \} \nonumber \\
& = &  \frac{1}{\nkt} \sumi  A_i  I_{i \in [k] } E \{ Y_i(1)   | B_i  \} \nonumber \\
& = & \frac{1}{\nkt}  \sumi  A_i  I_{i \in [k] }  E \{ Y_i(1)   | B_i = k   \}   \nonumber \\
& = & E \{ Y_i(1)   | B_i = k \},  
\end{eqnarray}
where the second equality is because of Assumption~\ref{assum::A1} that $\bm{W}^{(n)} \perp A^{(n)} | B^{(n)}$, the third equality is because $W_i = ( Y_i(1), Y_i(0), \bx_i)$ are i.i.d., and the last equality is due to
$
(1/\nkt)  \sumi  A_i  I_{i \in [k] } = 1.
$
Similarly, 
$$
E \{  \YkChat | B^{(n)}, A^{(n)} \} =   E \{ Y_i(0)   | B_i  = k \}.
$$ 
Thus,
$$
E \{ \YkThat - \YkChat  | B^{(n)}, A^{(n)} \} = \tauk.
$$
Therefore
$$
E ( \taufe ) = \sumk E \big\{ \omega_{[k]} E ( \YkThat - \YkChat  | B^{(n)}, A^{(n)} ) \big\} = \sumk  E ( \omega_{[k]} ) \tauk.
$$
Next, we prove the asymptotic property of $\hat \sigma^2_{\fe}$. Regression \eqref{reg::fe} is equivalent to the following regression:
\begin{equation}
\label{reg::fe2}
Y_i \sim \tau  A_i  + \sum_{k=1}^{K} \alpha_k  I_{i \in [k] }. \nonumber
\end{equation}
The $l_2$ loss function of the OLS is
$$
\sumi ( Y_i -  \tau A_i  -  \sum_{k=1}^{K} \alpha_k  I_{i \in [k] } ) ^2  = \sumk \sumik ( Y_i - \tau A_i - \alpha_k )^2.
$$
The OLS estimator $\taufe, \hat \alpha_{1,\fe},\dots, \hat \alpha_{K,\fe}$ has the following formula:
$$
\taufe = \sumk \omega_{[k]} ( \YkThat - \YkChat ), \quad \hat \alpha_{k,\fe} = \Ykbar - \pik, \quad k=1,\dots,K.
$$ 
Let $\hat r_{i,\fe} $ and $\bz_{\fe} $  be the residuals and  the $n \times (K + 1)$ design matrix  of regression~\eqref{reg::fe}, respectively.

\begin{lemma}
\label{lem::b9}
Under Assumptions~\ref{assum::Q}, \ref{assum::A1} and \ref{assum::A3}, 
\begin{equation}
\label{eqn::fe-resid}
\frac{1}{n} \sumi \hat r^2_{i,\fe}  \ \xrightarrow{P} \  \pi ( 1 - \pi )  \{ \varsigma^2_{\tilde Y} ( 1 - \pi ) +  \varsigma_{HY}^2   \}, \quad  \frac{1}{n} \bz_{\fe}^\T \bz_{\fe}  \xrightarrow{P} \Sigma_{\fe} , \nonumber
\end{equation}
where
\begin{eqnarray}
\label{eqn::fe-design}
\Sigma_{\fe}^{-1} =  \left[
\begin{array}{ccccc}
\frac{1}{\pi ( 1 - \pi )} & - \frac{1}{1 - \pi } & - \frac{1}{1 - \pi } & \cdots &  - \frac{1}{1 - \pi } \\
- \frac{1}{1 - \pi } & \frac{\pi}{ 1 - \pi }  + \frac{1}{p_{[1]}} & \frac{\pi}{ 1 - \pi } & \cdots & \frac{\pi}{ 1 - \pi }  \\
- \frac{1}{1 - \pi } &  \vdots & \ddots &  & \vdots \\
\vdots & \vdots &  & \ddots& \vdots \\
- \frac{1}{1 - \pi } & \frac{\pi}{ 1 - \pi }  & \cdots & \cdots &  \frac{\pi}{ 1 - \pi }  + \frac{1}{p_{[K]}}
\end{array}
\right]. \nonumber
\end{eqnarray}
\end{lemma}
The proof of Lemma~\ref{lem::b9} can be found in the proof of  Lemma B.9 in \cite{Bugni2018} (requiring minor modification when replacing Assumption~\ref{assum::A2} by Assumption~\ref{assum::A3}), so we omit it. By Lemma~\ref{lem::b9}, $ n \hat \sigma^2_{\fe}$ converges in probability to 
$$
\varsigma^2_{\tilde Y} ( 1 - \pi ) +  \varsigma_{HY}^2.
$$

\end{proof}

%The design matrix of the regression~\eqref{reg::fe2}, $\bz_{\fe} $, is an $n \times (K + 1)$ matrix with the $i$th row $(A_i, I_{i \in [1]}, \dots, I_{i \in [K]} )$. Since the stratification indicators are orthogonal, we have
%\begin{eqnarray}
%\frac{1}{n} \bz_{\fe}^\T \bz_{\fe} = \frac{1}{n} \left[
%\begin{array}{ccccc}
%\nt & n_{[1]1} & n_{[2]1} & \cdots & n_{[K]1} \\
%n_{[1]1} & n_{[1]} & 0 & \cdots & 0 \\
%n_{[2]1} &  \vdots & \ddots &  & \vdots \\
%\vdots & \vdots &  & \ddots& \vdots \\
%n_{[K]1} & 0 & \cdots & \cdots & n_{[K]} 
%\end{array}
%\right] \nonumber
%\end{eqnarray}
%By Lemma~\ref{lem::proportion}, $(1/n)  \bz_{\fe}^\T \bz_{\fe} \xrightarrow{P} \Sigma_{\fe} $, where 
%\begin{eqnarray}
%\Sigma_{\fe} = \left[
%\begin{array}{ccccc}
%\pi & \pi p(1) & \pi p(2) & \cdots & \pi p(K) \\
%\pi p(1) & p(1) & 0 & \cdots & 0 \\
%\pi p(2) &  \vdots & \ddots &  & \vdots \\
%\vdots & \vdots &  & \ddots& \vdots \\
%\pi p(K) & 0 & \cdots & \cdots & p(K) 
%\end{array}
%\right], \nonumber
%\end{eqnarray}
%and 
%\begin{eqnarray}
%\label{eqn::fe-design}
%\Sigma_{\fe}^{-1} =  \left[
%\begin{array}{ccccc}
%\frac{1}{\pi ( 1 - \pi )} & - \frac{1}{1 - \pi } & - \frac{1}{1 - \pi } & \cdots &  - \frac{1}{1 - \pi } \\
%- \frac{1}{1 - \pi } & \frac{\pi}{ 1 - \pi }  + \frac{1}{p(1)} & \frac{\pi}{ 1 - \pi } & \cdots & \frac{\pi}{ 1 - \pi }  \\
%- \frac{1}{1 - \pi } &  \vdots & \ddots &  & \vdots \\
%\vdots & \vdots &  & \ddots& \vdots \\
%- \frac{1}{1 - \pi } & \frac{\pi}{ 1 - \pi }  & \cdots & \cdots &  \frac{\pi}{ 1 - \pi }  + \frac{1}{p(K)}
%\end{array}
%\right].
%\end{eqnarray}

\subsection{Proof of Proposition~\ref{thm::str}}

\begin{proof}

We first prove the equivalence of the stratified difference-in-means estimator and the OLS estimator in the interactive regression \eqref{reg::str}. The OLS estimator of the coefficient of $A_i$ in regression~\eqref{reg::str} does not change if we run the following regression:
\begin{equation}
\label{eqn::reg-str2}
Y_i \sim \tau A_i + \sum_{k=1}^{K} \alpha_k  I_{i \in [k] } + \sumk \beta_k A_i [  I_{i \in [k] } - p_{n[k]} ]. \nonumber
\end{equation}
The $l_2$ loss function of the above regression is

\begin{equation}
\label{eqn::reg-str-k}
\sumk \sumik \Big \{ Y_i -  \alpha_k -  \Big( \tau + \beta_k - \sumk \beta_k \pnk \Big) A_i \Big\}^2 .
\end{equation}
By the property of OLS, the solution should satisfy
$$
\taustr + \hat \beta_k - \sumk \hat \beta_k \pnk = \taukhat.
$$
Taking average weighted by $\pnk$, we have
$$
\taustr = \sumk \pnk \taukhat.
$$

Second, we prove the unbiasedness. In the proof of Proposition~\ref{prop::fe}, we have shown in \eqref{eqn::Eykt} that
$$
E \{  \YkThat | B^{(n)}, A^{(n)} \} =   E \{ Y_i(1)   | B_i=k \},  \quad E \{ \YkChat | B^{(n)}, A^{(n)} \} =   E \{ Y_i(0)   | B_i = k \}.
$$ 
Thus,
$$
E \{  \YkThat - \YkChat  | B^{(n)}, A^{(n)} \} = \tauk.
$$
Since $E( \pnk ) = \pk$ and $\sumk \pk \tauk = \tau$, therefore,
$$
E ( \taustr ) = E \Big \{ \sumk \pnk ( \YkThat - \YkChat ) \Big \} = \sumk E ( \pnk ) \tauk = \sumk \pk \tauk = \tau.
$$

%& = & \sumk \frac{\nk}{n} \frac{1}{\nkt} \sumi  A_i  I_{i \in [k] } E \Big[ Y_i(1)   | B^{(n)}   \Big] \nonumber \\
%& = &  \sumk \frac{\nk}{n}  E \Big[ Y_i(1)   | B^{(n)}   \Big]  \frac{1}{\nkt}  \sumi  A_i  I_{i \in [k] } \nonumber \\
%& = & E \Big[ Y_i(1)   | B_i \Big],  \nonumber

Third, the asymptotic normality of $\taustr$ is a special case of Theorem 3.1 in \cite{Bugni2019}, so we omit the proof.

Finally, we derive the probability limit of the OLS variance estimator $\hat \sigma^2_{\str}$. By \eqref{eqn::reg-str-k}, the residual within stratum $k$ is
$$
 \hat r_{i, \str} = Y_i - \Ykbar - \taukhat ( A_i - \pik ),
$$
which is the residual of regressing $Y_i$ on $A_i$ within stratum $k$. Applying Lemma~\ref{lem::b9} within stratum $k$, we have
$$
\frac{1}{\nk } \sumik \hat r_{i, \str}^2 \xrightarrow{P} \pi ( 1 - \pi ) \left( \frac{ E[ \{ Y_i(1) - \mu_{[k]Y} (1) \}^2 | B_i = k]}{1 - \pi } +  \frac{ E[ \{ Y_i(0) - \mu_{[k]Y} (0) \}^2 | B_i = k]}{ \pi }  \right).
$$
Since $\nk / n \xrightarrow{P} \pk$, then,
\begin{eqnarray}
\label{eqn::fe-rss}
\frac{1}{n} \sumi  \hat r_{i, \str}^2 =  \sumk \frac{\nk}{n}  \frac{1}{\nk } \sumik \hat r_{i, \str}^2 \xrightarrow{P}  \pi ( 1 - \pi )  \varsigma^2_{\tilde Y} ( 1 - \pi ).
\end{eqnarray}
Regression~\eqref{reg::str} is equivalent to the following regression:
\begin{equation}
Y_i \sim \tau_1 A_i + \tau_0 ( 1 - A_i ) + \sum_{k=1}^{K-1} \alpha_k (I_{i \in [k] } - \pnk) + \sum_{k=1}^{K-1} \beta_k A_i (I_{i \in [k] } - \pnk). \nonumber
\end{equation}
Let $(\hat \tau_1, \hat \tau_0)$ be the OLS estimator of $(\tau_1, \tau_0)$ and let $\bz_{\str}$ be the $n \times  (2K) $ design matrix with the $i$th row 
$$
\big(A_i, 1 - A_i,  I_{i \in [1]} - p_{n[1]} , \dots, I_{i \in [K-1]} - p_{n[K-1]} ,  A_i ( I_{i \in [1]} - p_{n[1]}), \dots,  A_i ( I_{i \in [K-1]} - p_{n[K-1]})  \big).
$$
By the property of OLS, 
\begin{equation}
\label{eqn::var-fe-formula}
\taustr = \hat \tau_1 - \hat \tau_0, \quad n \hat \sigma^2_{\str} = \left( \frac{1}{n - 2K } \sumi  \hat r_{i, \str}^2 \right) \left(  \frac{1}{n}  \bz_{\str}^\T \bz_{\str} \right)^{-1}_{(1,1) + (2,2) - 2 (1,2)},
\end{equation}
where the subscript $(1,1) + (2,2) - 2 (1,2)$ denotes the $(1,1)$th element  plus the $(2,2)$th element minus 2 times the $(1,2)$th element of the matrix. Applying Lemma~\ref{lem::b3} to each element of the matrix $(1/n)  \bz_{\str}^\T \bz_{\str}$, we have $ (1/n)  \bz_{\str}^\T \bz_{\str}  \xrightarrow{P} \Sigma_{\str}$, where
\begin{eqnarray}
\Sigma_{\str} = \left[
\begin{array}{ccccc}
\pi & 0 & 0 & \cdots & 0 \\
0 & 1 - \pi & 0 & \cdots & 0 \\
0 & 0 &   & & \\
\vdots & \vdots & & * & \\
0 & 0 & & & 
\end{array}
\right]. \nonumber
\end{eqnarray}
Thus, 
\begin{equation}
\label{eqn::design-str}
(\Sigma_{\str}^{-1})_{ (1,1) + (2,2) - 2 (1,2) } = \frac{1}{\pi} + \frac{1}{1 - \pi} = \frac{1}{\pi ( 1 - \pi ) }.
\end{equation}
Combing \eqref{eqn::fe-rss}, \eqref{eqn::var-fe-formula} and \eqref{eqn::design-str}, we have
$$
n \hat \sigma^2_{\str} \xrightarrow{P}  \varsigma^2_{\tilde Y} ( 1 - \pi ).
$$

\end{proof}

% By carefully examining the proof of Theorem 4.1 of \cite{Bugni2018}, when $\pi = 1/2$, the asymptotic normality of $\taufe$ can be obtained under the weaker  Assumption~\ref{assum::A3} (replacing Assumption~\ref{assum::A2}). Moreover, 

\subsection{Proof of Corollary~\ref{cor::strfe}}
\begin{proof}
When $\pi = 1/2$, $\varsigma^2_{\pi Y} = 0$, thus Proposition~\ref{prop::fe} and Proposition \ref{thm::str} imply that $\taustr$ and $\taufe$ are asymptotically equivalent.  Since $\varsigma^2_{AY}(\pi) \geq 0$, thus both the asymptotic variances of  $\taustr$ and $\taufe$ are smaller than or equal to that of $\hat \tau$.  Finally, when $\pi = 1/2$, $\varsigma^2_{\pi Y} = 0$ and
$$
n \hat \sigma^2_{\fe} \xrightarrow{P} \varsigma^2_{\tilde Y} ( 1 - \pi ) + \varsigma^2_{HY} = \varsigma^2_{\tilde Y} (  \pi ) + \varsigma^2_{HY},
$$
therefore, $n \hat \sigma^2_{\fe}$ is a consistent variance estimator. 
\end{proof}

\subsection{Proof of Theorem~\ref{thm::simp}}

\begin{proof}
First, we obtain the formula of $\tausimp$. By definition, $\tausimp$ is the OLS estimator of the regression coefficient of $A_i$ when we regress $Y_i$ on $A_i$ and $\bx_i$ with intercept, which is equivalent to solving  the following equations:
\begin{eqnarray}
\label{eqn::simp}
\left \{ \begin{array}{l}
n^{-1}  \sumi  ( Y_i - \alpha - \tau A_i - \bx_i^\T \gamma ) =0  \\
n^{-1}  \sumi  A_i(Y_i - \alpha - \tau A_i  - \bx_i^\T \gamma) =0 \\ 
n^{-1}  \sumi   (\bx_i - \bar{\bx} ) (Y_i - \alpha - \tau A_i  - \bx_i^\T \gamma)] =0  
\end{array}  
\right.\nonumber
\end{eqnarray}
Let $\{ \hat \alpha^*, \tausimp, (\hat \gamma^*) ^\T \} ^\T$ be the solution of the above equations, then by the first equation, we have
\begin{equation}
\label{eqn::simp1}
 \bar{Y}  - \hat \alpha^* - \bar{A}  \tausimp - \bar{\bx} ^\T \hat \gamma^* = 0. \nonumber
\end{equation}
Thus,
$$
 \hat \alpha^* =  \bar{Y}   - \tausimp  \bar{A} - \bar{\bx} ^\T \hat \gamma^*.
$$
Taking it into the second and third equations, we have
\begin{equation}
\label{eqn::simp2} 
  \YT - \bar{Y} - \tausimp ( 1  - \bar{A} )   -  (\XT - \bar{\bx} ) ^\T \hat \gamma^* = 0, 
\end{equation}
\begin{equation}
\label{eqn::simp3}
S_{\bx Y}  - \tausimp  S_{\bx A} -  S_{\bx \bx} \hat \gamma^* =0 .  \nonumber
\end{equation}
where 
$$
S_{\bx Y} = \frac{1}{n} \sum_{i=1}^{n} (\bx_i - \bar{\bx} ) ( Y_i -  \bar{Y}), \quad S_{\bx \bx } = \frac{1}{n} \sumi (\bx_i - \bar{\bx} ) ( \bx_i - \bar{\bx})^\T,
$$
$$
S_{\bx A} = \frac{1}{n} \sumi (\bx_i - \bar{\bx} ) ( A_i - \bar{A} ) = \frac{\nt}{n} \Big(  \frac{1}{\nt} \sumi A_i \bx_i - \bar{\bx}  \Big) = \frac{\nt}{n} ( \XT - \bar{\bx} ) .
$$ 
Let $\hat \gamma = S^{-1}_{\bx \bx} S_{\bx Y}$, then
\begin{equation}
\label{eqn::simp3}
\hat \gamma^* = \hat \gamma - \tausimp S^{-1}_{\bx \bx} S_{\bx A}.
\end{equation}
Since $\bar{A} = \nt / n$, taking \eqref{eqn::simp3} into \eqref{eqn::simp2}, we have
$$
\tausimp = \frac{ \YT - \bar{Y} - ( \XT - \bar{X} )^\T \hat \gamma  }{  \nc / n - \nt / n  ( \XT - \bar{X} ) ^\T S^{-1}_{\bx \bx} ( \XT - \bar{X}  )  }.
$$
Since $\bar{Y} = (\nt / n) \YT + ( \nc / n ) \YC$, we have $ \YT - \bar{Y} = (\nc / n ) ( \YT - \YC ) $. Similarly 
$$
\XT - \bar{\bx} = (\nc / n ) ( \XT - \XC ) , \quad  \XC - \bar{\bx} = (\nt / n ) ( \XC - \XT ) = - (\nt / \nc ) ( \XT - \bar{\bx} ).
$$
Therefore, the numerator  of $\tausimp$ is
$$
(\nc / n )  [ \{ \YT - ( \XT - \bar{\bx} ) ^\T \hat \gamma \} -   \{ \YC - ( \XC - \bar{\bx} ) ^\T \hat \gamma \} ],
$$
and the denominator is
$$
(\nc / n) \{  1 +  ( \XT - \bar{X} ) ^\T S^{-1}_{\bx \bx} ( \XC - \bar{X}  )   \}
$$
Thus,
\begin{equation}
\label{eqn::tausimp}
\tausimp = \frac{ [ \YT - ( \XT - \bar{\bx} ) ^\T \hat \gamma ] -   [ \YC - ( \XC - \bar{\bx} ) ^\T \hat \gamma ]  }{  1 +  ( \XT - \bar{X} ) ^\T S^{-1}_{\bx \bx} ( \XC - \bar{X}  )   }.  \nonumber
\end{equation}
 
Second, we prove the asymptotic normality of $\tausimp$, for which we need the following lemma.
\begin{lemma}
\label{lem::beta}
Suppose that $( r_{i}(1), r_{i}(0)) \in  \mathcal{R}_2$ and Assumptions~\ref{assum::Q}, \ref{assum::A1} and \ref{assum::A3} hold, then 
$$
S_{\bx \bx} - \Sigma_{\bx \bx} = o_p(1), \quad \hat \gamma - \gamma = o_p(1), \quad
\XT - \bar{X}  = O_p(n^{-1/2}) , \quad   \XC - \bar{X}  = O_p(n^{-1/2}).
$$
\end{lemma}

Lemma~\ref{lem::beta} can be obtained directly from the strong law of large numbers, Lemmas B.1 and B.2 in \cite{Bugni2018}. We will prove it later. By Lemma~\ref{lem::beta}, the denominator of $\tausimp$ convergences to one in probability. Thus, $\sqrt{n} ( \tausimp - \tau )$ has the same asymptotic distribution as
\begin{eqnarray}
& & \sqrt{n} \big [  \{ \YT - ( \XT - \bar{\bx} ) ^\T \hat \gamma \} -   \{ \YC - ( \XC - \bar{\bx} ) ^\T \hat \gamma \} - \tau \big ] \nonumber \\
 & = & \sqrt{n} \big[ \{  \YT - ( \XT - \bar{\bx} ) ^\T \gamma \} -   \{ \YC - ( \XC - \bar{\bx} ) ^\T  \gamma \}  - \tau \big]  \nonumber \\ 
 && - \sqrt{n} ( \XT - \bar{\bx} ) ^\T ( \hat \gamma - \gamma ) + \sqrt{n} ( \XC - \bar{\bx} ) ^\T ( \hat \gamma - \gamma ). \nonumber 
\end{eqnarray}
By Lemma~\ref{lem::beta}, 
$$
 \sqrt{n} ( \XT - \bar{\bx} ) ^\T ( \hat \gamma - \gamma ) = o_p(1), \quad  \sqrt{n} ( \XC - \bar{\bx} ) ^\T ( \hat \gamma - \gamma ) = o_p(1).
$$
Thus, $\sqrt{n} ( \tausimp - \tau )$ has the same asymptotic distribution as
\begin{eqnarray}
&& \sqrt{n} \big[ \{  \YT - ( \XT - \bar{\bx} ) ^\T \gamma \} -   \{ \YC - ( \XC - \bar{\bx} ) ^\T  \gamma \}  - \tau \big]   \nonumber \\
& = &  \sqrt{n} \{  (\YT -  \XT^\T \gamma) -  (  \YC -  \XC  ^\T  \gamma ) - \tau \} \nonumber \\
& = & \sqrt{n} ( \bar{r}_{1} - \bar{r}_{0} - \tau ), \nonumber
\end{eqnarray}
which is $\sqrt{n}$ times the difference-in-means estimator minus $\tau$ for the transformed outcomes $r_{i}(a) = Y_i(a) - \bx_i ^\T \gamma$ with $E \{ r_{i}(1) - r_{i}(0) \} = \tau$. By Theorem 1 of \cite{Bugni2018} (Proposition~\ref{prop::difference-in-mean} in this paper),
$$
 \sqrt{n} ( \bar{r}_{1} - \bar{r}_{0} - \tau ) \xrightarrow{d} \mathcal{N} (0, \varsigma^2_{\tilde r}(\pi) + \varsigma^2_{H r} + \varsigma^2_{A r}(\pi)   ).
$$

% we compare the asymptotic variance of $\tausimp$ and simple randomization (without using covariates).
Finally, we compare the asymptotic variances of $\tausimp$ and $\hat \tau$. By Lemma~\ref{lem::equality}, 
\begin{eqnarray}
\label{eqn::simpvar}
&& \varsigma^2_{\tilde r}(\pi) + \varsigma^2_{H r} + \varsigma^2_{A r} (\pi) \nonumber \\
& = &  \varsigma^2_{ r }(\pi)  - \sumk \pk \{ \pi ( 1 - \pi ) - \qk \} \Big\{ \frac{ \mu_{[k]r}(1) - \mu_{r}(1) }{\pi}  + \frac{ \mu_{[k]r}(0) - \mu_{r}(0) }{ 1 - \pi}  \Big\}^2.  
%& = & \frac{1}{\pi} \sigma^2_{Y(1) - \bx^\T \gamma} + \frac{1}{1 - \pi} \sigma^2_{Y(0) - \bx^\T \gamma}  \nonumber \\
%&& - \sumk \pk \{ \pi ( 1 - \pi ) - \qk \} \Big\{ \frac{ \mu_{[k]r}(1) - \mu_{r}(1) }{\pi}  + \frac{ \mu_{[k]r}(0) - \mu_{r}(0) }{ 1 - \pi}  \Big\}^2.
\end{eqnarray}
By definition, $ \gamma(a) =  \Sigma_{ \bx \bx } ^{-1} \Sigma_{ \bx  Y(a)},$ $a=0,1$, then
\begin{eqnarray}
\sigma^2_{r(a)} & = &  \sigma^2_{Y(a) - \bx ^\T \gamma }  =  \Var [ Y_i(a) -   \bx_i ^\T \gamma  ]    =   \sigma^2_{Y(a)} + \gamma^\T \Sigma_{ \bx \bx } \gamma - 2 \gamma^\T \Sigma_{ \bx  Y(a)}, \nonumber \\
& = & \sigma^2_{Y(a)} + \gamma^\T \Sigma_{ \bx \bx } \gamma - 2 \gamma^\T \Sigma_{ \bx \bx } \gamma(a). \nonumber
%& = & \sigma^2_{Y(a)} + \gamma^\T \Sigma_{ \bx \bx } \gamma - 2 \gamma^\T   \Sigma_{ \bx \bx } \gamma(1),
\end{eqnarray}
Thus,
\begin{eqnarray}
\label{eqn::simpvar1}
 \varsigma^2_{ r } (\pi ) -  \varsigma^2_{ Y } (\pi) & = &\Big\{ \frac{\sigma^2_{ r(1)} }{\pi} +  \frac{\sigma^2_{ r(0) }}{ 1 - \pi} \Big\}  -  \Big\{ \frac{\sigma^2_{Y(1) }}{\pi} +  \frac{\sigma^2_{Y(0)}}{ 1 - \pi}  \Big\} \nonumber \\
 & = &  \frac{1}{\pi ( 1 - \pi ) }  \gamma^\T  \Sigma_{ \bx \bx } \gamma  -  \frac{2}{\pi ( 1 - \pi ) }  \gamma^\T  \Sigma_{ \bx \bx }  \{  ( 1 - \pi ) \gamma(1) + \pi \gamma(0) \} \nonumber \\
 & = &  \frac{1}{\pi ( 1 - \pi ) }  \gamma^\T  \Sigma_{ \bx \bx } \gamma  -  \frac{2}{\pi ( 1 - \pi ) }  \gamma^\T  \Sigma_{ \bx \bx }  [   \gamma +  ( 2 \pi - 1 ) \{ \gamma(0) - \gamma(1) \} ] \nonumber \\
&= & -   \frac{1}{\pi ( 1 - \pi ) }  \gamma^\T  \Sigma_{ \bx \bx } \gamma + \frac{2 ( 2\pi - 1 )}{\pi ( 1 - \pi )} \gamma^\T  \Sigma_{ \bx \bx }  \{ \gamma(1) - \gamma(0)  \} ,  % \nonumber \\
%& \triangleq & \Delta_{r_{\simp}-Y},
\end{eqnarray}
where the third equality is because $\gamma = \pi \gamma(1) + ( 1 - \pi ) \gamma(0)$. By definition, for $a=0,1$,
\begin{eqnarray}
 \mu_{[k]r}(a) - \mu_{r}(a)   = \{  \mu_{[k]Y}(a) - \mu_{Y}(a) \} -  ( \mu_{[k] \bx} - \mu_{\bx} ) ^\T \gamma. \nonumber 
%& = & \mu_{Ya}(k) - \mu_{Ya}  - \{ \mu_{[k] \bx }  - \mu_{\bx} \} ^\T \gamma,
\end{eqnarray}
Thus,
$$
\{ \mu_{[k]r}(1) - \mu_{r}(1) \}  -  \{ \mu_{[k]r}(0) - \mu_{r}(0) \}= \{ \mu_{[k]Y}(1) - \mu_{Y}(1) \}  -  \{ \mu_{[k]Y}(0) - \mu_{Y}(0) \} .
$$
Therefore,
\begin{equation}
\label{eqn::simpvar2}
\varsigma^2_{Hr} =  \varsigma^2_{HY},
\end{equation}
and
\begin{eqnarray}
&&\frac{ \mu_{[k]r}(1) - \mu_{r}(1) }{\pi}  + \frac{ \mu_{[k]r}(0) - \mu_{r}(0) }{ 1 - \pi}  \nonumber \\
&=& \frac{ \{ \mu_{[k]Y}(1) - \mu_{Y}(1) \}  }{\pi} + \frac{  \{ \mu_{[k]Y}(0) - \mu_{Y}(0) \}  }{ 1 - \pi } - \frac{1}{ \pi ( 1 - \pi ) }  \{ \mu_{[k] \bx }  - \mu_{\bx} \} ^\T \gamma, \nonumber
\end{eqnarray}
%\sumk \pk \qk \Big( \frac{ \{ \mu_{[k]Y}(1) - \mu_{Y}(1) \}  }{\pi} + \frac{ \{ \mu_{[k]Y}(0) - \mu_{Y}(0) \} }{ 1 - \pi }  \Big)^2  \nonumber \\
By Lemma~\ref{lem::equality}, 
\begin{eqnarray}
\label{eqn::simpvar4}
&&  \varsigma^2_{\tilde Y} (\pi) + \varsigma^2_{H Y} + \varsigma^2_{A Y} (\pi) = \varsigma^2_{ Y } (\pi)  - \sumk \pk \{ \pi ( 1 - \pi ) - \qk \} \Big\{ \frac{ \mu_{[k]Y}(1) - \mu_{Y}(1) }{ \pi } + \frac{ \mu_{[k]Y}(0) - \mu_{Y}(0) }{ 1 - \pi }  \Big\}^2. \nonumber \\
\end{eqnarray}
Combing \eqref{eqn::simpvar}, \eqref{eqn::simpvar1}, \eqref{eqn::simpvar2} and \eqref{eqn::simpvar4}, the difference of the asymptotic variances of $\tausimp$ and $\hat \tau$ is
\begin{eqnarray}
&& \Delta_{\diff * - \diff } \nonumber \\
 & = & \{  \varsigma^2_{\tilde r} (\pi) + \varsigma^2_{H r} + \varsigma^2_{A r} (\pi) \}  - \{  \varsigma^2_{\tilde Y} (\pi) + \varsigma^2_{H Y} + \varsigma^2_{A Y} (\pi) \}  \nonumber \\
 & = &  \varsigma^2_{ r} (\pi) -  \varsigma^2_{ Y} (\pi) - \sumk \pk \{ \pi ( 1 - \pi ) - \qk \} \Big\{ \frac{ \mu_{[k]r}(1) - \mu_{r}(1) }{\pi}  + \frac{ \mu_{[k]r}(0) - \mu_{r}(0) }{ 1 - \pi}  \Big\}^2 \nonumber \\
 & & + \sumk \pk \{ \pi ( 1 - \pi ) - \qk \} \Big\{ \frac{ \mu_{[k]Y}(1) - \mu_{Y}(1) }{ \pi } + \frac{ \mu_{[k]Y}(0) - \mu_{Y}(0) }{ 1 - \pi }  \Big\}^2 \nonumber \\
 & = & -  \frac{1}{\pi ( 1 - \pi ) }  \gamma^\T  \Sigma_{ \bx \bx } \gamma + \frac{2 ( 2\pi - 1 )}{\pi ( 1 - \pi )} \gamma^\T  \Sigma_{ \bx \bx }  \{ \gamma(1) - \gamma(0)  \} \nonumber \\
&&  - \frac{1}{\pi ( 1 - \pi ) }\gamma^\T   \sumk \pk \Big\{ 1 - \frac{\qk}{\pi ( 1 - \pi )} \Big\}   \{ \mu_{[k] \bx }  - \mu_{\bx} \}  \{ \mu_{[k] \bx }  - \mu_{\bx} \}^\T    \gamma \nonumber \\
&& +  \frac{2}{\pi ( 1 - \pi ) }  \gamma^\T  \sumk \pk \Big\{ 1 - \frac{\qk}{\pi ( 1 - \pi )} \Big\}   \{ \mu_{[k] \bx }  - \mu_{\bx} \} \{ ( 1 - \pi ) ( \mu_{[k] Y }(1) - \mu_{Y}(1) ) + \pi ( \mu_{[k] Y }(0) - \mu_{Y}(0) )   \}   . \nonumber 
\end{eqnarray}

\end{proof}

\subsection{Proof of Theorem~\ref{thm::ancova}}

%Before proving the theorem, we introduce the following lemma, whose proof will be given in the next section.

% Lemma~\ref{lem::b3} is in fact the Lemma B.3. in \cite{Bugni2018}, so we omit its proof. We will prove Lemma~\ref{lem::betaancova} in next section.

\begin{proof}
First, we derive the formula of $\tauancova$. Regression~\eqref{reg::ancova}: 
$$Y_i \sim \tau  A_i +  \sumk \alpha_k I_{i \in [k] } +  \bx_i^\T \beta $$  
is equivalent to first centering the outcome $Y_i$, treatment indicators $A_i$ and covariates $\bx_i$ at their stratum-specific means and then running the same regression but without the stratification indicators $ I_{i \in [k] } $, that is, the OLS estimator $(\tauancova,  \hat \beta_{\ancova})$ satisfy
\begin{equation}
\label{eqn::ancova-formula1}
(\tauancova,  \hat \beta_{\ancova}) = \argmin_{\tau, \beta} \sumk \sum_{i \in [k] } \big\{ Y_i  - \Ykbar - \tau ( A_i - \pik ) - (\bx_i - \Xkbar )^\T \beta \big\}^2,
\end{equation}
where $\pik = \nkt / \nk  = \bar{A}_{[k]}$. Solve the quadratic optimization problem, we can obtain (given $\hat \beta_{\ancova}$, $\tauancova$ is the fixed effect estimator (the second regression estimator considered in the main text) applying to the transformed outcomes $Y_i(a) - \bx_i ^\T \hat \beta_{\ancova}$
\begin{eqnarray*}
\tauancova & =  & \sumk      \omega_{[k]}  \Big\{ \YkThat - \YkChat - ( \XkThat - \XkChat ) ^\T \hat \beta_{\ancova}  \Big\} ,
\end{eqnarray*}
\begin{equation*}
\hat \beta_{\ancova} = \Big\{ \hat S^{\fe}_{\bx \bx} -    \hat \tau_{\fe}^{\bx}  ( \hat \tau_{\fe}^{\bx} ) ^\T  \sumk \pik ( 1 - \pik ) \pnk \Big\}^{-1}  \Big\{ \hat S^{\fe}_{\bx Y}  -  \taufe \hat \tau_{\fe}^{\bx} \sumk \pik ( 1 - \pik ) \pnk    \Big\},
\end{equation*}
where 
$$ 
\hat \tau_{\fe}^{\bx} = \sumk \omega_{[k]} (  \XkThat - \XkChat  ),
$$
\begin{equation}
\hat S^{\fe}_{\bx \bx} = \frac{1}{n} \sumk \sumik ( \bx_i - \Xkbar ) ( \bx_i - \Xkbar ) ^\T, \quad \hat S^{\fe}_{\bx Y} = \frac{1}{n} \sumk \sumik ( \bx_i - \Xkbar ) ( Y_i - \Ykbar ) . \nonumber
\end{equation}
%\begin{equation*}
% \hat \tau_{\fe}^{\bx} = \sumk \omega_{[k]} (  \XkThat - \XkChat  ).
%\end{equation*}

Second, we prove the asymptotic normality of $\tauancova$. We need the following lemma with its proof given later.
\begin{lemma}
\label{lem::betaancova}
Suppose that $(r_{i,\ancova}(1), r_{i,\ancova}(0), \bx_i) \in  \mathcal{R}_2$ and Assumptions~\ref{assum::Q}, \ref{assum::A1} and \ref{assum::A3} hold, then 
$$ 
\hat S^{\fe}_{\bx \bx} \xrightarrow{P} \Sigma_{\tilde \bx \tilde \bx}, \quad \hat S^{\fe}_{\bx Y}  \xrightarrow{P}  \pi \Sigma_{\tilde \bx \tilde Y(1)} + ( 1 - \pi ) \Sigma_{\tilde \bx \tilde Y(0)},  \quad  \hat \beta_{\ancova} - \beta_{\ancova} = o_p(1).
$$
\end{lemma}
To proceed, let 
$$
\check{\tau}_{\ancova} = \sumk    \omega_{[k]}  \big\{ \YkThat - \YkChat - ( \XkThat - \XkChat ) ^\T  \beta_{\ancova}  \big\},
$$
then,
\begin{eqnarray*}
\tauancova = \check{\tau}_{\ancova}  - \sumk \omega_{[k]} ( \XkThat - \XkChat ) ^\T  ( \hat \beta_{\ancova} - \beta_{\ancova}).
\end{eqnarray*}
Applying the asymptotic normality of the fixed effect estimator (Proposition~\ref{prop::fe}) to the potential outcomes $r_{i, \ancova} (a) = Y_i(a) - \bx_i^\T \beta_{\ancova}$ and the potential outcomes $\bx_i(1) = \bx_i(0) = \bx_i$, we obtain
$$
\sqrt{n} ( \check{\tau}_{\ancova}  - \tau  ) \xrightarrow{d} \mathcal{N} (0,  \varsigma^2_{\tilde r_{\ancova}}(\pi) + \varsigma^2_{Hr_{\ancova}} + \varsigma^2_{\pi r_{\ancova}} ),
$$
$$
\sqrt{n}  \sumk \omega_{[k]} ( \XkThat - \XkChat ) \xrightarrow{d} \mathcal{N} (0,  \varsigma^2_{\tilde \bx}(\pi) + \varsigma^2_{H\bx} + \varsigma^2_{\pi \bx} ).
$$
Note that, when $\pi = 1/2$, the above two asymptotic normality results still hold if we replace Assumption~\ref{assum::A2} by Assumption~\ref{assum::A3}. Together with Lemma~\ref{lem::betaancova}, we have
$$
\sqrt{n} ( \tauancova - \tau )  \xrightarrow{d} \mathcal{N} (0,  \varsigma^2_{\tilde r_{\ancova}}(\pi) + \varsigma^2_{Hr_{\ancova}} + \varsigma^2_{\pi r_{\ancova}} ).
$$

% under Assumptions~\ref{assum::Q}, \ref{assum::A1} and \ref{assum::A3}

Third, we investigate the OLS variance estimator $\hat \sigma^{2*}_{\ancova}$. From~\eqref{eqn::ancova-formula1}, $\tauancova$ is the fixed effect estimator (the second regression estimator considered in the main text) applying to the transformed outcomes $Y_i(a) - \bx_i ^\T \hat \beta_{\ancova}$, the residual sum of squares is
\begin{eqnarray}
&& \sumk \sumik \{ Y_i  - \Ykbar - \tauancova ( A_i - \pik ) - (\bx_i - \Xkbar )^\T \hat \beta_{\ancova}  \}^2 \nonumber \\
& = & \sumk \sumik \{ Y_i  - \Ykbar - \check{\tau}_{\ancova} ( A_i - \pik ) - (\bx_i - \Xkbar )^\T  \beta_{\ancova}  \}^2 \nonumber \\
&&  + \sumk \sumik \{   ( \tauancova -  \check{\tau}_{\ancova}  ) (A_i - \pik )  + (\bx_i - \Xkbar )^\T ( \hat \beta_{\ancova} -   \beta_{\ancova} ) \}^2  \label{eqn::ancova-term2} \\
&&  - 2 \sumk \sumik   \{ Y_i  - \Ykbar - \check{\tau}_{\ancova} ( A_i - \pik ) - (\bx_i - \Xkbar )^\T  \beta_{\ancova}  \} \nonumber \\
&& \quad \quad  \quad \quad \{   ( \tauancova -  \check{\tau}_{\ancova}  )   (A_i - \pik ) +   (\bx_i - \Xkbar )^\T ( \hat \beta_{\ancova} -   \beta_{\ancova} ) \}, \label{eqn::ancova-term3}
\end{eqnarray}
where $\check{\tau}_{\ancova}$ is the fixed effect estimator applying to the transformed outcomes $r_{i,\ancova}(a) = Y_i(a) -  \bx_i^\T  \beta_{\ancova} $, and the corresponding residual sum of squares is
$$
 \sumk \sumik \{ Y_i  - \Ykbar - \check{\tau}_{\ancova} ( A_i - \pik ) - (\bx_i - \Xkbar )^\T  \beta_{\ancova}  \}^2.
$$
By Lemma~\ref{lem::b9} with $Y_i(a)$ replaced by $r_{i,\ancova}(a)$, we have
\begin{equation}
\label{eqn::ancova-term1-limit}
\frac{1}{n}  \sumk \sumik \{ Y_i  - \Ykbar - \check{\tau}_{\ancova} ( A_i - \pik ) - (\bx_i - \Xkbar )^\T  \beta_{\ancova}  \}^2 \xrightarrow{P}  \pi ( 1 - \pi )  \{ \varsigma^2_{\tilde r_{\ancova} } (\pi) + \varsigma^2_{H r_{\ancova}}  \}.
\end{equation}
For the term in \eqref{eqn::ancova-term2}, by Lemma~\ref{lem::proportion} and Lemma~\ref{lem::betaancova}, we have
$$
\frac{1}{n} \sumk \sumik (A_i - \pik )^2 = \sumk \pik ( 1 - \pik ) \pnk \xrightarrow{P} \pi ( 1 - \pi ),
$$
$$
\frac{1}{n} \sumk \sumik  (\bx_i - \Xkbar ) (\bx_i - \Xkbar )^\T = \hat S^{\fe}_{\bx \bx}  \xrightarrow{P} \Sigma_{\tilde \bx \tilde \bx}.
$$
Since $\tauancova$ and $\check{\tau}_{\ancova}$ have the same asymptotic normality, and $\hat \beta_{\ancova} -   \beta_{\ancova} = o_p(1)$, we have
\begin{equation}
\label{eqn::ancova-term2-limit}
\frac{1}{n} \sumk \sumik \{   ( \tauancova -  \check{\tau}_{\ancova}  ) (A_i - \pik )  + (\bx_i - \Xkbar )^\T ( \hat \beta_{\ancova} -   \beta_{\ancova} ) \}^2 = o_p(1).
\end{equation}
Combing \eqref{eqn::ancova-term1-limit} and \eqref{eqn::ancova-term2-limit}, and applying Cauchy-Schwarz inequality, the term in \eqref{eqn::ancova-term3} converges to zero in probability. Therefore, the mean residual sum of squares satisfies
\begin{equation}
\label{eqn::ancova-resid-limit}
\frac{1}{n} \sumk \sumik \{ Y_i  - \Ykbar - \tauancova ( A_i - \pik ) - (\bx_i - \Xkbar )^\T \hat \beta_{\ancova}  \}^2 \xrightarrow{P}  \pi ( 1 - \pi )  \{ \varsigma^2_{\tilde r_{\ancova} }( 1 - \pi ) + \varsigma^2_{H r_{\ancova}}  \}.
\end{equation}
Moreover, regression~\eqref{reg::ancova} is equivalent to the following regression:
\begin{equation}
Y_i \sim  \tau_1 A_i  + \tau_0 ( 1 - A_i ) + \sum_{k=1}^{K-1} \alpha_k  ( I_{i \in [k] } - \pnk )   + ( \bx_i  - \bar{\bx} )^\T \beta, \nonumber
\end{equation}
with an $n \times (K + p + 1) $ design matrix $\bz_{\ancova}$ whose $i$th row is
$$
( A_i, 1 - A_i ,   I_{i \in [1] } - p_{n[1]} , \dots,  I_{i \in [K-1] } - p_{n[K-1]} , \bx_i^\T - \bar{\bx}^\T ).
$$
By the property of OLS,  $n \hat \sigma^{2*}_{\ancova}$ is the $(1,1) + (2,2) - 2 (1,2)$ element of the inverse of the matrix $(1/n) \bz_{\ancova}^\T \bz_{\ancova} $ times 
$$
\frac{1}{n - K - p - 1} \sumk \sumik \{ Y_i  - \Ykbar - \tauancova ( A_i - \pik ) - (\bx_i - \Xkbar )^\T \hat \beta_{\ancova}  \}^2.
$$
Similar to the proof of Proposition~\ref{thm::str}, the $(1,1) + (2,2) - 2 (1,2)$ element of the inverse of the matrix $(1/n) \bz_{\ancova}^\T \bz_{\ancova} $ converges in probability to $\{\pi ( 1 - \pi )\}^{-1}$, which together with \eqref{eqn::ancova-resid-limit} gives
$$
n \hat \sigma^{2*}_{\ancova}  \xrightarrow{P}      \varsigma^2_{\tilde r_{\ancova} } ( 1 - \pi ) + \varsigma^2_{H r_{\ancova}}.
$$
Note that,  the above convergence in probability holds if we replace Assumption~\ref{assum::A2} by Assumption~\ref{assum::A3} when $\pi = 1/2$ (only the asymptotic normality of $\tauancova$ and $\check{\tau}_{\ancova}$ requires Assumption~\ref{assum::A2}, which still holds under Assumption~\ref{assum::A3} when $\pi = 1/2$).

% , where $r^{\es}$ is the residual vector of regression~\eqref{reg::ancova}, that is,
%$$
%r_{i, \es}= \hat r_{i,\ancova} = Y_i  - \Ykbar - \tauancova ( A_i - \pik ) - (\bx_i - \Xkbar )^\T \hat \beta_{\ancova} , \quad i \in [k].
%$$

Fourth, we prove the consistency of the variance estimator $\hat  \varsigma^2_{\tilde r_{\es} }( \pi ) + \hat \varsigma^2_{H r_{\es} } + \hat \varsigma^2_{\pi r_{\es}}  $. By our definition of notation, 
\begin{equation}
\hat  \varsigma^2_{\tilde r_{\es} }( \pi ) =  \frac{1}{\pi}  \sumk    \frac{  \pnk }{\nkt} \sumik  A_i  ( r_{i, \es}-  \bar{r}_{ [k]1,\es} ) ^2  + \frac{1}{1 - \pi} \sumk    \frac{\pnk}{\nkc} \sumik (1 - A_i ) ( r_{i, \es}-  \bar{r}_{ [k]0,\es} )^2   , \nonumber
\end{equation}
\begin{equation}
 \hat \varsigma^2_{H r_{\es} } =  \sumk \pnk \big\{ ( \bar{r}_{ [k]1,\es}  - \bar{r}_{1,\es}  )  - (  \bar{r}_{ [k]0,\es}  - \bar{r}_{0,\es} )  \big\}^2, \nonumber
\end{equation}
\begin{equation}
\hat \varsigma^2_{\pi r_{\es}} = \frac{( 1 - 2\pi ) ^2 }{ \pi^2 ( 1 - \pi )^2 }  \sumk \pnk \qk \Big[  \{ \bar{r}_{ [k]1,\es}  - \bar{r}_{1,\es}  \}  -  \{  \bar{r}_{ [k]0,\es}  - \bar{r}_{0,\es} \}  \Big]^2. \nonumber
\end{equation}
Recall that
$$
r_{i, \es} (a) = Y_i(a) - \bx_i^ \T \hat \beta_{\ancova} = r_{i,\ancova}(a) -  \bx_i ^\T (  \hat \beta_{\ancova} - \beta_{\ancova} ),
$$
thus,
$$
r_{i, \es}-  \bar{r}_{ [k]1,\es} = r_{i,\ancova} - \bar{r}_{[k]1,\ancova} - ( \bx_i - \XkThat )^\T (  \hat \beta_{\ancova} - \beta_{\ancova} ).
$$
Therefore, 
\begin{eqnarray}
&& \sumk \frac{  \pnk }{\nkt} \sumik  A_i  ( r_{i, \es}-  \bar{r}_{ [k]1,\es} ) ^2  \nonumber \\
&=& \sumk   \frac{  \pnk }{\nkt} \sumik  A_i  \{ r_{i,\ancova} -  \bar{r}_{ [k]1, \ancova} - ( \bx_i - \XkThat )^\T (  \hat \beta_{\ancova} - \beta_{\ancova} )  \} ^2  \nonumber \\
& = &  \sumk   \frac{  \pnk }{\nkt} \sumik  A_i  ( r_{i,\ancova} -  \bar{r}_{ [k]1, \ancova} )^2 + \label{eqn::term1-cons-var} \\
&&  (  \hat \beta_{\ancova} - \beta_{\ancova} ) ^\T \sumk    \frac{  \pnk }{\nkt} \sumik  A_i ( \bx_i - \XkThat )  ( \bx_i - \XkThat )^\T  (  \hat \beta_{\ancova} - \beta_{\ancova} )  -  \label{eqn::term2-cons-var}  \\
&& 2 \sumk   \frac{  \pnk }{\nkt} \sumik  A_i  ( r_{i,\ancova} -  \bar{r}_{ [k]1, \ancova} ) ( \bx_i - \XkThat  ) ^\T  (  \hat \beta_{\ancova} - \beta_{\ancova} ).   \label{eqn::term3-cons-var} 
\end{eqnarray}
For the first term \eqref{eqn::term1-cons-var}, applying the consistency of the variance estimator part of Proposition~\ref{prop::fe} to $r_{i,\ancova}$, we have
$$
  \sumk   \frac{  \pnk }{\nkt} \sumik  A_i  ( r_{i,\ancova} -  \bar{r}_{ [k]1, \ancova} )^2   \xrightarrow{P} \sigma^2_{\tilde r_{\ancova}(1)}.
$$
For the second term \eqref{eqn::term2-cons-var}, by similar argument in the proof of Lemma~\ref{lem::gamma},
$$
\sumk    \frac{  \pnk }{\nkt} \sumik  A_i ( \bx_i - \XkThat )  ( \bx_i - \XkThat )^\T \xrightarrow{P} \Sigma_{\tilde \bx \tilde \bx}.
$$
Since  $ \hat \beta_{\ancova} - \beta_{\ancova}   = o_p(1)$, thus, 
$$
 (  \hat \beta_{\ancova} - \beta_{\ancova} ) ^\T \sumk    \frac{  \pnk }{\nkt} \sumik   ( \bx_i - \XkThat )  ( \bx_i - \XkThat )^\T  (  \hat \beta_{\ancova} - \beta_{\ancova} ) = o_p(1).
$$
For the third term \eqref{eqn::term3-cons-var}, using Cauchy-Schwarz inequality, we have
$$
 \sumk   \frac{  \pnk }{\nkt} \sumik  A_i  ( r_{i,\ancova} -  \bar{r}_{ [k]1, \ancova} ) ( \bx_i - \XkThat  ) ^\T  (  \hat \beta_{\ancova} - \beta_{\ancova} ) = o_p(1).
$$
Therefore,
\begin{equation}
\label{eqn::term-r1}
\sumk \frac{  \pnk }{\nkt} \sumik  A_i  ( r_{i, \es}-  \bar{r}_{ [k]1,\es} ) ^2   = \sigma^2_{\tilde r_{\ancova}(1)} + o_p(1).
\end{equation}
Similarly,
\begin{equation}
\label{eqn::term-r0}
\sumk \frac{  \pnk }{\nkc} \sumik  (1 - A_i )  ( r_{i, \es}-   \bar{r}_{ [k]0, \es} ) ^2   = \sigma^2_{\tilde r_{\ancova}(0)} + o_p(1).
\end{equation}
Combing~\eqref{eqn::term-r1} and \eqref{eqn::term-r0}, we have
$$
\hat  \varsigma^2_{\tilde r_{\es} }( \pi )  \xrightarrow{P}  \varsigma^2_{\tilde r_{\ancova} }( \pi ).
$$
Using similar arguments, we can prove that
$$
 \hat \varsigma^2_{H r_{\es} }  \xrightarrow{P}  \varsigma^2_{H r_{\ancova} }, \quad \hat \varsigma^2_{\pi r_{\es}}  \xrightarrow{P}  \varsigma^2_{\pi r_{\ancova} }.
$$
Therefore, $\hat  \varsigma^2_{\tilde r_{\es} }( \pi ) + \hat \varsigma^2_{H r_{\es} } + \hat \varsigma^2_{\pi r_{\es}} $ is a consistent estimator of the asymptotic variance $ \varsigma^2_{\tilde r_{\ancova}}(\pi) + \varsigma^2_{Hr_{\ancova}} + \varsigma^2_{\pi r_{\ancova}}$. Note that, the above arguments are valid under Assumptions~\ref{assum::Q}, \ref{assum::A1} and \ref{assum::A3}.

% Again, the consistency of variance estimator holds if we  replace Assumption~\ref{assum::A2} by Assumption~\ref{assum::A3} when $\pi = 1/2$.

Finally, we compare the asymptotic variances of $\tauancova$ and  $\taufe$. Note that $r_{i, \ancova} (a) = Y_i(a) - \bx_i^\T \beta_{\ancova}$, by definition,
\begin{eqnarray}
&& \mu_{ [k] r_{\ancova} }(a) - \mu_{ r_{\ancova}  }(a)  =  \mu_{[k]Y}(a) - \mu_{Y}(a)  - ( \mu_{ [k] \bx }  - \mu_{ \bx }   )^\T \beta_{\ancova} . \nonumber
\end{eqnarray}
Thus,
$$
\{ \mu_{ [k] r_{\ancova} }(1) - \mu_{ r_{\ancova}  }(1) \}   -  \{ \mu_{ [k] r_{\ancova} }(0) - \mu_{ r_{\ancova}  }(0)  \} = \{ \mu_{[k]Y}(1) - \mu_{Y}(1) \}  -  \{ \mu_{[k]Y}(0) - \mu_{Y}(0) \} .
$$
Therefore,
\begin{equation}
\label{eqn::hrancova}
\varsigma^2_{Hr_{\ancova}} =  \varsigma^2_{HY} , \quad  \varsigma^2_{\pi r_{\ancova}} = \varsigma^2_{\pi Y}.
\end{equation}
The difference of asymptotic variances of $\tauancova$ and $\taufe$ is
\begin{eqnarray}
\label{eqn::ancovavar}
\Delta_{\ancova * - \fe} = \{ \varsigma^2_{\tilde r_{\ancova}}(\pi) + \varsigma^2_{Hr_{\ancova}} + \varsigma^2_{\pi r_{\ancova}}  \} - \{ \varsigma^2_{\tilde Y}(\pi) + \varsigma^2_{HY} + \varsigma^2_{\pi Y}  \} =   \varsigma^2_{\tilde r_{\ancova}}(\pi) -  \varsigma^2_{\tilde Y}(\pi).
\end{eqnarray}
By definition and simple calculation,
\begin{eqnarray}
\sigma^2_{\tilde r_{\ancova}(a)} & = &  \Var[  Y_i(a) - E \{ Y_i(a) | B_i \} - ( \bx_i - E \{ \bx_i | B_i \} ) ^\T \beta_{\ancova}  ] \nonumber \\
& = & \Var[  \tilde Y(a) ] + \beta_{\ancova}^\T \Sigma_{\tilde \bx \tilde \bx} \beta_{\ancova}  - 2 \beta_{\ancova}^\T \Sigma_{\tilde \bx \tilde Y(a)}  \nonumber
\end{eqnarray}
Thus,
\begin{eqnarray}
 \varsigma^2_{\tilde r_{\ancova}}(\pi) & = & \frac{ \sigma^2_{\tilde r_{\ancova}(1)} }{ \pi } + \frac{ \sigma^2_{\tilde r_{\ancova}(0)} }{ 1 - \pi } \nonumber \\
 & = &   \varsigma^2_{\tilde Y }(\pi) + \Big(  \frac{1}{\pi}  + \frac{1}{ 1 - \pi } \Big) \beta_{\ancova}^\T \Sigma_{\tilde \bx \tilde \bx} \beta_{\ancova} - \frac{2}{\pi}  \beta_{\ancova}^\T \Sigma_{\tilde \bx \tilde Y(1)}  - \frac{2}{1 - \pi }  \beta_{\ancova}^\T \Sigma_{\tilde \bx \tilde Y(0)}  \nonumber \\
 & = &  \varsigma^2_{\tilde Y }(\pi) + \frac{1}{ \pi ( 1 - \pi ) } \Big( \beta_{\ancova}^\T \Sigma_{\tilde \bx \tilde \bx} \beta_{\ancova} - 2 ( 1 - \pi )  \beta_{\ancova}^\T \Sigma_{\tilde \bx \tilde Y(1)}  -  2 \pi  \beta_{\ancova}^\T \Sigma_{\tilde \bx \tilde Y(0)}    \Big). \nonumber
\end{eqnarray} 
By definition, $ \beta_{\ancova} = \pi \beta(1) + ( 1 - \pi ) \beta(0), $ where
$$
\beta(1) =  \Sigma^{-1}_{\tilde \bx \tilde \bx}  \Sigma_{\tilde \bx \tilde Y(1)} , \quad \beta(0) =  \Sigma^{-1}_{\tilde \bx \tilde \bx}  \Sigma_{\tilde \bx \tilde Y(0)}.
$$
By simple calculation (similar to the argument for $ \varsigma^2_{\tilde r_{\simp}}(\pi) $, just replacing $\gamma$, $\gamma(1)$ and $\gamma(0)$  by $\beta_{\ancova}$, $\beta(1)$ and $\beta(0)$, respectively),
\begin{eqnarray}
\label{eqn::ancovavar2}
 \varsigma^2_{\tilde r_{\ancova}}(\pi)  =    \varsigma^2_{\tilde Y }(\pi) - \frac{1}{ \pi ( 1 - \pi ) } \beta_{\ancova}^\T \Sigma_{\tilde \bx \tilde \bx} \beta_{\ancova} +  \frac{ 2 (2 \pi - 1 ) }{ \pi ( 1 - \pi ) } \beta_{\ancova}^\T \Sigma_{\tilde \bx \tilde \bx}  \{  \beta(1)  - \beta(0)   \} .    
\end{eqnarray} 
Combing \eqref{eqn::ancovavar} and \eqref{eqn::ancovavar2}, the difference of asymptotic variances of $\tauancova$ and $\taufe$ is
\begin{eqnarray}
\Delta_{\ancova * - \fe} =    - \frac{1}{ \pi ( 1 - \pi ) } \beta_{\ancova}^\T \Sigma_{\tilde \bx \tilde \bx} \beta_{\ancova} +  \frac{ 2 (2 \pi - 1 ) }{ \pi ( 1 - \pi ) } \beta_{\ancova}^\T \Sigma_{\tilde \bx \tilde \bx}  \{  \beta(1)  - \beta(0)   \} . \nonumber 
\end{eqnarray}

\end{proof}

\subsection{Proof of Theorem~\ref{thm::inter}}
%Since $\sumk  I_{i \in [k] } - p_{n[k]} = 0 $ and $\sumk   I_{i \in [k] } = 1$, the coefficient of $A_i$ in the second-order interactive regression \eqref{eqn::reg_inter} will not change if we run
%$$
%Y_i \sim   A_i \tau + \sum_{k=1}^{K} \alpha_k  I_{i \in [k] } + \sum_{k=1}^{K} \nu_k A_i [  I_{i \in [k] } - p_{n[k]} ] + \sumk \lambda_k  I_{i \in [k] }   \bx_i ^\T \beta  + \sumk \mu_k  A_i I_{i \in [k] }  ( \bx_i - \Xkbar ) ^\T \tilde \beta.
%$$
%Let $(\tauinter, \hat \alpha_{k,\inter}, \hat \nu_{k, \inter}, \hat \beta_{\inter}, \hat \beta_{\inter})$ be the OLS estimator, then they are the minimizer of the following quadratic loss (not unique for $\hat \nu_{k,\inter}$):
%$$
% \sumi \Big( Y_i  - A_i \tau - \sum_{k=1}^{K} \alpha_k  I_{i \in [k] } - \sum_{k=1}^{K} \nu_k A_i [  I_{i \in [k] } - p_{n[k]} ] - \bx_i ^\T \beta  - A_i ( \bx_i - \Xkbar ) ^\T \tilde \beta \Big)^2.
%$$
%Taking partial derivative with respect to $\tau$, we have
First, we obtain the formula of $\tauinter$. The coefficient of $A_i$ in regression \eqref{eqn::reg_inter} is the OLS estimator of $\tau_1 - \tau_0$ in the following regression:
\begin{eqnarray}
\label{eqn::reg_inter_1}
Y_i  & \sim &   A_i \tau_1 + ( 1 - A_i ) \tau_0 + \sum_{k=1}^{K} \alpha_{k1} A_i [  I_{i \in [k] } - p_{n[k]} ] +  \sum_{k=1}^{K} \alpha_{k0} ( 1 - A_i ) [  I_{i \in [k] } - p_{n[k]} ] \nonumber \\
&& +   A_i \Big( \bx_i  - \sumk I_{i \in [k]} \Xkbar \Big) ^\T \beta(1)  +   ( 1 - A_i )   \Big( \bx_i  - \sumk I_{i \in [k]} \Xkbar \Big) ^\T  \beta (0).
\end{eqnarray}
Note that the over-parameterization does not affect the OLS estimator of $\tau_1 - \tau_0$. The above regression is equivalent to running two regressions, in the treatment and control groups, separately:
\begin{eqnarray}
\label{reg::treatment}
Y_i(1) \sim \tau_1 - \sumk \alpha_{k1} \pnk + \sumk \alpha_{k1} I_{i \in [k]} +  \Big( \bx_i  - \sumk I_{i \in [k]} \Xkbar \Big) ^\T  \beta(1), \quad  A_i = 1
\end{eqnarray}
\begin{eqnarray}
\label{reg::control}
Y_i(0) \sim \tau_0 - \sumk \alpha_{k0} \pnk + \sumk \alpha_{k0} I_{i \in [k]} +  \Big( \bx_i  - \sumk I_{i \in [k]} \Xkbar \Big) ^\T  \beta(0), \quad A_i = 0. \nonumber
\end{eqnarray}
The OLS estimator of regression \eqref{reg::treatment} satisfies, 
\begin{equation}
\label{eqn::tautreat}
\hat \tau_1 - \sumk \hat \alpha_{k1} \pnk  - \hat \alpha_{k1}  = \YkThat - ( \XkThat -  \XkT )^\T \hat  \beta (1), \quad  k= 1, \dots, K,  
\end{equation}
\begin{equation}
\label{eqn::beta1hat}
 \hat \beta(1) = S^{-1}_{\tilde \bx  \tilde \bx}(1)S_{\tilde \bx \tilde Y(1)},  \nonumber
\end{equation}
where
\begin{equation}
\label{eqn::hatS1}
S_{\tilde \bx \tilde \bx  }(1) = \frac{1}{\nt} \sumk \sumik A_i ( \bx_i - \XkThat ) ( \bx_i - \XkThat )^\T,  \nonumber
\end{equation}
\begin{equation}
\label{eqn::hatS1}
S_{\tilde \bx \tilde Y(1) } = \frac{1}{\nt} \sumk \sumik A_i ( \bx_i - \XkThat ) \{ Y_i (1) - \YkThat \}. \nonumber
\end{equation}
Taking average weighted by $\pnk$ of both hands sides of \eqref{eqn::tautreat}, we have (note that $\sumk \pnk = 1$)
\begin{equation}
\label{eqn::tau1}
\hat \tau_1 = \sumk \pnk \Big \{   \YkThat - ( \XkThat -  \XkT )^\T \hat \beta(1) \Big\}. \nonumber
\end{equation}
Similarly,
\begin{equation}
\label{eqn::tau0}
\hat \tau_0 = \sumk \pnk \Big \{   \YkChat - ( \XkChat -  \XkC )^\T \hat \beta(0) \Big\}, \nonumber
\end{equation}
where
\begin{equation}
\label{eqn::beta0hat}
 \hat \beta(0) =  S^{-1}_{\tilde \bx  \tilde \bx }(0) S_{\tilde \bx \tilde Y(0)}, \nonumber
\end{equation}
\begin{equation}
\label{eqn::hatSx0}
S_{\tilde \bx \tilde \bx  } (0) = \frac{1}{\nc} \sumk \sumik ( 1 - A_i ) ( \bx_i - \XkChat ) ( \bx_i - \XkChat )^\T, \nonumber
\end{equation}
\begin{equation}
\label{eqn::hatSy0}
S_{\tilde \bx \tilde Y(0) } = \frac{1}{\nc} \sumk \sumik ( 1 - A_i ) ( \bx_i - \XkChat ) \{ Y_i (0) - \YkChat ). \nonumber
\end{equation}
Thus,
\begin{equation}
\label{eqn::tauinter}
\tauinter = \hat \tau_1 - \hat \tau_0  = \sumk \pnk \Big[  \Big\{   \YkThat - ( \XkThat -  \XkT )^\T \hat \beta(1)   \Big\}  - \Big\{  \YkChat - ( \XkChat -  \XkC )^\T \hat \beta(0)   \Big\}  \Big].
\end{equation}

Second, we prove the asymptotic normality of $\tauinter$. We introduce the following lemma with proof given later.
\begin{lemma}
\label{lem::gamma}
Suppose that $(r_{i,\inter}(1), r_{i,\inter}(0)) \in  \mathcal{R}_2$ and Assumptions~\ref{assum::Q}, \ref{assum::A1} and \ref{assum::A3} hold, then 
$$  
S_{\tilde \bx \tilde \bx  }(a) \xrightarrow{P} \Sigma_{\tilde \bx \tilde \bx}, \quad S_{\tilde \bx \tilde Y(a) } \xrightarrow{P} \Sigma_{\tilde \bx \tilde Y(a)},   \quad \hat  \beta(a) \xrightarrow{P} \beta(a), \quad a=0,1.
$$
\end{lemma}
It is easy to see that 
$$ 
\XkT = \pik \XkThat + ( 1 - \pik ) \XkChat.
$$ 
Thus,
$$
\XkThat -  \XkT = ( 1 - \pik ) ( \XkThat - \XkChat), \quad \XkChat -  \XkC = - \pik  ( \XkThat - \XkChat).
$$
Taking them into \eqref{eqn::tauinter}, we have
\begin{eqnarray}
& & \tauinter  \nonumber \\
& = &\sumk \pnk \Big[  \Big\{   \YkThat - ( \XkThat -  \XkT )^\T \hat \beta(1)   \Big\}  - \Big\{  \YkChat - ( \XkChat -  \XkC )^\T \hat \beta(0)   \Big\}  \Big] \nonumber \\
& = & \sumk \pnk \Big[    \YkThat -  \YkChat  -  ( \XkThat -  \XkChat )^\T  \big \{ ( 1 - \pik ) \hat \beta(1) + \pik   \hat \beta(0) \big \} \Big]  \nonumber \\
& = & \sumk \pnk \Big[    \YkThat -  \YkChat  -  ( \XkThat -  \XkChat )^\T \beta_{\inter}  \Big]  -    \sumk \pnk (  \XkThat -  \XkChat )^\T \{ \hat \beta_{\inter} -  \beta_{\inter} \}, \nonumber 
\end{eqnarray}
where $ \hat \beta_{\inter} = ( 1 - \pik ) \hat \beta(1) + \pik   \hat \beta(0)  $. By Lemma~\ref{lem::proportion}, $\pnk  \xrightarrow{P} \pk$, and applying the asymptotic normality of stratified difference-in-means estimator (Proposition~\ref{thm::str}) to the transformed outcomes $R_i(a) = \bx_i  I_{i \in [k]}$, $a=0,1$, we have $  \XkThat - \XkChat   = O_p( n^{-1/2} )$, thus,
$$
 \sumk \pnk ( \XkThat -  \XkChat )  = O_p( n^{-1/2} ).
$$
By Lemma~\ref{lem::gamma} and $\pik = \nkt / \nk \xrightarrow{P} \pi$, we have
$$
\hat \beta_{\inter} - \beta_{\inter} = o_p(1).
$$
Thus,
$$
\sqrt{n} \sumk \pnk ( \XkThat -  \XkChat )^\T \{ \hat \beta_{\inter} -  \beta_{\inter} \}  = o_p(1).
$$
Therefore, $ \tauinter $ has the same asymptotic distribution as
$$
\sumk \pnk \Big[    \YkThat -  \YkChat  -  ( \XkThat -  \XkChat )^\T \beta_{\inter}  \Big],
$$
which is the stratified difference-in-means estimator (the third regression estimator in the main text) applied to the transformed outcomes $r_{i, \inter}(a) = Y_i(a) - \bx_i ^\T  \beta_{\inter} $. By the asymptotic normality result of Proposition~\ref{thm::str}, 
$$
\sqrt{n} ( \tauinter - \tau ) \xrightarrow{d} \mathcal{N} (0,  \varsigma^2_{\tilde r_{\inter}}(\pi) + \varsigma^2_{Hr_{\inter}}  ).
$$

%Third, we compare the asymptotic variance of $\tauinter$ and $\taufe$. The difference is 
%\begin{eqnarray}
%\Delta_{\inter - \fe}
%\end{eqnarray}

%Regression~\ref{eqn::reg_inter} is equivalent to running the following regression:
%\begin{eqnarray}
%\label{eqn::reg_inter2}
%Y_i  & \sim &  \tau_1 A_i + \tau_0 ( 1 - A_i )  + \sum_{k=1}^{K-1} \alpha_k  ( I_{i \in [k] }  - \pnk ) + \sum_{k=1}^{K-1} \nu_k A_i [  I_{i \in [k] } - p_{n[k]} ] \nonumber \\
%&&  +  \Big( \bx_i  - \sumk I_{i \in [k]} \Xkbar \Big)  ^\T \beta  + A_i  \Big( \bx_i  - \sumk I_{i \in [k]} \Xkbar \Big)   ^\T \beta. \nonumber
%\end{eqnarray}
%Then 

Third, we study the asymptotic property of the OLS variance estimator $\hat \sigma^{2*}_{\inter}$. By the properties of OLS, $n \hat \sigma^{2*}_{\inter}$ equals the $(1,1) + (2,2) - 2 (1,2)$ element of the inverse of the Gram matrix $(1/n) \bz_{\inter}^\T \bz_{\inter} $ in regression \eqref{eqn::reg_inter_1}  times $    \sumi \hat r_{i, \inter}^2 / ( n - 2K - 2p ),$ where $\hat r_{i, \inter}$ is the residual of the regression which satisfies
\begin{eqnarray}
&&\frac{1}{n } \sumi \hat r_{i, \inter}^2 \nonumber \\
&= & \frac{1}{ n } \sumk \sumik A_i \Big\{ Y_i - \YkThat - ( \bx_i - \XkThat ) ^\T \hat \beta(1)  \Big\}^2 +  (1 - A_i ) \Big\{ Y_i - \YkChat - ( \bx_i - \XkChat ) ^\T \hat \beta(0)  \Big\}^2. \nonumber
\end{eqnarray}
For the first term, note that
\begin{eqnarray}
Y_i - \YkThat - ( \bx_i - \XkThat ) ^\T \hat \beta(1) & = & Y_i - \YkThat - ( \bx_i - \XkThat ) ^\T \beta(1) +    ( \bx_i - \XkThat ) ^\T \{ \beta(1) - \hat \beta(1) \}, \nonumber 
%& = & r_{i,\inter} - \bar{r}_{\inter,[k]1} +  ( \bx_i - \XkThat ) ^\T \{ \beta(1) - \hat \beta(1) \}. \nonumber
\end{eqnarray}
thus,
\begin{eqnarray}
\label{eqn::term1-inter-var}
&& \frac{1}{ n } \sumk \sumik A_i \Big\{ Y_i - \YkThat - ( \bx_i - \XkThat ) ^\T \hat \beta(1)  \Big\}^2 \nonumber \\
&=& \frac{1}{n} \sumk \sumik A_i   \Big\{ Y_i - \YkThat - ( \bx_i - \XkThat ) ^\T \beta(1)   \Big\}^2 + \frac{\nt}{n} \{ \beta(1) - \hat \beta(1) \}^\T  S_{\tilde \bx \tilde \bx  }(1)  \{ \beta(1) - \hat \beta(1) \}  + \nonumber \\
&& \frac{2}{n} \sumk \sumik A_i \{ Y_i - \YkThat - ( \bx_i - \XkThat ) ^\T \beta(1)  \}  ( \bx_i - \XkThat ) ^\T \{ \beta(1) - \hat \beta(1) \}
\end{eqnarray}
By Lemma~\ref{lem::gamma},
$$
\frac{\nt}{n} \{ \beta(1) - \hat \beta(1) \}^\T  S_{\tilde \bx \tilde \bx  }(1)  \{ \beta(1) - \hat \beta(1) \}  = o_p(1).
$$
Let $u_i(a) = Y_i(a) - \bx_i ^\T \beta(a)$, $a=0,1$, then
\begin{eqnarray}
& & \frac{1}{n} \sumk \sumik A_i   \Big\{ Y_i - \YkThat - ( \bx_i - \XkThat ) ^\T \beta(1)   \Big\}^2 \nonumber \\
& = & \frac{1}{n} \sumk \sumik A_i   \big\{ u_i(1) - \bar{u}_{[k]1}    \big\}^2  \nonumber \\
& = & \frac{1}{n} \sumi A_i u_i^2(1)  - \sumk \frac{\nkt}{n}  (  \bar{u}_{[k]1}  )^2 \nonumber \\
& \xrightarrow{P} & \pi E \{ u_i^2(1) \} - \pi E [ E \{ u_i(1) | B_i  \} ]^2 \nonumber \\
& = & \pi \sigma^2_{\tilde u(1)}, \nonumber
\end{eqnarray}
where the convergence in probability is due to Lemma~\ref{lem::b3} and Lemma~\ref{lem::proportion}. Using Cauchy-Schwarz inequality for the third term of \eqref{eqn::term1-inter-var}, we have it converges to zero in probability. Thus,
$$
 \frac{1}{ n } \sumk \sumik A_i \Big\{ Y_i - \YkThat - ( \bx_i - \XkThat ) ^\T \hat \beta(1)  \Big\}^2  \xrightarrow{P}  \pi \sigma^2_{\tilde u(1)}.
$$
Similarly,
$$
 \frac{1}{n} \sumk \sumik (1 - A_i )  \Big\{ Y_i - \YkChat - ( \bx_i - \XkChat ) ^\T \hat \beta(0)  \Big\}^2 \xrightarrow{P} ( 1 - \pi ) \sigma^2_{\tilde u(0)}.
$$
Thus,
$$
\frac{1}{n } \sumi \hat r_{i, \inter}^2 \xrightarrow{P} \pi \sigma^2_{\tilde u(1)} + ( 1 - \pi ) \sigma^2_{\tilde u(0)}.
$$
Similar to the third part of the proof of Theorem~\ref{thm::ancova},  the $(1,1) + (2,2) - 2 (1,2)$ element of the inverse of the matrix $(1/n) \bz_{\inter}^\T \bz_{\inter} $ converges in probability to $\{ \pi ( 1 - \pi ) \}^{-1}$. Therefore, 
\begin{eqnarray}
n \hat \sigma^{2*}_{\inter} \xrightarrow{P} \frac{ \sigma^2_{\tilde u(1)} }{ 1 - \pi } + \frac{ \sigma^2_{\tilde u(0)} }{ \pi }. \nonumber
\end{eqnarray}
%Since  $ r_i(a) = Y_i(a) - \bx_i ^\T \beta(a) $, then
%\begin{eqnarray}
%\sigma^2_{\tilde r(a)} = \sigma^2_{\tilde Y(a)}  +  \beta(a)   ^\T \Sigma_{\tilde \bx \tilde \bx}  \beta(a)  - 2  \beta(a) ^\T  \Sigma_{\tilde \bx \tilde Y(a)} =  \sigma^2_{\tilde Y(a)}  -  \beta(a)   ^\T \Sigma_{\tilde \bx \tilde \bx}  \beta(a). \nonumber
%\end{eqnarray}
Since
$$
u_i(1) = Y_i(1) - \bx_i ^\T \beta(1) = r_{i,\inter}(1) - \bx_i ^\T \{ \beta(1)  -  \beta_{\inter} \},
$$ 
then
\begin{eqnarray}
\sigma^2_{\tilde u(1)} &= & \sigma^2_{\tilde r_{\inter}(1)}  + \{ \beta(1)  -  \beta_{\inter} \} ^\T \Sigma_{\tilde \bx \tilde \bx} \{ \beta(1)  -  \beta_{\inter} \} - 2 \{ \beta(1)  -  \beta_{\inter} \} ^\T  \Sigma_{\tilde \bx \tilde r_{\inter}(1)} \nonumber \\
& = &  \sigma^2_{\tilde r_{\inter}(1)} - \{ \beta(1)  -  \beta_{\inter} \} ^\T \Sigma_{\tilde \bx \tilde \bx} \{ \beta(1)  -  \beta_{\inter} \}, \nonumber
\end{eqnarray}
where the last equality holds because
$$
 \Sigma_{\tilde \bx \tilde r_{\inter}(1)} = \Sigma_{\tilde \bx \tilde Y(1) } - \Sigma_{\tilde \bx \tilde \bx} \beta_{\inter} = \Sigma_{\tilde \bx \tilde \bx} \{ \beta(1) -  \beta_{\inter}  \}.
$$
Therefore,
\begin{eqnarray}
&& \frac{ \sigma^2_{\tilde u(1)} }{ 1 - \pi } + \frac{ \sigma^2_{\tilde u(0)} }{ \pi }  \nonumber \\
& = & \varsigma^2_{\tilde r_{\inter}}( 1 - \pi )  - \frac{1}{1 - \pi}  \{ \beta(1)  -  \beta_{\inter} \} ^\T \Sigma_{\tilde \bx \tilde \bx} \{ \beta(1)  -  \beta_{\inter} \} - \nonumber \\
&&  \frac{1}{\pi}  \{ \beta(0)  -  \beta_{\inter} \} ^\T \Sigma_{\tilde \bx \tilde \bx} \{ \beta(0)  -  \beta_{\inter} \} \nonumber \\
& = &  \varsigma^2_{\tilde r_{\inter}}( 1 - \pi )  - \Big \{  \frac{1}{\pi ( 1 - \pi )} - 3  \Big \}   \{ \beta(1)  -  \beta(0) \} ^\T \Sigma_{\tilde \bx \tilde \bx} \{ \beta(1)  -  \beta(0) \}. \nonumber
\end{eqnarray}

%Both $\tauinter$ and $\hat \sigma^{2*}_{\inter}$ will not change if we 

Fourth, we prove the consistency of the variance estimator $ \hat  \varsigma^2_{\tilde r_{\essed} }(\pi) + \hat \varsigma^2_{Hr_{\essed}} $. Note that
$$
r_{i,\essed}(a) = Y_i(a) - \bx_i ^\T \hat \beta_{\inter} = r_{i, \inter}(a) - \bx_i ^\T ( \hat \beta_{\inter} - \beta_{\inter} ).
$$
We have shown that $ \hat \beta_{\inter} - \beta_{\inter}  = o_p(1)$, thus, using similar arguments as in the fourth part of the proof of Theorem~\ref{thm::ancova}, we can obtain the consistency of the variance estimator, that is,
$$
\hat \varsigma^2_{\tilde r_{\essed} } (\pi) + \hat \varsigma^2_{H r_{\essed}} \xrightarrow{P}  \hat \varsigma^2_{\tilde r_{\inter} } (\pi) + \hat \varsigma^2_{H r_{\inter}}.
$$

Fifth, we compare the asymptotic variances of $\tauinter$ and $\taustr$.  
Simple calculation gives
\begin{eqnarray}
\sigma^2_{\tilde r_{\inter}(a)} & = & \Var[ Y_i(a) - E \{ Y_i(a) | B_i \} - ( \bx_i - E \{ \bx_i | B_i \}^\T \beta_{\inter} ]  \nonumber \\
& = & \sigma^2_{\tilde Y(a)} +  \beta_{\inter}^\T \Sigma_{\tilde \bx \tilde \bx} \beta_{\inter} - 2 \beta_{\inter}^\T \Sigma_{\tilde \bx \tilde Y(a)} \nonumber \\
& = &  \sigma^2_{\tilde Y(a)} +  \beta_{\inter}^\T \Sigma_{\tilde \bx \tilde \bx} \beta_{\inter} - 2 \beta_{\inter}^\T \Sigma_{\tilde \bx \tilde \bx} \beta(a), \nonumber
\end{eqnarray}
where the last equality is because $\beta(a) = \Sigma_{\tilde \bx \tilde \bx}^{-1} \Sigma_{\tilde \bx \tilde Y(a)}$. Therefore,
\begin{eqnarray}
\label{eqn::intervar1}
&&  \varsigma^2_{\tilde r_{\inter}}(\pi) - \varsigma^2_{\tilde Y}(\pi) \nonumber \\
  & = & \frac{ \sigma^2_{\tilde r_{\inter} (1)}  - \sigma^2_{\tilde Y(1)}  }{\pi} +  \frac{\sigma^2_{\tilde r_{\inter} (0)}  - \sigma^2_{\tilde Y(0)}  }{ 1 - \pi } \nonumber \\
 & = & \frac{1}{\pi }  \beta_{\inter}^\T \Sigma_{\tilde \bx \tilde \bx} \beta_{\inter} + \frac{1}{1 - \pi } \beta_{\inter}^\T \Sigma_{\tilde \bx \tilde \bx} \beta_{\inter}  - \frac{2}{\pi } \beta_{\inter}^\T   \Sigma_{\tilde \bx \tilde \bx}  \beta(1) - \frac{2}{ 1 - \pi }  \beta_{\inter}^\T   \Sigma_{\tilde \bx \tilde \bx}  \beta(0) \nonumber \\
 & = &   \frac{1}{\pi ( 1 - \pi ) }  \beta_{\inter}^\T \Sigma_{\tilde \bx \tilde \bx} \beta_{\inter} - \frac{2}{ \pi ( 1 - \pi ) } \beta_{\inter}^\T \Sigma_{\tilde \bx \tilde \bx} \{ ( 1 - \pi ) \beta(1) + \pi \beta(0) \} \nonumber \\
 & = & -  \frac{1}{\pi ( 1 - \pi ) }  \beta_{\inter}^\T \Sigma_{\tilde \bx \tilde \bx} \beta_{\inter},
\end{eqnarray}
where the last equality is because $\beta_{\inter}  = ( 1 - \pi ) \beta(1) + \pi \beta(0) $. Since 
$$
\mu_{ [k] r_{\inter} }(a) - \mu_{ r_{\inter}  }(a)= \mu_{ [k] Y }(a) - \mu_{ Y  }(a) - ( \mu_{ [k] \bx } - \mu_{ \bx  } )^\T \beta_{\inter},
$$
then,
\begin{equation}
\{ \mu_{ [k] r_{\inter} }(1)  - \mu_{ r_{\inter}  }(1) \} -  \{ \mu_{ [k] r_{\inter} }(0)  - \mu_{ r_{\inter}  }(0) \}   = \{ \mu_{ [k] Y }(1)  - \mu_{ Y  }(1) \} -  \{ \mu_{ [k] Y }(0)  - \mu_{ Y  }(0) \}  . \nonumber
\end{equation}
Therefore,
\begin{equation}
\label{eqn::intervar2}
 \varsigma^2_{Hr_{\inter}} =  \varsigma^2_{HY}.
\end{equation}
Combing \eqref{eqn::intervar1} and \eqref{eqn::intervar2}, the difference of the asymptotic variances of $\tauinter$ and $\taustr$ is
\begin{eqnarray}
\Delta_{\inter * - \str } & = &   \{ \varsigma^2_{\tilde r_{\inter}}(\pi) + \varsigma^2_{Hr_{\inter}}  \} - \{ \varsigma^2_{\tilde Y}(\pi) + \varsigma^2_{HY}  \}  \nonumber \\
& = &  \varsigma^2_{\tilde r_{\inter}}(\pi)  - \varsigma^2_{\tilde Y}(\pi)  \nonumber \\
& = &  -  \frac{1}{\pi ( 1 - \pi ) }  \beta_{\inter}^\T \Sigma_{\tilde \bx \tilde \bx} \beta_{\inter} \leq 0. \nonumber
\end{eqnarray}

Finally, we compare the asymptotic variances of $\tauinter$ and  $\tauancova$.  We have showed in \eqref{eqn::ancovavar2} that
\begin{eqnarray}
\label{eqn::ancovavar22}
 \varsigma^2_{\tilde r_{\ancova}}(\pi)  =    \varsigma^2_{\tilde Y }(\pi) - \frac{1}{ \pi ( 1 - \pi ) } \beta_{\ancova}^\T \Sigma_{\tilde \bx \tilde \bx} \beta_{\ancova} +  \frac{ 2 (2 \pi - 1 ) }{ \pi ( 1 - \pi ) } \beta_{\ancova}^\T \Sigma_{\tilde \bx \tilde \bx}  \{  \beta(1)  - \beta(0)   \} .  
\end{eqnarray} 
Combing ~\eqref{eqn::intervar1} -- \eqref{eqn::ancovavar22} and \eqref{eqn::hrancova}, the difference of asymptotic variances of $\tauinter$ and  $\tauancova$ is
\begin{eqnarray}
& & \Delta_{\inter * - \ancova *} \nonumber \\
& = & \{  \varsigma^2_{\tilde r_{\inter}}(\pi) + \varsigma^2_{Hr_{\inter}}  \}  - \{  \varsigma^2_{\tilde r_{\ancova}}(\pi) + \varsigma^2_{Hr_{\ancova}} + \varsigma^2_{\pi r_{\ancova}}  \} \nonumber \\
& = &  -  \frac{1}{\pi ( 1 - \pi ) }  \beta_{\inter}^\T \Sigma_{\tilde \bx \tilde \bx} \beta_{\inter}  + \frac{1}{ \pi ( 1 - \pi ) } \beta_{\ancova}^\T \Sigma_{\tilde \bx \tilde \bx} \beta_{\ancova} -  \frac{ 2 (2 \pi - 1 ) }{ \pi ( 1 - \pi ) } \beta_{\ancova}^\T \Sigma_{\tilde \bx \tilde \bx}  \{  \beta(1)  - \beta(0) \}  - \varsigma^2_{\pi r_{\ancova}}  \nonumber \\
& = &  -  \frac{1}{\pi ( 1 - \pi ) } \Big [   \beta_{\inter}^\T \Sigma_{\tilde \bx \tilde \bx} \beta_{\inter}  - \beta_{\ancova}^\T \Sigma_{\tilde \bx \tilde \bx} \beta_{\ancova}  + 2   \beta_{\ancova}^\T \Sigma_{\tilde \bx \tilde \bx}  ( 2 \pi - 1) \{  \beta(1)  - \beta(0) \}   \Big]   - \varsigma^2_{\pi r_{\ancova}}  \nonumber \\
& = &   -  \frac{1}{\pi ( 1 - \pi ) } \Big [   \beta_{\inter}^\T \Sigma_{\tilde \bx \tilde \bx} \beta_{\inter}  - \beta_{\ancova}^\T \Sigma_{\tilde \bx \tilde \bx} \beta_{\ancova}  + 2   \beta_{\ancova}^\T \Sigma_{\tilde \bx \tilde \bx}  \{  \beta_{\ancova}  - \beta_{\inter} \}   \Big]   - \varsigma^2_{\pi r_{\ancova}}  \nonumber \\
& = &  -  \frac{1}{\pi ( 1 - \pi ) }    ( \beta_{\inter} -  \beta_{\ancova} ) ^\T \Sigma_{\tilde \bx \tilde \bx}  ( \beta_{\inter} -  \beta_{\ancova} ) ^\T   - \varsigma^2_{\pi r_{\ancova}}.  \nonumber 
\end{eqnarray}

\subsection{Proof of Corollary~\ref{cor::ancovainter}}
\begin{proof}
By Theorem~\ref{thm::inter}, the difference of the asymptotic variances of $\tauinter$ and  $\tauancova$ is
$$
 \Delta_{\inter * - \ancova *} =  -  \frac{1}{\pi ( 1 - \pi ) }    ( \beta_{\inter} -  \beta_{\ancova} ) ^\T \Sigma_{\tilde \bx \tilde \bx}  ( \beta_{\inter} -  \beta_{\ancova} ) ^\T   - \varsigma^2_{\pi r_{\ancova}}.
$$
When $\pi = 1/2$, it holds that
$$
 \beta_{\inter} = ( 1 - \pi ) \beta(1) + \pi \beta (0) = \pi \beta(1) + ( 1 - \pi ) \beta(0) =  \beta_{\ancova} , \quad  \varsigma^2_{\pi r_{\ancova}} = 0.
$$
Therefore, $ \Delta_{\inter * - \ancova *} = 0$, i.e., $\tauinter$ and  $\tauancova$ are asymptotically equivalent. By Theorem~\ref{thm::inter}, both of them are generally more efficient than $\taustr$. For the consistency of the variance estimator, by Theorem~\ref{thm::ancova}, 
$$
n \hat \sigma^{2*}_{\ancova}  \xrightarrow{P}      \varsigma^2_{\tilde r_{\ancova} } ( 1 - \pi ) + \varsigma^2_{H r_{\ancova}},
$$
and the asymptotic variance of $\tauancova$ is $  \varsigma^2_{\tilde r_{\ancova}}(\pi) + \varsigma^2_{Hr_{\ancova}} + \varsigma^2_{\pi r_{\ancova}}  $. When $\pi = 1/2$, it holds that
$$
\varsigma^2_{\tilde r_{\ancova} } ( 1 - \pi ) = \varsigma^2_{\tilde r_{\ancova} } ( \pi ) , \quad \varsigma^2_{\pi r_{\ancova}}  = 0, \quad  \hat \varsigma^2_{\pi r_{\es}} = 0.
$$
Thus, both $n \hat \sigma^{2*}_{\ancova}$ and $\hat  \varsigma^2_{\tilde r_{\essed} }( \pi ) + \hat \varsigma^2_{H r_{\essed} }$ are consistent variance estimators.
\end{proof}

\subsection{Proof of Theorem~\ref{thm::optimal}}

\begin{proof}
Since
$$
 \XT - \bar{\bx} = \XT - \frac{\nt}{n} \XT - \frac{ \nc }{ n } \XC = \frac{\nc}{n} ( \XT - \XC ), \quad \XC - \bar{\bx}  = - \frac{\nt}{n}  ( \XT - \XC ),
$$
then, by definition and simple calculation, we have
\begin{eqnarray}
\hat \tau^* ( \hat \eta(1), \hat \eta(0)  ) & = &  \{ \YT -  ( \XT - \bar{\bx} ) ^\T \hat \eta(1)  \}   -  \{ \YC -  ( \XC - \bar{\bx} ) ^\T \hat \eta(0) \} \nonumber \\
& = &  \YT - \YC -  ( \XT - \XC ) ^\T  \Big\{  \frac{\nc}{n} \hat \eta(1) + \frac{\nc}{n} \hat \eta(0)   \Big \}  \nonumber \\
& = &  \YT - \YC -  ( \XT - \XC ) ^\T \eta -   ( \XT - \XC ) ^\T  \Big\{  \frac{\nc}{n} \hat \eta(1) + \frac{\nc}{n} \hat \eta(0) - \eta   \Big \}, \nonumber
\end{eqnarray}
where $\eta = (1 - \pi ) \eta(1) + \pi \eta(0)$. Since 
$$
\XT - \XC = O_p( n^{-1/2}), \quad \frac{\nc}{n} \hat \eta(1) + \frac{\nc}{n} \hat \eta(0) \xrightarrow{P} \eta,
$$
then $\hat \tau^* ( \hat \eta(1), \hat \eta(0)  ) $ has the same asymptotic distribution as
$$
 \YT - \YC -  ( \XT - \XC ) ^\T \eta ,
$$
which is the difference-in-means estimator for the transformed outcomes $r_{i,\text{gen}}(a) = Y_i(a) - \bx_i ^\T \eta$, $a=0,1$. Applying Proposition~\ref{prop::difference-in-mean}, we have
$$
\sqrt{n} \big( \hat \tau^* ( \hat \eta(1), \hat \eta(0)  )   - \tau  \big) \xrightarrow{d} \mathcal{N}(0, \varsigma^2_{\tilde r_{\text{gen}} }(\pi) + \varsigma^2_{H r_{\text{gen}} } + \varsigma^2_{A r_{\text{gen}}}(\pi) ).
$$
Similarly, we can show that
$$
\sqrt{n} \big( \hat \tau^*_{\fe} ( \hat \eta(1), \hat \eta(0)  )   - \tau  \big) \xrightarrow{d} \mathcal{N}(0, \varsigma^2_{\tilde r_{\text{gen}} }(\pi) + \varsigma^2_{H r_{\text{gen}} } + \varsigma^2_{\pi r_{\text{gen}}}(\pi) ),
$$
$$
\sqrt{n} \big( \hat \tau^*_{\inter} ( \hat \eta(1), \hat \eta(0)  )   - \tau  \big) \xrightarrow{d} \mathcal{N}(0, \varsigma^2_{\tilde r_{\text{gen}} }(\pi) + \varsigma^2_{H r_{\text{gen}} } ).
$$
Similar to the proof of Theorem~\ref{thm::inter}, 
$$
\varsigma^2_{H r_{\text{gen}} } = \varsigma^2_{H Y }.
$$
Since $ \varsigma^2_{A r_{\text{gen}}}, \varsigma^2_{\pi r_{\text{gen}}} \geq 0$, thus if $\eta_{\text{opt}}$ minimizes $ \varsigma^2_{\tilde r_{\text{gen}}}(\pi) $, the corresponding estimator $\tauinter (\hat \eta(1), \hat \eta(0) )$ has the smallest asymptotic variance. By definition and simple calculus, 
\begin{eqnarray}
 \varsigma^2_{\tilde r_{\text{gen}}}(\pi) & = & \frac{ \sigma^2_{\tilde r_{\text{gen}}(1) } }{ \pi } +  \frac{ \sigma^2_{\tilde r_{\text{gen}}(0) } }{ 1 -  \pi } \nonumber \\
 & = &  \frac{ \Var \{ \tilde Y_i(1) - \tilde \bx_i^\T \eta \} }{ \pi } +  \frac{ \Var \{ \tilde Y_i(0) - \tilde \bx_i^\T \eta \} }{ 1 -  \pi }  \nonumber \\
 & = & \frac{ \Var \{ \tilde Y_i(1) \} }{ \pi } +  \frac{ \Var \{ \tilde Y_i(0) \} }{ 1 -  \pi } + \frac{1}{\pi ( 1 - \pi )} \eta^\T \Sigma_{\tilde \bx \tilde \bx} \eta - \frac{2}{\pi ( 1 - \pi ) } \eta^\T \Big \{ ( 1 - \pi ) \Sigma_{\tilde \bx \tilde Y(1) } + \pi \Sigma_{\tilde \bx \tilde Y(0) } \Big \} \nonumber \\
 & = &  \frac{ \Var \{ \tilde Y_i(1) \} }{ \pi } +  \frac{ \Var \{ \tilde Y_i(0) \} }{ 1 -  \pi } + \frac{1}{\pi ( 1 - \pi )} \eta^\T \Sigma_{\tilde \bx \tilde \bx} \eta - \frac{2}{\pi ( 1 - \pi ) } \eta^\T  \Sigma_{\tilde \bx \tilde \bx} \Big\{  ( 1 - \pi ) \beta(1) + \pi  \tilde \beta (0) \Big\} \nonumber \\
 & = &  \frac{ \Var \{ \tilde Y_i(1) \} }{ \pi } +  \frac{  \Var \{ \tilde Y_i(0) \} }{ 1 -  \pi } + \frac{1}{\pi ( 1 - \pi )} \eta^\T \Sigma_{\tilde \bx \tilde \bx} \eta - \frac{2}{\pi ( 1 - \pi ) } \eta^\T  \Sigma_{\tilde \bx \tilde \bx} \beta_{\inter},  \nonumber
\end{eqnarray}
where the last but second equality is because $\beta(a) = \Sigma_{\tilde \bx \tilde \bx}^{-1} \Sigma_{\tilde \bx \tilde Y(a)}$, $a = 0,1$, and the last equality is because $\beta_{\inter} =  ( 1 - \pi ) \beta(1) + \pi  \beta (0)$. Taking derivative with respect to $\eta$, the optimal coefficient vector $\eta_{\text{opt}}$ should satisfy
$$
2  \Sigma_{\tilde \bx \tilde \bx} \eta_{\text{opt}} - 2  \Sigma_{\tilde \bx \tilde \bx} \beta_{\inter} = 0.
$$
That is, $\eta_{\text{opt}} = \beta_{\inter}$. Since $\tauinter = \tauinter( \hat \beta(1), \hat \beta(0) )$ and 
$$
\hat \beta(1)  \xrightarrow{P} \beta(1), \quad \hat \beta(0) \xrightarrow{P} \beta(0), \quad \hat \beta_{\inter}  \xrightarrow{P} \beta_{\inter},
$$
then, $\tauinter$ achieves the smallest asymptotic variance. The conclusion follows immediately. 

\end{proof}

% \section{Proof of lemmas}

\subsection{Proof of Lemma~\ref{lem::beta}}
By the strong low of large numbers,  for $j,k=1,\dots,p$, 
$$
\bar{\bx} \xrightarrow{P} E ( \bx ), \quad \frac{1}{n} \sumk \sumik x_{ij} x_{ik}  \xrightarrow{P} E ( x_{ij} x_{ik} ), 
$$
thus,
$$
S_{\bx \bx} =  \frac{1}{n} \sumi (\bx_i - \bar{\bx} ) ( \bx_i - \bar{\bx})^\T =  \frac{1}{n} \sumk \sumik \bx_i  \bx_i ^\T - \bar{\bx} \bar{\bx}^\T  \xrightarrow{P} E ( \bx \bx ^\T ) - E ( \bx ) E ( \bx^\T ) = \Sigma_{\bx \bx}.
$$
Since $Y_i = A_i  Y_i(1) + ( 1 - A_i ) Y_i(0)$, by Lemma~\ref{lem::b3},
$$
\bar{Y} = \frac{1}{n} \sumi A_i Y_i(1) + \frac{1}{n} \sumi ( 1 - A_i ) Y_i(0) \xrightarrow{P} \pi E\{Y(1)\} + ( 1 - \pi ) E\{Y(0) \}.
$$
Similarly,
$$
\frac{1}{n} \sumi \bx_i Y_i  =  \frac{1}{n} \sumi  A_i  \bx_i Y_i(1) + \frac{1}{n} \sumi ( 1 -  A_i )  \bx_i Y_i(0) \xrightarrow{P} \pi E\{ \bx Y(1) \} + ( 1 - \pi ) E \{ \bx Y(0) \}.
$$
Thus,
\begin{eqnarray}
S_{\bx Y} & = & \frac{1}{n} \sumi (\bx_i - \bar{\bx} ) ( Y_i -  \bar{Y}) = \frac{1}{n} \sumi \bx_i Y_i - \bar{\bx} \bar{Y} \nonumber \\
& \xrightarrow{P} &  \pi E \{ \bx - E ( \bx ) \} [ Y(1) - E\{ Y(1) \}  ] + ( 1 - \pi )  E \{ \bx - E(\bx) \} [  Y(0) - E \{Y(0) \} ] . \nonumber
\end{eqnarray}
Therefore,
$$
\hat \gamma =  S_{\bx \bx}^{-1} S_{\bx Y}   \xrightarrow{P} \Sigma_{\bx \bx} ( \pi \Sigma_{\bx Y(1) } + ( 1 - \pi ) \Sigma_{\bx Y(0)}  ) = \pi \gamma(1) + ( 1 - \pi ) \gamma(0) = \gamma.
$$
In the proof of Theorem~\ref{thm::simp}, we have shown that
$$
\XT - \bar{\bx} = (\nc / n ) ( \XT - \XC )  .
$$
Since the covariates $\bx$ are not affected by the treatment assignment, the  treatment effect for $\bx$ is $\tau^{\bx} = 0$, then applying the asymptotic theory (Proposition~\ref{prop::difference-in-mean} or Theorem 4.1 of \cite{Bugni2018}) to the difference-in-means estimator $ \XT - \XC $, we have $  \XT - \XC = O_p( n^{-1/2} ) $ under Assumptions~\ref{assum::Q} -- \ref{assum::A2}. Note that, since $\bx_i(1) = \bx_i(0) = \bx_i$, the third term of the asymptotic variance of $ \XT - \XC $ (requiring Assumption~\ref{assum::A2}) vanishes even when $\pi \neq 1/2$. The conclusion  that $  \XT - \XC = O_p( n^{-1/2} ) $ still holds if Assumption~\ref{assum::A2} is replaced by the weaker Assumption~\ref{assum::A3}.

% to obtain the asymptotic normality result of Theorem 4.1 of \cite{Bugni2018}, the authors assumed Assumptions~\ref{assum::Q} -- \ref{assum::A2}. However, when we carefully examined the proof of Theorem 4.1, we found that $  \XT - \XC = O_p( n^{-1/2} ) $ still holds if Assumption~\ref{assum::A2} is replaced by the weaker Assumption~\ref{assum::A3}.

By Lemma~\ref{lem::proportion}, $\nc / n \xrightarrow{P} 1 - \pi$, therefore,
$$
\XT - \bar{\bx}  = O_p( n^{-1/2} ).
$$
Similarly,
$$
\XC - \bar{\bx}  = O_p( n^{-1/2} ).
$$

%S_{\bx Y} = \frac{1}{n} \sum_{i=1}^{n} (\bx_i - \bar{\bx} ) ( Y_i -  \bar{Y}), \quad S_{\bx \bx } = \frac{1}{n} \sumi (\bx_i - \bar{\bx} ) ( \bx_i - \bar{\bx})^\T,

\subsection{Proof of Lemma~\ref{lem::betaancova}}
\begin{proof}
Since the covariates $\bx$ are not affected by the treatment assignment, the treatment effect for $\bx$ is $\tau^{\bx} = 0$, then applying the asymptotic theory (Proposition~\ref{prop::fe}) to the fixed effect estimator $ \hat \tau_{\fe}^{\bx}$, we have $ \hat \tau_{\fe}^{\bx} = O_p( n^{-1/2} ) $. Again, note that, to obtain the $O_p$ result, we only require Assumption~\ref{assum::A3} instead of Assumption~\ref{assum::A2} since the third term of the asymptotic variance of $  \hat \tau_{\fe}^{\bx} $ vanishes. Applying the strong low of large number,  
$$
\frac{1}{n} \sumk \sumik x_{ij} x_{ik}  \xrightarrow{P} E ( x_{ij} x_{ik} ), \quad  \frac{1}{n} \sumi \bx_i I_{i \in [k]}  \xrightarrow{P} E (  \bx_i I_{i \in [k]} ),
$$
thus,
$$
\frac{1}{n} \sumk \sumik \bx_i  \bx_i ^\T  \xrightarrow{P} E ( \bx \bx ^\T ),
$$
and
$$
\Xkbar = \frac{1}{\nk } \sumik \bx_i  = \frac{n}{\nk} \frac{1}{n} \sumi \bx_i I_{i \in [k]}   \xrightarrow{P}  \frac{ E (  \bx_i I_{i \in [k]} ) }{\pk} = \mu_{ [k] \bx } .
$$
Therefore
\begin{eqnarray}
\hat S^{\fe}_{\bx \bx} & = & \frac{1}{n} \sumk \sumik ( \bx_i - \Xkbar ) ( \bx_i - \Xkbar ) ^\T  \nonumber \\
& = & \frac{1}{n} \sumk \sumik \bx_i  \bx_i ^\T - \sumk \frac{\nk}{n} \Xkbar  \Xkbar ^\T \nonumber \\
&  \xrightarrow{P} & E ( \bx \bx ) - \sumk \pk \mu_{ [k] \bx }  \mu_{ [k] \bx }^\T  \nonumber \\
& = & E \{ \bx - E ( \bx | B ) \}  \{ \bx - E ( \bx | B ) \}^\T = \Sigma_{\tilde \bx \tilde \bx}. \nonumber
\end{eqnarray} 
For $\hat S^{\fe}_{\bx Y}$, by  Lemma~\ref{lem::b3},
\begin{eqnarray}
\frac{1}{n} \sumi \bx_i Y_i =   \frac{1}{n} \sumi A_i \bx_i Y_i(1) + \frac{1}{n} \sumi ( 1 - A_i ) \bx_i Y_i(0)  \xrightarrow{P}    \pi E \{ \bx Y(1) \} + ( 1 - \pi ) E \{ \bx Y(0) \}, \nonumber 
\end{eqnarray}
\begin{eqnarray}
\Ykbar & = & \frac{1}{\nk } \sumik Y_i  = \frac{n}{\nk} \frac{1}{n} \sumi A_i Y_i(1)  I_{i \in [k]} +  \frac{n}{\nk} \frac{1}{n} \sumi ( 1 - A_i ) Y_i(0)  I_{i \in [k]} \nonumber \\
& \xrightarrow{P} & \pi  E \{ Y(1) | B = k \} + ( 1 - \pi )  E \{ Y(0) | B = k \}. \nonumber 
\end{eqnarray}
Therefore,
\begin{eqnarray}
\hat S^{\fe}_{\bx Y} & = & \frac{1}{n} \sumi \bx_i Y_i - \sumk \frac{\nk}{n} \Xkbar \Ykbar \nonumber \\
& \xrightarrow{P} &   \pi E \{ \bx Y(1) \} + ( 1 - \pi ) E \{ \bx Y(0) \} -  \pi E ( \bx | B )  E \{ Y_i(1) | B \} - ( 1 - \pi ) E ( \bx | B )   E \{ Y_i(0) | B \}     \nonumber \\
& = & \pi E \{ \bx - E ( \bx | B ) \}  [ Y(1) - E \{ Y(1) | B \} ] + ( 1 - \pi )  E \{ \bx - E ( \bx | B ) \}  [ Y(0) - E \{ Y(0) | B \} ] \nonumber \\
& = & \pi \Sigma_{\tilde \bx  \tilde Y(1)} + ( 1 - \pi ) \Sigma_{\tilde \bx  \tilde Y(0)}. \nonumber
\end{eqnarray}
Finally, from the formula of $\taufe$ and applying Lemma~\ref{lem::b3}, we can obtain $\taufe = O_p(1)$. Together with $\hat \tau^{\bx}_{\fe} = O_p(n^{-1/2}) $ and 
$$
\sumk \pik ( 1 - \pik ) \pnk  \xrightarrow{P} \sumk \pi ( 1 - \pi ) \pk = \pi ( 1 - \pi ),
$$
we have
$$
\hat \beta_{\adj} -  \beta_{\ancova} = o_p(1),
$$
where 
$$  
\beta_{\ancova} = \pi \beta(1) + ( 1 - \pi ) \beta(0) = \pi \Sigma_{\tilde \bx \tilde \bx}^{-1} \Sigma_{\tilde \bx \tilde Y(1)} + ( 1 - \pi ) \Sigma_{\tilde \bx \tilde \bx}^{-1} \Sigma_{\tilde \bx \tilde Y(0)}  .
$$

\end{proof}

\subsection{Proof of Lemma~\ref{lem::gamma}}
\begin{proof}
It is enough to show that 
$$
S_{\tilde \bx  \tilde \bx}(a) \xrightarrow{P} \Sigma_{\tilde \bx \tilde \bx}, \quad S_{\tilde \bx  Y(a)} \xrightarrow{P} \Sigma_{\tilde \bx Y(a)}, \quad a= 0, 1.
$$
By definition
\begin{eqnarray}
S_{\tilde \bx \tilde \bx  }(1) & = &  \frac{1}{\nt} \sumk \sumik A_i ( \bx_i - \XkThat ) ( \bx_i - \XkThat )^\T \nonumber \\
& = &  \frac{1}{\nt} \sumk \sumik A_i  \bx_i \bx_i^\T -  \frac{1}{\nt} \sumk  \nkt ( \XkThat ) ( \XkThat )^\T. \nonumber
\end{eqnarray}
Applying Lemma~\ref{lem::b3} to each element of the matrix $\bx_i \bx_i^\T$, we have
$$
 \frac{1}{\nt} \sumk \sumik A_i  \bx_i \bx_i^\T  = \frac{n}{\nt} \frac{1}{n} \sumk \sumik A_i  \bx_i \bx_i^\T \xrightarrow{P} E ( \bx \bx^\T ).
$$
Moreover,
\begin{eqnarray}
\XkThat = \frac{1}{\nkt} \sumi  A_i I_{i \in [k]} \bx_i = \frac{n}{\nkt} \frac{1}{n} \sumi  A_i I_{i \in [k]} \bx_i  \xrightarrow{P} \frac{1 }{ \pi \pk } \pi E ( \bx_i I_{i \in [k]} ) = E ( \bx_i | B_i = k ). \nonumber
\end{eqnarray}
Since $\nkt / \nt \xrightarrow{P}  \pk$, thus,
$$
 \frac{1}{\nt} \sumk  \nkt ( \XkThat ) ( \XkThat )^\T \xrightarrow{P}  \sumk \pk E ( \bx_i | B_i = k ) E ( \bx_i^\T | B_i = k ) = E \{ E ( \bx | B  ) E ( \bx^\T | B )  \}.
$$
Recall that $\tilde \bx  = \bx - E( \bx | B )$, therefore, 
$$
S_{\tilde \bx \tilde \bx  }(1) \xrightarrow{P}  E ( \bx \bx^\T ) - E \{ E ( \bx | B  ) E ( \bx^\T | B )  \} = E ( \tilde \bx \tilde \bx^\T ) =  \Sigma_{\tilde \bx \tilde \bx}.
$$
Similarly, we can prove that
$$
S_{\tilde \bx \tilde \bx  }(0) \xrightarrow{P} \Sigma_{\tilde \bx \tilde \bx},  \quad S_{\tilde \bx  \tilde Y(1)} \xrightarrow{P} \Sigma_{\tilde \bx \tilde Y(1)}, \quad  S_{\tilde \bx  \tilde Y(0)} \xrightarrow{P} \Sigma_{\tilde \bx \tilde Y(0)}.
$$

\end{proof}

\section{Additional Simulation Results}

Table~\ref{equalappendix} and Table~\ref{unequalappendix} present the simulation results obtained under stratified randomization with Efron's biased-coin design \citep{Efron1971} and Wei's urn design  \citep{Wei1978}, and the covariate-adaptive randomization proposed by \citet{Hu2012}. The data-generating models are the same as those described in section~\ref{SimStudy} of the main text. First, the six treatment-effect estimators still have small finite-sample biases, but $\hat{\tau}$ and $\tausimp$ appear to have larger biases with unequal allocation. Second, the standard deviations of $\hat{\tau}, \taufe$, and $\taustr$ are almost identical under a stratified biased-coin design and the randomization method proposed by \citet{Hu2012}, both of which achieve strong balance. For randomization methods that satisfy Assumption~\ref{assum::A2}, we can see that the larger is $q_{[k]}$, the larger are the standard deviations of $\hat{\tau}$. More precisely, the standard deviations of $\hat{\tau}$ under complete randomization and the stratified Wei's urn design are $2.27\%-69.41\%$ and $0.57\%-30.59\%$, respectively, which are larger than those obtained using the stratified biased-coin design and stratified block randomization. Other relationships between the six estimators are consistent with those reported in the main text. Third, the OLS and Huber--White variance estimators are only valid for $\taufe$ and $\tauancova$ under equal allocation, whereas the proposed variance estimators are always valid and exhibit the expected $95\%$ coverage probability under all scenarios.

\begin{table}[H]
	\centering
	\caption{Simulated biases, standard deviations, standard errors, and coverage probabilities for different estimators and randomization methods under equal allocation ($\pi = 1/2$)}\label{equalappendix}
	\vskip 5mm
	\begin{threeparttable}
		\setlength{\tabcolsep}{2pt}
\resizebox{\textwidth}{55mm}{
		\begin{tabular}{llcccccccccccccccccccccccc}
		\cline{1-26}
		&  & \multicolumn{8}{c}{Stratified Biased-Coin Design} & \multicolumn{8}{c}{Stratified Wei's Urn Design} & \multicolumn{8}{c}{Hu and Hu's Randomization} \\ \cline{3-26}
		&
		&
		\multicolumn{1}{c}{Bias} &
		\multicolumn{1}{c}{SD} &
		\multicolumn{3}{c}{SE} &
        \multicolumn{3}{c}{CP} &
		\multicolumn{1}{c}{Bias} &
		\multicolumn{1}{c}{SD} &
		\multicolumn{3}{c}{SE} &
        \multicolumn{3}{c}{CP} &
		\multicolumn{1}{c}{Bias} &
		\multicolumn{1}{c}{SD} &
		\multicolumn{3}{c}{SE} &
        \multicolumn{3}{c}{CP}  \\ \cline{5-7} \cline{8-10} \cline{13-15} \cline{16-18} \cline{21-23} \cline{24-26}
		Model &
		Estimator &
		\multicolumn{1}{c}{} &
		\multicolumn{1}{c}{} &
		\multicolumn{1}{c}{NEW} &
		\multicolumn{1}{c}{OLS} &
		\multicolumn{1}{c}{HW} &
		\multicolumn{1}{c}{NEW} &
		\multicolumn{1}{c}{OLS} &
		\multicolumn{1}{c}{HW} &
		\multicolumn{1}{c}{} &
		\multicolumn{1}{c}{} &
		\multicolumn{1}{c}{NEW} &
		\multicolumn{1}{c}{OLS} &
		\multicolumn{1}{c}{HW} &
		\multicolumn{1}{c}{NEW} &
		\multicolumn{1}{c}{OLS} &
		\multicolumn{1}{c}{HW} &
		\multicolumn{1}{c}{} &
		\multicolumn{1}{c}{} &
		\multicolumn{1}{c}{NEW} &
		\multicolumn{1}{c}{OLS} &
		\multicolumn{1}{c}{HW} &
		\multicolumn{1}{c}{NEW} &
		\multicolumn{1}{c}{OLS} &
		\multicolumn{1}{c}{HW}  \\ \hline
1&$\hat{\tau}$ & 0.00 & 0.85 & 0.83 & 1.01 & 1.01 & 0.94 & 0.98 & 0.98 & -0.01 & 0.89 & 0.90 & 1.02 & 1.02 & 0.95 & 0.98 & 0.98 & -0.07 & 0.85 & 0.83 & 1.02 & 1.01 & 0.95 & 0.98 & 0.98 \\
  &$\taufe$ & 0.00 & 0.85 & 0.83 & 0.83 & 0.83 & 0.94 & 0.95 & 0.94 & -0.00 & 0.82 & 0.83 & 0.84 & 0.83 & 0.96 & 0.96 & 0.96 & -0.07 & 0.85 & 0.83 & 0.84 & 0.83 & 0.95 & 0.96 & 0.95 \\
  &$\taustr$ & 0.00 & 0.85 & 0.83 & 0.83 & 0.82 & 0.94 & 0.94 & 0.94 & -0.00 & 0.82 & 0.83 & 0.84 & 0.83 & 0.96 & 0.96 & 0.96 & -0.07 & 0.85 & 0.83 & 0.84 & 0.83 & 0.95 & 0.95 & 0.95 \\
  &$\tausimp$ & 0.01 & 0.42 & 0.40 & 0.71 & 0.70 & 0.94 & 1.00 & 1.00 & -0.02 & 0.52 & 0.52 & 0.71 & 0.70 & 0.95 & 0.99 & 0.99 & -0.01 & 0.41 & 0.40 & 0.71 & 0.70 & 0.95 & 1.00 & 1.00 \\
  &$\tauancova$ & 0.01 & 0.41 & 0.40 & 0.41 & 0.40 & 0.94 & 0.95 & 0.94 & -0.00 & 0.40 & 0.40 & 0.41 & 0.40 & 0.95 & 0.95 & 0.95 & -0.02 & 0.41 & 0.40 & 0.41 & 0.40 & 0.94 & 0.95 & 0.95 \\
  &$\tauinter$ & 0.01 & 0.41 & 0.40 & 0.41 & 0.40 & 0.94 & 0.95 & 0.94 & -0.00 & 0.40 & 0.40 & 0.41 & 0.40 & 0.95 & 0.95 & 0.95 & -0.02 & 0.41 & 0.40 & 0.41 & 0.40 & 0.94 & 0.95 & 0.94 \\
 
  2&$\hat{\tau}$ & -0.01 & 0.84 & 0.84 & 1.12 & 1.12 & 0.95 & 0.99 & 0.99 & -0.02 & 0.94 & 0.94 & 1.12 & 1.12 & 0.95 & 0.98 & 0.98 & -0.02 & 0.87 & 0.84 & 1.12 & 1.12 & 0.94 & 0.99 & 0.99 \\
  &$\taufe$ & -0.02 & 0.84 & 0.84 & 0.84 & 0.84 & 0.95 & 0.95 & 0.95 & -0.04 & 0.83 & 0.84 & 0.84 & 0.84 & 0.95 & 0.95 & 0.95 & -0.03 & 0.85 & 0.84 & 0.84 & 0.84 & 0.94 & 0.94 & 0.94 \\
  &$\taustr$ & -0.01 & 0.84 & 0.84 & 0.41 & 0.41 & 0.95 & 0.67 & 0.66 & -0.02 & 0.84 & 0.84 & 0.41 & 0.41 & 0.95 & 0.67 & 0.67 & -0.02 & 0.85 & 0.84 & 0.41 & 0.41 & 0.94 & 0.66 & 0.66 \\
  &$\tausimp$ & 0.00 & 0.82 & 0.82 & 1.09 & 1.09 & 0.96 & 0.99 & 0.99 & -0.01 & 0.92 & 0.92 & 1.09 & 1.09 & 0.94 & 0.98 & 0.98 & -0.01 & 0.86 & 0.82 & 1.09 & 1.09 & 0.94 & 0.99 & 0.99 \\
  &$\tauancova$ & 0.01 & 0.82 & 0.82 & 0.83 & 0.82 & 0.95 & 0.95 & 0.95 & -0.02 & 0.82 & 0.82 & 0.83 & 0.83 & 0.95 & 0.96 & 0.96 & -0.01 & 0.84 & 0.82 & 0.83 & 0.82 & 0.94 & 0.95 & 0.95 \\
  &$\tauinter$ & -0.00 & 0.82 & 0.82 & 0.35 & 0.35 & 0.95 & 0.60 & 0.60 & -0.02 & 0.82 & 0.82 & 0.35 & 0.35 & 0.95 & 0.61 & 0.61 & -0.01 & 0.84 & 0.82 & 0.35 & 0.35 & 0.95 & 0.59 & 0.59 \\
  3&$\hat{\tau}$ & -0.04 & 1.92 & 1.94 & 2.00 & 2.00 & 0.96 & 0.96 & 0.96 & -0.04 & 1.94 & 1.96 & 2.00 & 2.00 & 0.95 & 0.95 & 0.96 & -0.01 & 1.90 & 1.94 & 2.00 & 2.00 & 0.95 & 0.95 & 0.95 \\
  &$\taufe$ & -0.04 & 1.92 & 1.94 & 1.94 & 1.94 & 0.96 & 0.96 & 0.96 & -0.02 & 1.92 & 1.94 & 1.94 & 1.94 & 0.95 & 0.95 & 0.95 & -0.01 & 1.89 & 1.94 & 1.94 & 1.94 & 0.95 & 0.95 & 0.95 \\
  &$\taustr$ & -0.04 & 1.92 & 1.94 & 1.77 & 1.76 & 0.96 & 0.94 & 0.94 & -0.03 & 1.92 & 1.94 & 1.76 & 1.76 & 0.95 & 0.92 & 0.92 & -0.01 & 1.89 & 1.94 & 1.76 & 1.76 & 0.95 & 0.92 & 0.92 \\
  &$\tausimp$ & -0.09 & 1.49 & 1.49 & 1.54 & 1.54 & 0.95 & 0.96 & 0.95 & -0.04 & 1.48 & 1.50 & 1.54 & 1.54 & 0.95 & 0.96 & 0.95 & -0.03 & 1.47 & 1.48 & 1.54 & 1.54 & 0.95 & 0.96 & 0.95 \\
  &$\tauancova$ & -0.09 & 1.49 & 1.49 & 1.49 & 1.49 & 0.95 & 0.95 & 0.95 & -0.02 & 1.44 & 1.48 & 1.49 & 1.49 & 0.95 & 0.95 & 0.96 & -0.03 & 1.47 & 1.48 & 1.49 & 1.49 & 0.95 & 0.95 & 0.95 \\
  &$\tauinter$ & -0.06 & 1.49 & 1.49 & 0.70 & 0.70 & 0.95 & 0.65 & 0.65 & -0.00 & 1.44 & 1.48 & 0.70 & 0.70 & 0.96 & 0.65 & 0.65 & -0.01 & 1.47 & 1.48 & 0.70 & 0.70 & 0.95 & 0.67 & 0.67 \\
   \hline
		\end{tabular}}
\begin{tablenotes}
\item Note: SD, standard deviation; SE, standard error; CP, coverage probability; HW,  the \\ Huber--White  variance estimator; NEW: the proposed non-parametric variance estimator.
%\item Abbreviation: NA, not available.
\end{tablenotes}
\end{threeparttable}
\end{table}

\begin{table}[H]
	\centering
	\caption{Simulated biases, standard deviations, standard errors, and coverage probabilities for different estimators and randomization methods under unequal allocation ($\pi = 2/3$)}\label{unequalappendix}
	\vskip 5mm
	\begin{threeparttable}
		\setlength{\tabcolsep}{2pt}
\resizebox{\textwidth}{55mm}{
		\begin{tabular}{llcccccccccccccccccccccccc}
		\cline{1-26}
		&  & \multicolumn{8}{c}{Stratified Biased-Coin Design} & \multicolumn{8}{c}{Stratified Wei's Urn Design} & \multicolumn{8}{c}{Hu and Hu's Randomization} \\ \cline{3-26}
		&
		&
		\multicolumn{1}{c}{Bias} &
		\multicolumn{1}{c}{SD} &
		\multicolumn{3}{c}{SE} &
        \multicolumn{3}{c}{CP} &
		\multicolumn{1}{c}{Bias} &
		\multicolumn{1}{c}{SD} &
		\multicolumn{3}{c}{SE} &
        \multicolumn{3}{c}{CP} &
		\multicolumn{1}{c}{Bias} &
		\multicolumn{1}{c}{SD} &
		\multicolumn{3}{c}{SE} &
        \multicolumn{3}{c}{CP}  \\ \cline{5-7} \cline{8-10} \cline{13-15} \cline{16-18} \cline{21-23} \cline{24-26}
		Model &
		Estimator &
		\multicolumn{1}{c}{} &
		\multicolumn{1}{c}{} &
		\multicolumn{1}{c}{NEW} &
		\multicolumn{1}{c}{OLS} &
		\multicolumn{1}{c}{HW} &
		\multicolumn{1}{c}{NEW} &
		\multicolumn{1}{c}{OLS} &
		\multicolumn{1}{c}{HW} &
		\multicolumn{1}{c}{} &
		\multicolumn{1}{c}{} &
		\multicolumn{1}{c}{NEW} &
		\multicolumn{1}{c}{OLS} &
		\multicolumn{1}{c}{HW} &
		\multicolumn{1}{c}{NEW} &
		\multicolumn{1}{c}{OLS} &
		\multicolumn{1}{c}{HW} &
		\multicolumn{1}{c}{} &
		\multicolumn{1}{c}{} &
		\multicolumn{1}{c}{NEW} &
		\multicolumn{1}{c}{OLS} &
		\multicolumn{1}{c}{HW} &
		\multicolumn{1}{c}{NEW} &
		\multicolumn{1}{c}{OLS} &
		\multicolumn{1}{c}{HW}  \\ \hline
1&$\hat{\tau}$ & -0.03 & 0.91 & 0.88 & 1.07 & 1.06 & 0.94 & 0.98 & 0.98 & -0.03 & 0.91 & 0.95 & 1.07 & 1.07 & 0.96 & 0.98 & 0.98 & -0.02 & 0.88 & 0.87 & 1.08 & 1.07 & 0.95 & 0.98 & 0.98 \\
  &$\taufe$ & -0.01 & 0.89 & 0.88 & 0.88 & 0.87 & 0.95 & 0.95 & 0.95 & -0.04 & 0.83 & 0.87 & 0.89 & 0.87 & 0.96 & 0.96 & 0.96 & -0.00 & 0.88 & 0.87 & 0.89 & 0.88 & 0.95 & 0.96 & 0.96 \\
  &$\taustr$ & -0.01 & 0.89 & 0.88 & 0.88 & 0.86 & 0.95 & 0.95 & 0.94 & -0.04 & 0.83 & 0.87 & 0.89 & 0.87 & 0.96 & 0.96 & 0.95 & -0.00 & 0.88 & 0.87 & 0.89 & 0.87 & 0.95 & 0.96 & 0.95 \\
  &$\tausimp$ & -0.03 & 0.44 & 0.42 & 0.75 & 0.74 & 0.93 & 1.00 & 1.00 & 0.02 & 0.54 & 0.56 & 0.75 & 0.74 & 0.96 & 0.99 & 0.99 & -0.02 & 0.45 & 0.42 & 0.75 & 0.74 & 0.93 & 1.00 & 1.00 \\
  &$\tauancova$ & -0.01 & 0.40 & 0.42 & 0.43 & 0.42 & 0.96 & 0.97 & 0.96 & 0.01 & 0.42 & 0.42 & 0.43 & 0.42 & 0.95 & 0.96 & 0.95 & -0.00 & 0.43 & 0.42 & 0.44 & 0.42 & 0.94 & 0.95 & 0.94 \\
  &$\tauinter$ & -0.01 & 0.40 & 0.42 & 0.43 & 0.41 & 0.96 & 0.97 & 0.96 & 0.01 & 0.42 & 0.42 & 0.44 & 0.41 & 0.95 & 0.96 & 0.95 & -0.00 & 0.43 & 0.42 & 0.44 & 0.42 & 0.94 & 0.95 & 0.93 \\
  2&$\hat{\tau}$ & -0.15 & 0.91 & 0.88 & 1.29 & 1.05 & 0.94 & 0.99 & 0.98 & -0.03 & 0.91 & 0.95 & 1.30 & 1.06 & 0.96 & 0.99 & 0.98 & -0.05 & 0.91 & 0.88 & 1.30 & 1.06 & 0.95 & 1.00 & 0.98 \\
  &$\taufe$ & 0.03 & 0.89 & 0.88 & 0.82 & 1.01 & 0.95 & 0.93 & 0.97 & 0.05 & 0.91 & 0.94 & 0.82 & 1.02 & 0.96 & 0.92 & 0.97 & -0.00 & 0.91 & 0.88 & 0.82 & 1.02 & 0.94 & 0.92 & 0.97 \\
  &$\taustr$ & -0.01 & 0.89 & 0.88 & 0.36 & 0.49 & 0.96 & 0.56 & 0.71 & 0.03 & 0.86 & 0.89 & 0.36 & 0.49 & 0.95 & 0.58 & 0.74 & -0.00 & 0.89 & 0.88 & 0.36 & 0.49 & 0.95 & 0.57 & 0.72 \\
  &$\tausimp$ & -0.11 & 0.89 & 0.86 & 1.27 & 1.00 & 0.94 & 1.00 & 0.98 & -0.01 & 0.87 & 0.92 & 1.28 & 1.01 & 0.96 & 0.99 & 0.97 & -0.04 & 0.88 & 0.86 & 1.28 & 1.01 & 0.95 & 1.00 & 0.98 \\
  &$\tauancova$ & 0.05 & 0.87 & 0.86 & 0.81 & 0.99 & 0.95 & 0.93 & 0.98 & 0.06 & 0.88 & 0.92 & 0.81 & 1.00 & 0.96 & 0.93 & 0.98 & 0.01 & 0.89 & 0.86 & 0.81 & 1.00 & 0.95 & 0.93 & 0.97 \\
  &$\tauinter$ & -0.01 & 0.86 & 0.86 & 0.31 & 0.42 & 0.95 & 0.50 & 0.65 & 0.03 & 0.82 & 0.86 & 0.31 & 0.42 & 0.96 & 0.53 & 0.69 & 0.00 & 0.86 & 0.86 & 0.31 & 0.43 & 0.95 & 0.52 & 0.66 \\
  3&$\hat{\tau}$ & -0.00 & 1.76 & 1.75 & 2.38 & 1.80 & 0.95 & 0.99 & 0.96 & 0.01 & 1.80 & 1.77 & 2.39 & 1.80 & 0.95 & 0.99 & 0.95 & -0.01 & 1.78 & 1.75 & 2.39 & 1.80 & 0.95 & 0.99 & 0.96 \\
  &$\taufe$ & -0.00 & 1.76 & 1.75 & 2.27 & 1.85 & 0.95 & 0.99 & 0.97 & 0.03 & 1.82 & 1.79 & 2.28 & 1.85 & 0.95 & 0.98 & 0.96 & -0.00 & 1.77 & 1.75 & 2.28 & 1.85 & 0.95 & 0.98 & 0.96 \\
  &$\taustr$ & 0.05 & 1.76 & 1.75 & 2.13 & 1.56 & 0.95 & 0.98 & 0.92 & 0.04 & 1.79 & 1.76 & 2.14 & 1.56 & 0.94 & 0.98 & 0.91 & 0.02 & 1.77 & 1.75 & 2.13 & 1.55 & 0.94 & 0.98 & 0.93 \\
  &$\tausimp$ & -0.01 & 1.62 & 1.62 & 1.71 & 1.63 & 0.94 & 0.96 & 0.95 & -0.09 & 1.61 & 1.63 & 1.71 & 1.64 & 0.96 & 0.96 & 0.96 & 0.04 & 1.62 & 1.62 & 1.71 & 1.64 & 0.94 & 0.96 & 0.95 \\
  &$\tauancova$ & -0.04 & 1.62 & 1.62 & 1.56 & 1.71 & 0.95 & 0.94 & 0.96 & -0.08 & 1.65 & 1.66 & 1.57 & 1.72 & 0.95 & 0.94 & 0.96 & 0.03 & 1.63 & 1.62 & 1.56 & 1.71 & 0.95 & 0.93 & 0.96 \\
  &$\tauinter$ & 0.05 & 1.45 & 1.45 & 0.84 & 0.62 & 0.95 & 0.73 & 0.59 & -0.01 & 1.45 & 1.45 & 0.85 & 0.62 & 0.95 & 0.76 & 0.62 & 0.06 & 1.46 & 1.45 & 0.84 & 0.61 & 0.94 & 0.76 & 0.59 \\
   \hline
		\end{tabular}}
\begin{tablenotes}
\item Note: SD, standard deviation; SE, standard error; CP, coverage probability; HW, the \\ Huber--White  variance estimator; NEW: the proposed non-parametric variance estimator.
%\item Abbreviation: NA, not available.
\end{tablenotes}
\end{threeparttable}
\end{table}

\section{Synthetic Data of Nefazodone CBASP Trial}
To generate the synthetic data, we first fit a non-parametric spline using the function bigssa in the R package bigspline with six selected covariates: AGE, HAMD17, HAMD24, HAMD\_COGIN, Mstatus2 and TreatPD, which are detailed in Table~\ref{tab:description}. The fitted model can be loaded from the file spline.RData.

\begin{table}[H]
\centering
\caption{Description of selected covariates}\label{tab:description}
\begin{tabular}{l|l}
\hline
 Variable & Description \\
 \hline
 AGE  & Age of patients in years \\
 HAMD17 & Total HAMD-17 score \\
 HAMD24  & Total HAMD-24 score \\
 HAMD\_COGIN & HAMD cognitive disturbance score  \\
 Mstatus2 & Marriage status: 1 if married or living with someone and 0 otherwise \\
 TreatPD & Treated past depression: 1 yes and 0 no \\
 \hline
\end{tabular}
\begin{tablenotes}
\item Note: HAMD, Hamilton Rating Scale for Depression.
\end{tablenotes}
\end{table}

Then, we implement simple randomization and stratified block randomization to obtain the treatment assignments for both the equal ($\pi=1/2$) and unequal ($\pi=2/3$) allocations. Among the six covariates, we use the stratified HAMD17 and Mstatus2 in stratified block randomization, where the stratified HAMD17 is determined by the relative values of 18 and 21. Once the treatment assignments are produced, we generate patient outcomes through the fitted model. For data analysis, AGE, HAMD24, HAMD\_COGIN, and TreatPD are used as the additional covariates.

\end{singlespace}

\end{document}